\documentstyle[aps,rmp,psfig]{revtex}
\def\pmb#1{\setbox0=\hbox{$#1$}%
\kern-.025em\copy0\kern-\wd0
\kern.05em\copy0\kern-\wd0
\kern-.025em\raise.0433em\box0}
\begin{document}
\title{Modern optical astronomy: technology and impact of interferometry}
\author{Swapan K Saha} 
\address{Indian Institute of Astrophysics, Koramangala, Bangalore-560 034, India.}
%e-mail: sks@iiap.ernet.in
\maketitle
 
\begin{abstract}
The present `state of the art' and the path to future progress in high spatial 
resolution imaging interferometry is reviewed. The review begins with a
treatment of the fundamentals of stellar optical interferometry, the origin, 
properties, optical effects of turbulence in the Earth's atmosphere, the
passive methods that are applied on a single telescope to overcome 
atmospheric image degradation such as speckle interferometry, and various other 
techniques. These topics include differential speckle interferometry, speckle 
spectroscopy and polarimetry, phase diversity, wavefront shearing 
interferometry, phase-closure methods, dark speckle imaging, as well as the 
limitations imposed by the detectors on the performance of speckle imaging. A 
brief account is given of the technological innovation of adaptive-optics (AO) 
to compensate such atmospheric effects on the image in real time. A
major advancement involves the transition from single-aperture to the 
dilute-aperture interferometry using multiple telescopes. Therefore, the 
review deals with recent developments involving 
ground-based, and space-based optical arrays. Emphasis is placed on the problems
specific to delay-lines, beam recombination, polarization, dispersion,
fringe-tracking, bootstrapping, coherencing and cophasing, and recovery of the 
visibility functions. The role of AO in enhancing visibilities is also 
discussed. The applications of interferometry, such as imaging, astrometry, and 
nulling are described. The mathematical intricacies of the various 
`post-detection' image-processing techniques are examined critically. The 
review concludes with a discussion of the astrophysical importance and the 
perspectives of interferometry. 
\end{abstract}
%\narrowtext

\tableofcontents

\section{INTRODUCTION}
 
\label{sec:introduc}
 
Optical interferometry provides physicists and astronomers with an 
exquisite set of probes of the micro and macrocosmos.  From the 
laboratory to the observatory over the past few decades, there has 
been a surge of activity in developing new tools for ground-based 
optical astronomy, of which interferometry is one of the most 
powerful.

An optical interferometer is a device that combines two or more light 
waves emitted from the same source at the same time to produce 
interference fringes.  The implementation of interferometry in 
optical astronomy began more than a century ago with the work of 
Fizeau (1868).  Michelson and Pease (1921) measured successfully the 
angular diameter of ($\alpha$~Ori),
using an interferometer based on two flat mirrors, which allowed them 
to measure the fringe visibility in the interference pattern formed 
by starlight at the detector plane.  However, progress was hindered 
by the severe image degradation produced by atmospheric turbulence in 
the optical spectrum.  The field remained dormant until the 
development of intensity interferometry by Hanbury Brown and Twiss 
(1958), a technique that employs two adjacent sets of mirrors and 
photoelectric correlation.

Turbulence and the concomitant development of thermal convection in 
the atmosphere distort the phase and amplitude of an incoming 
wavefront of starlight.  The longer the path, the greater the 
degradation that the image suffers.  Light  reaching the entrance 
pupil of an imaging system is coherent only within patches of 
diameters of order r$_\circ$ , Fried's parameter (Fried, 1966).  This 
limited coherence causes blurring of the image, blurring that is 
modeled by a convolution with the point-spread function (PSF).  Both 
the sharpness of astronomical images and the signal-to-noise (S/N) 
ratio (hence faintness of the objects that can be studied) depend on 
angular resolution, the latter because noise comes from as much of 
the sky as is in the resolution element.  Thus reducing the beam 
width from, say, 1 arcsecond ($^{\prime\prime}$) to 0.5$^{\prime\prime}$ 
reduces sky noise by a 
factor of 4.  Two physical phenomena limit the minimum resolvable 
angle at optical and infrared (IR) wavelengths - diameter of the collecting 
area and turbulence in the atmosphere.  The crossover between 
domination by aperture size ($\lambda$/aperture diameter) and domination by 
atmospheric turbulence (`seeing') occurs when the aperture becomes 
larger than the size of a characteristic turbulent element.

The image of a star obtained through a large telescope looks 
`speckled' or grainy because different parts of the image are blurred 
by small areas of turbulence in the earth's atmosphere.  Labeyrie 
(1970) proposed speckle interferometry (SI), a process that deciphers 
the diffraction-limited Fourier spectrum and image features of 
stellar objects by taking a large number of very-short-exposure 
images of the same field.  Computer assistance is then used to 
reconstruct from these many images a single image that is free of 
turbulent areas-in essence, an image of the object as it might appear 
from space.

The success of speckle interferometry in measuring the diameters of 
stars encouraged astronomers to develop further image-processing 
techniques.  These techniques are, for the most part, post-detection 
processes.  Recent advances in technology have produced the hardware 
to compensate for wave-front distortion in real time.  Adaptive 
optics (AO; Babcock, 1953; Rousset et al., 1990) is based on this 
hardware-oriented approach, which sharpens telescope images blurred 
by the earth's atmosphere.  It employs a combination of deformation 
of reflecting surfaces (i.e., flexible mirrors) and post-detection 
image restoration (Roddier, 1999).  One of its most successful 
applications has been in imaging of Neptune's ring arcs. 
Adaptive optical imaging systems have been treated in depth by the 
review of Roggemann et al. (1997), which includes discussion of 
wavefront compensation devices, wavefront sensors, control systems, 
performance of AO systems, and representative experimental results. 
It also deals with the characterization of atmospheric turbulence, 
the SI technique, and deconvolution techniques for wavefront sensing.

Long-baseline optical interferometry (LBOI) uses two or more optical 
telescopes to synthesize the angular resolution of a much larger 
aperture (aperture synthesis) than would be possible with a single 
mirror.  Labeyrie (1975) extended the concept of speckle 
interferometry  to a pair of telescopes in a North-South baseline 
configuration, and subsequently astronomers have created larger 
ground-based arrays.  A few of these arrays, e.g., the Keck 
interferometer and the Very Large Telescope Interferometer (VLTI), 
employ large telescopes fitted with AO systems.  The combination of 
long-baseline interferometry, mimicking a wide aperture, and AO 
techniques to improve the images offers the best of both approaches 
and shows great promise for applications such as the search for 
extra-solar planets.  At this point it seems clear that interferometry 
and AO are complementary, and neither can reach its full potential 
without the other.

The present review addresses the aims, methods, scientific 
achievements, and future prospects of long-baseline interferometry 
(LBI) at optical and infrared wavelengths, carried out with two or 
more apertures separated by more than their own sizes.  In order to 
embark on such a subject, we first review the basic principles of 
interferometry and its applications, the theoretical aspects of SI as 
a statistical analysis of a speckle pattern, and the limitations 
imposed by the atmosphere and the detectors on the performance of 
single-aperture diffraction-limited imaging.  Other related concerns, 
such as the relationship between image-plane techniques and 
pupil-plane interferometry, phase-closure methods, and dark speckle 
imaging, will also be treated.  Adaptive optics as a pre-detection 
compensation technique is described in brief, as are the strengths 
and weaknesses of pre and post detection.

The final part of this review deals with the applications of 
multiple-telescope interferometry to imaging, astrometry, and 
nulling.  These applications entail specific problems having to do 
with delaylines, beam recombination, polarization, dispersion, fringe 
tracking, and the recovery of visibility functions.  Various image 
restoration techniques are discussed as well, with emphasis on the 
deconvolution methods used in aperture-synthesis mapping.  

\section{BASIC PRINCIPLES}
 
\label{sec:preamble}
 
Astronomical sources emit incoherent light consisting of the random
superposition of numerous successive short-lived waves sent out from many 
elementary emitters, and therefore, the optical coherence is related to the 
various forms of correlations of the random process. For a monochromatic 
wave field, the amplitude of vibration at any point is constant and the phase 
varies linearly with time. Conversely, the amplitude and 
phase in the case of quasi-monochromatic wave field, undergo irregular 
fluctuations (Born and Wolf, 1984). The rapidity of fluctuations depends on the
light crossing time of the emitting region. Interferometers based on (i) 
wavefront division (Young's experiment that is sensitive to the size and
bandwidth of the source), (ii) amplitude division (Michelson's 
interferometer) are generally used to measure spatial coherence and
temporal coherence, respectively. In what follows, some of the fundamental 
mathematical steps pertinent to the interferometry are illustrated.

\subsection{Mathematical framework}
 
\label{subsec:framework}

A monochromatic plane wave, $V{\bf (r}, t)$ at a point, ${\bf r}$, is 
expressed as, 

\begin{equation}
V{\bf (r}, t) = \Re\left\{{\it a}({\bf r})
e^{-i[2\pi\nu_\circ t - \psi_j({\bf r})]}\right\}.
\end{equation}

\noindent
Here the symbol $\Re$ is the `real part of', ${\bf r}$ the position vector of a 
point $(x, y, z)$, ${\it a}({\bf r})$ is the amplitude of the wave, $t$ the 
time, $\nu_\circ$ the frequency of the wave, $\psi_j({\bf r})$ the phase 
functions that are of the form ${\bf k\cdot r} - \delta_j$, in which ${\bf k}$
is the propagation vector, and $\delta_j$ the phase constants which specify 
the state of polarization, Denoting the complex vector function of position by 
$\Psi({\bf r}) = {\it a}({\bf r})e^{i\psi_j({\bf r})}$, 
the complex representation of the analytic signal,
${\cal U}{\bf (r}, t)$, associated with $V{\bf (r}, t)$ becomes,

\begin{eqnarray}
{\cal U}{\bf (r}, t) &=& {\it a}({\bf r})e^{-i[2\pi\nu_\circ t - \psi_j({\bf r})
]}\\  
&&= \Psi({\bf r})e^{-i2\pi\nu_\circ t}.  
\end{eqnarray}

\noindent
This complex representation is preferred for linear time invariant systems,
because the eigenfunctions of such systems are of the form 
$e^{-i\omega_\circ t}$, where $\omega_\circ = 2\pi\nu_\circ$ is the angular 
frequency. From equations (1) and (2), the relationship translates into,  

\begin{eqnarray}
V{\bf (r}, t) &=& \Re\left\{\Psi({\bf r})e^{-i\omega_\circ t}\right\} 
\nonumber\\
&&= \frac{1}{2}\left[\Psi({\bf r})e^{-i\omega_\circ t} + 
\Psi^\ast({\bf r})e^{i\omega_\circ t} \right],   
\end{eqnarray}
 
The intensity of light is defined as the time average of the amount of energy,
therefore, taking the latter over an interval much greater than the period, 
${\tt T_\circ} = 2\pi/\omega_\circ$, the intensity ${\cal I}$ at the same point 
is calculated as,

\begin{equation}
{\cal I} \propto <V^2> = \frac{1}{2}\Psi\Psi^\ast,   
\end{equation}
 
\noindent 
where $< \ >$ stands for the ensemble average of the quantity within the 
bracket and $\Psi^\ast$ represents for the complex conjugate of $\Psi$.
 
Since the complex amplitude is a constant phasor in the monochromatic case, the 
Fourier transform (FT) of the complex representation of the signal, 
${\cal U}{\bf (r}, t)$, is given by,

\begin{equation}
\widehat{\cal U}{\bf (r}, \nu) = {\it a}({\bf r})e^{i\psi}\delta(\nu - 
\nu_\circ). 
\end{equation}

\noindent 
It is equal to twice the positive part of the instantaneous spectrum,
$\widehat{V}{\bf (r}, \nu)$. In the polychromatic case, the  
complex amplitude becomes,

\begin{equation}
{\cal U}{\bf (r}, t) = 2\int_0^\infty\widehat{V}{\bf (r}, \nu) 
e^{-i2\pi\nu t}d\nu. 
\end{equation}

The disturbance produced by a real physical source is calculated by the 
integration of the monochromatic signals over an optical band pass. 
In the case of quasi-monochromatic approximation (if the width of the 
spectrum, $\Delta\nu \ll \nu_\circ$), the expression modifies as,

\begin{equation}
{\cal U}{\bf (r}, t) = |\Psi({\bf r}, t)| e^{i[\psi(t) - 
2\pi\nu_\circ t]},  
\end{equation}

\noindent
where the field is characterized by the complex amplitude, $\Psi{(\bf r}, t)$, 
{\it i.e.},  

\begin{equation}
\Psi{(\bf r}, t) = |\Psi({\bf r}, t)| e^{i\psi(t)}.
\end{equation}

\noindent 
This phasor is time dependent,
although it varies slowly with respect to $e^{-i2\pi\nu_\circ t}$.

The complex amplitude, $\Psi({\pmb \alpha})$ diffracted at angle 
${\pmb \alpha}$ in the telescope focal-plane is given by,

\begin{equation}
\Psi({\pmb \alpha}) \propto \int{\cal P}_\circ({\bf x})\Psi_\circ
({\bf x}) e^{-i2\pi{\pmb \alpha}.{\bf x}/\lambda} d{\bf x},  
\end{equation}
 
\noindent 
where ${\pmb \alpha} = ({\bf x}/{f})$ is a two-dimensional (2-d) space vector, 
${f}$ the focal length, $\Psi_\circ({\bf x})$ the complex amplitude at the 
telescope aperture, and ${\cal P}_\circ({\bf x})$ the pupil transmission 
function of the telescope aperture. For an ideal telescope, we have 
${\cal P}_\circ({\bf x}) =1$ inside the aperture and 
${\cal P}_\circ({\bf x}) = 0$ outside the aperture. In the space-invariant case,

\begin{eqnarray}
\Psi({\pmb \alpha}) & \propto & \int{\cal P}({\bf u})\Psi({\bf u})
e^{-i2\pi{\pmb \alpha}{\bf u}}d{\bf u}, \\  
&& = {\cal F}[\Psi({\bf u}) \cdot {\cal P}({\bf u})],
\end{eqnarray}

\noindent 
where ${\cal F}$ represents for the complex FT and the dimensionless variable 
${\bf u}$ is equal to ${\bf x}/\lambda$ and hence, $\Psi({\bf u})$ can be 
replaced by $\Psi_\circ(\lambda{\bf u})$ and so with ${\cal P}({\bf u})$ by 
${\cal P}_\circ(\lambda{\bf u})$. The irradiance diffracted in the direction 
${\pmb \alpha}$ is the PSF of the telescope and the atmosphere and its FT, 
$\widehat{\cal S}({\bf f})$ is the optical transfer function (OTF):

\begin{equation}
\widehat{\cal S}({\bf f}) = \int{\cal S}({\pmb \alpha}) 
e^{[-2i\pi{\pmb \alpha}.{\bf f}]} d{\pmb \alpha},
\end{equation}

\noindent 
where ${\bf f}$ is the spatial frequency expressed in radian$^{-1}$, and 
$|\widehat{\cal S}({\bf f})|$ the modulation transfer function (MTF).

\subsubsection{Convolution} 

The convolution of two functions simulates phenomena such as a blurring of a 
photograph that may be caused by various reasons, e.g., (a) poor focus, 
(b) motion of a photographer during the exposure. In such a blurred picture
each point of object is replaced by a spread function. For the 2-d incoherent 
source, the complex amplitude in the image-plane is the convolution of complex 
amplitude of the pupil plane and the pupil transmission function.

\begin{equation}
{\cal S}({\pmb \alpha}) = {\cal P}({\pmb \alpha}) \star \Psi({\pmb \alpha}).  
\end{equation}
 
\noindent 
In the Fourier plane, the effect of the convolution becomes a multiplication, 
point by point of the OTF of the pupil, ${\cal P}({\bf u})$, with 
the transform of the object $\Psi({\bf u})$. i.e.,
 
\begin{equation}
\widehat{\cal S}({\bf f}) = {\cal P}({\bf u}) \cdot \Psi({\bf u}).
\end{equation}

The illumination at the focal-plane of the telescope observed as a function of
image-plane is,

\begin{eqnarray}
{\cal S}({\pmb \alpha}) &=& <\Psi({\pmb \alpha}, t)\Psi^\ast({\pmb \alpha}, t)>,
 \\
&& \propto |{\cal F}[\Psi({\bf u}) \cdot \Psi^\ast({\bf u})]|^2.
\end{eqnarray}

\noindent 
The autocorrelation of this function, ${\cal S}({\pmb \alpha})$, is expressed 
as,

\begin{equation}
{\cal F}[{\cal S}({\pmb \alpha}) \otimes {\cal S}({\pmb \alpha})] = 
\widehat{\cal S}({\bf f})
\widehat{\cal S}^\ast ({\bf f}) = |\widehat{\cal S}{\bf (f)}|^2, 
\end{equation}

\noindent 
where $\otimes$ stands for correlation.

\subsubsection{Resolution}

In an ideal condition, the resolution that 
can be achieved in an imaging experiment, ${\cal R}$, is limited only by the 
imperfections in the optical system and according to Strehl's criterion, 
the resolving power, ${\cal R}$, of any telescope of diameter $D$ is 
given by the integral of its transfer function, 

\begin{equation}
{\cal R} = \int\widehat{\cal S}({\bf u})d{\bf u}.  
\end{equation}

\noindent 
Therefore, ${\cal R} = {\cal S}({\bf 0})$. The Strehl ratio ${\cal S}_r$ is 
defined as the ratio of the intensity at the centroid of the observed PSF,
${\cal S}({\bf 0)}$, to the intensity of the peak of the diffraction-limited
image or `Airy spike', ${\cal S}({\bf 0})_{{\cal A}s}$, i.e., 

\begin{eqnarray}
{\cal S}_r &=& \frac{{\cal S}({\bf 0})}{{\cal S}({\bf 0})_{{\cal A}s}} 
\nonumber\\
&& \approx e^{-k^2\sigma^2_{OPD}},
\end{eqnarray}

\noindent 
where $k = 2\pi/\lambda$ is the wave number and $\sigma_{OPD}$ the rms 
optical path difference (OPD) error.
Typical ground-based observations with large telescopes in the visible
wavelength range are made with a Strehl ratio $\le 0.01$ (Babcock, 1990), while 
a diffraction-limited telescope would, by definition, have a Strehl ratio of 1.
 
\subsection{Principles of interference and its applications}
 
\label{subsec:of}

When two light beams from a single source are superposed, the intensity at
the place of superposition varies from point to point between maxima, which
exceed the sum of the intensities in the beams, and minima, which may be zero. 
This sum or difference is known as interference; the correlated fluctuation can 
be partially or completely coherent (Born and Wolf, 1984). 

Let the two monochromatic waves   
$V_1({\bf r}, t)$ and $V_2({\bf r}, t)$ be superposed at the recombination 
point. The correlator sums the instantaneous amplitudes of the fields. 
The total field at the output is,

\begin{eqnarray}
V &=& V_1 + V_2, \\
V^2 &=&  V^2_1 + V^2_2 + 2 V_1 \cdot V_2.
\end{eqnarray}

Then if $\Psi_1$ and $\Psi_2$ are the complex amplitudes of the two waves, 
with the corresponding phases $\psi_1$ and $\psi_2$, these two waves are 
propagating in $z$ direction and linearly polarized with electric field vector 
in $x$ direction. (A general radiation field is generally described by four 
Stokes parameters $I$, $Q$, $U$, and $V$, that specify intensity, the degree of 
polarization, the plane of polarization and the ellipticity of the 
radiation at each point and in any given direction). Therefore, 
the total intensity, (see equation 4), at the same point can be determined as,

\begin{eqnarray}
{\cal I} &=& {\cal I}_1 + {\cal I}_2 + {\cal J}_{12} \nonumber\\ 
&&= {\cal I}_1 + {\cal I}_2 + \frac{1}{2}\left(\Psi_1\Psi_2^\ast + 
\Psi_1^\ast\Psi_2\right)\nonumber\\
&&= {\cal I}_1 + {\cal I}_2 + 2 \sqrt{{\cal I}_1 {\cal I}_2} 
\ \cos\delta,   
\end{eqnarray}
 
\noindent 
where ${\cal I}_1 = <V_1^2>$, and ${\cal I}_2 = <V_2^2> $, are the 
intensities of the two terms, and ${\cal J}_{12} = 2 <V_1 \cdot V_2> 
= 2 \sqrt{{\cal I}_1 {\cal I}_2}$, is the interference term that depends on the
amplitude components, as well as on the phase-difference between the two waves,
$\delta = 2\pi\Delta\varphi/\lambda_\circ$, ($\Delta \varphi$, is the 
OPD between the two waves 
from the common source to the intersecting point and $\lambda_\circ$ is the 
wavelength in vacuum). In general, two light beams are not correlated but the
correlation term, $\Psi_1\Psi_2^\ast$, takes on significant values for
a short period of time and $<\Psi_1\Psi_2^\ast>$ = 0. Time variations of
$\Psi({\bf r})$ are statistical in nature (Mendel and Wolf, 1995). Hence,
one seeks a statistical description of the field (correlations) as the
field is due to a partially coherent source. Depending upon the correlations
between the phasor amplitudes at different object points, one would expect a 
definite correlation between the two points of the field emitted by the object.
The maximum and minimum intensity occur, when $|\delta| = 
0, 2\pi, 4\pi$ and $|\delta| = \pi, 3\pi, 5\pi$, respectively. If 
${\cal I}_1 = {\cal I}_2 = {\cal I}$, the intensity varies between 
4${\cal I}$, and 0.

In the case of quasi-monochromatic wave, the analytical signal, 
${\cal U}(t)$, obtained at the observation point is expressed as,

\begin{equation}
{\cal U}(t) = K_1{\cal U}(r_1, t-\tau_1) + K_2{\cal U}(r_2, t-\tau_2),  
\end{equation}
 
\noindent 
where $K_j$'s are constants, $r_j$'s the positions of two pinholes 
in the wave field, j = 1, 2, ${\it s}_j$'s the distances of a meeting point of 
the two beams from the two pinholes, $\tau_j = {\it s}_j/c$, the time needed to
travel from the respective pinholes to the meeting point, and $c$ the 
velocity of light. 
 
If the pinholes are small and the diffracted fields are considered to be 
uniform, the value $K_j$ of the constants, $|K_j|$ turns out to be,
$K_1^\ast K_2 = K_1 K^\ast_2 = K_1K_2$ and 
noting, ${\cal I}_j = |K_j|^2<|{\cal U}(r_j, t - \tau_j)|^2>$,
and therefore, the intensity at the output is found to be,

\begin{equation}
{\cal I} = {\cal I}_1 + {\cal I}_2 + 2K_1K_2 
\Re\left[{\bf \Gamma}_{12}\left(\frac{s_2 - s_1}{c}\right)\right].  
\end{equation}

The Van Cittert-Zernike theorem states that the modulus of the complex degree 
of coherence (describes the correlation of vibrations at a fixed point and a 
variable point) in a plane illuminated by a incoherent quasi-monochromatic 
source is equal to the modulus of the normalized spatial FT of 
its brightness distribution (Born and Wolf, 1984, Mendel and Wolf, 1995). The 
observed image is the FT of the mutual coherence function or the 
correlation function. The complex degree of (mutual) coherence, 
${\pmb \gamma}_{12}(\tau)$, of the observed source is defined as, 

\begin{equation}
{\pmb \gamma}_{12}(\tau) = \frac{{\bf \Gamma}_{12}(\tau)}{\sqrt{{\bf \Gamma}_
{11}(0){\bf \Gamma}_{22}(0)}} = \frac{{\bf \Gamma}_{12}(\tau)}
{\sqrt{{\cal I}_1 {\cal I}_2}},  
\end{equation}

\noindent 
where ${\cal I}_1 = K_1^2 {\bf \Gamma}_{11}(0)$ and 
${\cal I}_2 = K_2^2 {\bf \Gamma}_{22}(0)$.
The function, ${\bf \Gamma}_{ij}(\tau) = 
<{\cal U}(r_i, t + \tau){\cal U}^\ast(r_j, t)>$, is measured at two points. 
At a point where both the points coincide, the self coherence, 
${\bf \Gamma}_{11}(\tau) = <{\cal U}(r_1, t + \tau){\cal U}^\ast(r_1, t)>$, 
reduces to ordinary intensity. When $\tau = 0$, ${\bf \Gamma}_{11}(0) = 
{\cal I}_1; {\bf \Gamma}_{22}(0) = {\cal I}_2$. 
The ensemble average can be replaced by a time average due to the 
assumed ergodicity (a random process that is strictly stationary) of the fields.
If both the fields are directed on a 
quadratic detector, it yields the desired cross-term (time average due to the
finite time response). The measured intensity at the detector would be,

\begin{equation}
{\cal I} = {\cal I}_1 + {\cal I}_2 + 2\sqrt{{\cal I}_1. {\cal I}_2} 
\Re\left[{\pmb \gamma}_{12}\left(\frac{s_2 - s_1}{c}\right)\right].  
\end{equation}

In order to keep the time correlation close to unity, the delay, $\tau$, must
be limited to a small fraction of the temporal width or coherence time, $\tau_c
= 1/\Delta\nu$; here $\Delta\nu$, is the spectral width.  
The relative coherence of the two beams diminishes with the increase of
path length difference, culminating in a drop in the visibility (a dimensionless
number between zero and one that indicates the extent to which a source is
resolved on the baseline being used) of the fringes. For $\tau \ll \tau_c$, the 
function, ${\pmb \gamma}_{12}(\tau)$, can be approximated to,
${\pmb \gamma}_{12}(0) e^{-2\pi i\nu_\circ\tau}$. The exponential term is 
nearly constant and ${\pmb \gamma}_{12}(0)$, measures the spatial coherence. 
Let $\psi_{12}$, be the argument of ${\pmb \gamma}_{12}(\tau)$, then,

\begin{equation}
{\cal I} = {\cal I}_1 + {\cal I}_2 + 2\sqrt{{\cal I}_1, {\cal I}_2}\Re\left
[|{\pmb \gamma}_{12}(0)| e^{i(\psi_{12}-2\pi\nu_\circ \tau)}\right].    
\end{equation}

The measured intensity at a distance $x$ from the origin (point at zero OPD) on 
a screen at a distance, $z$, from the apertures is

\begin{eqnarray}
{\cal I}(x) &=& {\cal I}_1 + {\cal I}_2 + 2\sqrt{{\cal I}_1,{\cal I}_2}
|{\pmb \gamma}_{12}(0)| \nonumber\\
&& \cos\left\{\frac{2\pi d(x)}{\lambda}-\psi_{12}\right\},   
\end{eqnarray}
 
\noindent 
where $d(x) = bx/z$, is the OPD corresponding to $x$, 
and $b$ the distance between the two apertures.
  
The modulus of the fringe visibility is estimated as the ratio of high frequency
to low frequency energy in the average spectral density; the visibility of 
fringes, ${\cal V}$, is estimated as, 

\begin{equation}
{\cal V} = \frac{{\cal I}_{max}-{\cal I}_{min}}{{\cal I}_{max} + {\cal I}_{min}}
= |{\pmb \gamma}_{12}(0)|\frac{2\sqrt{{\cal I}_1{\cal I}_2}}{{\cal I}_1 + 
{\cal I}_2}.  
\end{equation}

\subsubsection{Fizeau interferometer}

Fizeau (1868) suggested that installing a screen with two holes in front of a 
telescope would allow measurements of stellar diameters with diffraction-limited
resolution. In this set up, the beams are diffracted by the 
sub-apertures and the telescope acts as both collector
and correlator. Therefore, temporal coherence is automatically obtained due to
the built-in zero OPD. The spatial modulation frequency, as well as the required
sampling of the image change with the separation of sub-apertures. 
The maximum resolution in this case depends on the separation between the 
sub-apertures; the maximum spacings that can be explored are limited by the 
physical diameter of the telescope. The number of stellar sources for measuring 
diameters is also limited. One of the first significant results
was the measurement of the diameter of the satellites of Jupiter with a Fizeau 
interferometer on the 40-inch Yerkes refractor by Michelson (1891). With the 
100-inch telescope on Mt. Wilson Anderson (1920) determined the angular 
separation ($\rho$) of spectroscopic binary star Capella. 

\subsubsection{Michelson interferometer}

Results from the classical Michelson interferometer were used to formulate 
special relativity. They are also being used in gravity-wave detection.
Gravitational radiation produced by coalescing binaries, or exploding stars,
for example, changes the metric of spacetime. This effect causes 
a differential change of the path length of the arms of the interferometer,
thereby introducing a phase-shift. Today, there are several ground-based 
long baseline laser-interferometric detectors based on this principle 
under construction, and within the 
next several years these detectors should be in operation (Robertson, 2000).
The proposed laser interferometer space antenna, consisting of three satellites
in formation about 50 million kilometers (km) above the Earth in a heliocentric 
orbit, may detect gravitational waves by measuring fluctuations 
in the distances between test masses carried by the satellites.

The essence of the Michelson's stellar interferometer is to determine the 
covariance $<\Psi_1\Psi_2>$ of the complex amplitudes $\Psi_1, \Psi_2$, at two 
different points of the wavefronts. This interferometer was equipped with four 
flat mirrors that fold the beams by installing a 7-meter (m) steel beam on top 
of the Mt. Wilson 100-inch telescope. Michelson and Pease (1921) resolved the 
supergiant $\alpha$~Ori and a few other stars. In this case, the spatial 
modulation frequency in the focal-plane is independent of the distance between 
the collectors. In the Fizeau mode, the ratio of aperture diameter/separation is
constant from light collection to recombination in the image-plane (homothetic 
pupil). In the Michelson mode, this ratio is not constant since the collimated 
beams have the same diameter from the output of the telescope to the 
recombination lens. The distance between pupils is equal to the baseline at the 
collection mirrors (the resolution is limited by the baseline) and to a much 
smaller value just before the recombination lens. The disadvantage of the 
Michelson mode is a very narrow field of view compared to the Fizeau mode. 
Unfortunately the project was abandoned due to various difficulties, including 
(i) effect of atmospheric turbulence, (ii) variations of refractive index above 
the small sub-aperture, (iii) inadequate separation of outer mirrors, and (iv) 
mechanical instability.

\subsubsection{Intensity interferometer}
 
Intensity interferometry considers the quantum theory of photon 
detection and correlation. It computes the fluctuations of the 
intensities ${\cal I}_1, {\cal I}_2$, at two different points of the 
wavefronts. The fluctuations of the electrical signals from the two detectors 
are compared by a multiplier. The current output of each photoelectric detector
is proportional to the instantaneous intensity ${\cal I}$ of the incident light,
which is the squared modulus of the amplitude $\Psi$. The fluctuation of the
current output is proportional to $\Delta {\cal I} = |\Psi|^2 - <|\Psi
|^2>$. The covariance of the fluctuations, according to Goodman (1985), can 
be expressed as, 

\begin{equation}
<\Delta{\cal I}_1\Delta{\cal I}_2> = <|\Psi_1\Psi_2^\ast|^2>. 
\end{equation}

\noindent
This expression indicates that the covariance of the intensity fluctuations is
the squared modulus of the covariance of the complex amplitude. 
 
Having succeeded in completing the intensity interferometer at radio wavelengths
(Hanbury Brown et al. 1952), Hanbury Brown and Twiss (1958) demonstrated its 
potential at optical wavelengths by measuring the angular diameter of Sirius. 
Subsequent development with a pair of 6.5~m light collector on 
a circular railway track spanning 188~m, provided the measurements of 32 
southern binary stars (Hanbury Brown, 1974) with angular resolution limit of 
0.5 milli-arcseconds (mas). In this arrangement, starlight  
collected by two concave mirrors is focused on to two photoelectric cells
and the correlation of fluctuations in the photocurrents is measured as
a function of mirror separation. The advantages of such a system over 
Michelson's interferometer are that it does not require high mechanical 
stability and remains unaffected by seeing. 
Another noted advantage is that the alignment tolerances are extremely relaxed 
since the pathlengths need to be maintained to a fraction of 
$c/{\it b}_e$, where ${\it b}_e$ is the electrical bandwidth of the
post-detection electronics. The significant effect comes from scintillation 
induced by the atmosphere. The sensitivity of this interferometer was found to 
be very low; it was limited by the narrow band-width filters that are used to 
increase the speckle life time. The correlated fluctuations can be obtained if 
the detectors are spaced by less than a speckle width. Theoretical calculations 
(Roddier, 1988) show that the limiting visual magnitude (mag), $m_v$ that can 
be observed with such a system is of the order of 2 (the 
faintest stars visible in the naked eye are 6th magnitude. The
magnitude scale is defined as $m_1 - m_2$ = -2.5 log $F_1/F_2$,
where $m_1$ and $m_2$ are the apparent magnitudes of two objects of 
fluxes $F_1$ and $F_2$, respectively).

\section{ATMOSPHERIC TURBULENCE}
  
\label{sec:turbulence}

The density inhomogeneities appear to be created and maintained by the 
parameters, viz., thermal gradients, humidity fluctuations, and wind shears, 
which produce atmospheric turbulence and therefore refractive index 
inhomogeneities. The gradients caused by these environmental parameters warp 
the wavefront incident on the telescope pupil. The image quality is directly
related to the statistics of the perturbations of the incoming wavefront.
The theory of seeing combines the theory of atmospheric turbulence with
the theory of optical physics to predict the modifications to the 
diffraction-limited image that the refractive index gradients produce (Young,
1974, Roddier, 1981, Coulman, 1985). Atmospheric turbulence has a significant 
effect on the propagation of radio-waves, sound-waves and on the flight of
aircraft as well. This section is devoted to the descriptions of the 
atmospheric turbulence theory, metrology of seeing and its impact on stellar 
images.

\subsection{Formation of eddies}
 
The random fluctuations in the atmospheric motions occur predominantly due to 
(i) the friction encountered by the air flow at the Earth's surface and 
consequent formation of a wind-velocity profile with large vertical gradients,
(ii) differential heating of different portions of the Earth's surface and the 
concomitant development of thermal convection, (iii) processes associated with
formation of clouds involving release of heat of condensation and
crystallization, and subsequent changes in the nature of temperature and wind 
velocity fields, (iv) convergence and interaction of air-masses with various
atmospheric fronts, and (v) obstruction of air-flows by mountain barriers that 
generate wave-like disturbances and rotor motions on their lee-side.

The atmosphere is difficult to study due to the high Reynolds number 
($Re \sim10^6$), a dimensionless quantity, that characterizes the turbulence.  
When the average velocity, $v_a$, of a viscous fluid of 
characteristic size, $l$, is gradually increased, two distinct 
states of fluid motion are observed (Tatarski, 1967, Ishimaru, 1978), 
viz., (i) laminar (regular and smooth in space and time), at very low $v_a$, and
(ii) unstable and random fluid motion at $v_a$ greater than some critical value.
The Reynolds number, obtained by equating the inertial and viscous forces, is 
given by, 

\begin{equation}
Re = v_a l/\nu_v. 
\end{equation}

\noindent
where $Re$ is a function of the flow geometry, $v_a, 
l$, and kinematic viscosity of the fluid, $\nu_v$. When $Re$ exceeds critical 
value in a pipe (depending on its geometry), a transition of the flow from 
laminar to turbulent or chaotic occurs. Between these two extreme conditions, 
the flow passes through a series of unstable states. High $Re$ turbulence is 
chaotic in both space and time and exhibits considerable spatial structure.
  
The velocity fluctuations occur on a wide range of space and time scales.
According to the atmospheric turbulence model (Taylor, 1921, Kolmogorov, 1941b, 
1941c), the energy enters the flow at low frequencies at scale length,
$L_\circ$ and spatial frequency, $k_{L_{\circ}} = 2\pi/L_\circ$, as a direct 
result of the non-linearity of the Navier-Stokes equation governing fluid 
motion. The large-scale fluctuations, referred to as large eddies, have a size
of the geometrically imposed outer scale length $L_\circ$. These eddies are not 
universal with respect to flow geometry; they vary according to the local 
conditions. Conan et al. (2000) derived a mean value $L_\circ =$ 24~m for 
a von K\'arm\'an spectrum from the data obtained at Cerro Paranal, Chile. 

The energy is transported to smaller and smaller loss-less eddies until at a
small enough Reynolds number, the kinetic energy of the flow is converted into 
heat by viscous dissipation resulting in a rapid drop in  power spectral 
density, $\Phi_n({\bf k})$ for $k > k_\circ$, where $k_\circ$ is critical wave 
number. These changes are characterized by the inner scale length, $l_\circ$, 
and spatial frequency, $k_{l_{\circ}} = 2\pi/l_\circ$, where $l_\circ$ varies 
from a few millimeter near the ground to a centimeter (cm) high in the 
atmosphere. The small-scale fluctuations with sizes $l_\circ < r < L_\circ$, 
known as the inertial subrange, where $r$ is the magnitude of ${\bf r}$,  
have universal statistics (scale-invariant behavior) independent of the flow 
geometry. The value of inertial subrange would be 
different at various locations on the site. The statistical distribution of the 
size and number of these eddies is characterized by $\Phi_n({\bf k})$, of 
$n_1({\bf r}, t)$, where $n_1({\bf r}, t)$ is a randomly fluctuating term, and 
$t$ the time. The dependence of the refractive index of air $n({\bf r}, t)$, 
upon pressure, $P$ (millibar) and temperature, $T$ (Kelvin), at optical 
wavelengths is given by $n_1 \cong n - 1 = 77.6~\times~10^{-6}P/T$ 
(Ishimaru, 1978). 
 
\subsection{Kolmogorov turbulence model}

The optically important property of the Kolmogorov law is that the refractive
index fluctuations are largest for the largest turbulent elements up to the
outer scale of the turbulence. At sizes below the outer scale, the 
one-dimensional (1-d) power spectrum of the refractive index fluctuations falls 
off with -(5/3) power of frequency and is independent of the direction along 
the fluctuations are measured, i.e., the small-scale fluctuations are isotropic 
(Young, 1974). The three-dimensional (3-d) power spectrum, $\Phi_n$, for the 
wave number, $k > k_\circ$, in the case of inertial subrange, can be equated as,

\begin{equation}
\Phi_n({\bf k}) = 0.033{\cal C}_n^2 k^{-11/3},  
\end{equation}
 
\noindent 
where ${\cal C}_n^2$ is known as the structure constant of the refractive index 
fluctuations. 
  
This Kolmogorov-Obukhov model of turbulence, describing 
the power-law spectrum for the inertial intervals of wave numbers, 
is valid within the inertial subrange and is widely 
used for astronomical purposes (Tatarski 1993). The refractive index 
structure function, ${\cal D}_n({\bf r})$, is defined as,

\begin{equation}
{\cal D}_n({\bf r}) = <| n\left({\pmb \rho} + {\bf r}\right) - n\left(
{\pmb \rho} \right)|^2>,  
\end{equation}
 
\noindent 
which expresses its variance at two points $r_1$, and $r_2$. 
  
The structure functions
are related to the covariance function, ${\cal B}_n({\bf r})$, through 

\begin{equation}
{\cal D}_n({\bf r)} = 2[{\cal B}_n({\bf 0}) - {\cal B}_n({\bf r})],  
\end{equation}
 
\noindent 
where ${\cal B}_n({\bf r}) = <n\left({\pmb \rho}\right)n\left({\pmb \rho} + 
{\bf r} \right)>$ and the covariance is the 3-d FT of the spectrum, 
$\Phi_n({\bf k)}$ (Roddier, 1981). The structure function in the inertial range
(homogeneous and isotropic random field), according to Kolmogorov (1941a) 
depends on the magnitude of ${\bf r}$, as well as on the values of the rate of 
production or dissipation of turbulent energy $\epsilon_\circ$ and the rate of 
production or dissipation of temperature inhomogeneities $\eta_\circ$. 

The refractive index $n$ is a function of $n(T, {\cal H})$ of the temperature, 
$T$ and humidity, ${\cal H}$. and therefore, the expectation value of the 
variance of the fluctuations about the average of the refractive index is given 
by,

\begin{eqnarray}
<dn>^2 &=& \left(\frac{\partial n}{\partial T}\right)^2<dT>^2 \nonumber\\
&& + 2\left(\frac{\partial n}{\partial T}\right)
\left(\frac{\partial n}{\partial {\cal H}}\right)<dT><d{\cal H}> \nonumber\\
&& + \left(\frac {\partial n}{\partial {\cal H}}\right)^2<d{\cal H}>^2.
\end{eqnarray}

\noindent
It has been argued that in optical propagation, the last term is negligible,
and that the second term is negligible for
most astronomical observations. It could be significant, however, in
high humidity situation, e.g., a marine boundary layer (Roddier, 1981).
Most treatments ignore the contribution from humidity and express
the refractive index structure function (Tatarski, 1967) as,

\begin{equation}
{\cal D}_n({\bf r}) = {\cal C}_n^2 r^{2/3}.  
\end{equation}

\noindent
Similarly, the velocity structure function, ${\cal D}_v ({\bf r}) = 
{\cal C}_v^2 r^{2/3}$, and temperature structure function, ${\cal D}_T({\bf r})
= {\cal C}_T^2 r^{2/3}$, can also be derived; the same form holds for the 
humidity structure function. The two structure coefficients ${\cal C}_n$ and
${\cal C}_T$ are related by, ${\cal C}_n = \frac{\partial n}{\partial T} 
{\cal C}_T$, and assuming pressure equilibrium, ${\cal C}_n = 80 \times \frac{P}
{T^2} {\cal C}_T$ (Roddier, 1981).

Several experiments confirm this two-thirds power law in the atmosphere 
(Wyngaard et al. 1971, Coulman, 1974, Hartley et al. 1981, Lopez, 1991). 
Robbe et al. (1997) reported from the observations using a long baseline
optical interferometer (LBOI), Interf\'erom\`etre \`a deux t\'elescopes (I2T;
Labeyrie, 1975) that most of the measured temporal spectra of the angle of 
arrival exhibit a behavior compatible with the said power law. Davis and Tango 
(1996) have measured the value of atmospheric coherence time that varied between
$\sim$1 and $\sim$7~ms with the Sydney University stellar interferometer 
(SUSI).

The value of ${\cal C}_n^2$ (in equation 33) depends on local conditions, as 
well as on the planetary boundary layer. The significant scale lengths, in the 
case of the former, depend on the local objects which primarily introduces 
changes in the inertial subrange and temperature differentials. The 
latter can be attributed to (i) surface boundary layer due to the ground 
convection, extending up to a few km height of the atmosphere, 
(${\cal C}_T^2 \propto z^{-2/3}$), (ii) the free convection layer associated 
with orographic disturbances, where the scale lengths are height dependent, 
(${\cal C}_T^2 \propto z^{-4/3}$), and (iii) in the tropopause and above, where 
the turbulence is due to the wind shear as the temperature gradient vanishes 
slowly. In real turbulent flows, turbulence is usually generated at solid 
boundaries. Near the boundaries, shear is the 
dominant source (Nelkin, 2000), where scale lengths are roughly constant. In an 
experiment, conducted by Cadot et al. (1997), it was found that Kolmogorov 
scaling is a good approximation for the energy dissipation, as well as for the 
torque due to viscous stress. They measured the energy dissipation and the 
torque for circular Couette flow with and without small vanes attached to the 
cylinders to break up the boundary layer. The theory of the turbulent flow in 
the neighborhood of a flat surface applies to the atmospheric surface layer. 
Masciadri et al. (1999) have noticed that the value of ${\cal C}_n^2$ 
increases about 11~km over Mt. Paranal, Chile. The turbulence concentrates into 
a thin layer of 100$-$200~m thickness, where the value of ${\cal C}_n^2$ 
increases by more than an order of magnitude over its background level. 
 
\subsection{Wave propagation through turbulence}
 
\label{subsec:propagation}

The spatial correlational properties of the 
turbulence-induced field perturbations are evaluated by combining the basic 
turbulence theory with the stratification and phase screen approximations. 
The variance of the ray can be translated into a variance of the phase 
fluctuations. For calculating the same, Roddier (1981) used
the correlation properties for propagation through a single (thin) turbulence 
layer and then extended the procedure to account for many such layers. Several 
investigators (Goodman, 1985, Troxel et al. 1994) have argued that 
individual layers can be treated as independent provided the separation of the 
layer centers is chosen large enough so that the fluctuations of the log 
amplitude and phase introduced by different layers are uncorrelated. 

\subsubsection{Effect of turbulent layers}
 
Let a monochromatic plane wave of wavelength $\lambda$
from a distant star at the zenith, propagates towards the ground-based observer;
the complex amplitude at co-ordinate, $({\bf x}, h)$, is given by,

\begin{equation}
\Psi_h({\bf x}) = |\Psi_h({\bf x})| e^{i\psi_h({\bf x})}.  
\end{equation}
 
\noindent 
The average value of the phase, $<\psi_h({\bf x})> = 0$ for height h, 
and the unperturbed complex amplitude outside the atmosphere is normalized
to unity $[\Psi_\infty({\bf x}) = 1]$. 
When this wave is allowed to pass through a thin layer of turbulent air 
of thickness $\delta h_j$, which is considered to be large compared to the
scale of turbulent eddies but small enough for the phase screen approximation
(diffraction effects is negligible over the distance, $\delta h_j$), 
the complex amplitude of the plane wavefront after passing through the layer 
is expressed as,

\begin{equation}
\Psi_j({\bf x}) = e^{i\psi_j({\bf x})}. 
\end{equation}

\noindent 
Here the phase-shift $\psi_j({\bf x})$ introduced by the refractive index 
fluctuations, $n(x, z)$ inside the layer can be written as,

\begin{equation} 
\psi_j({\bf x}) = k\int_{h_j}^{h_j+\delta h_j}n(x, z) dz,
\end{equation} 

\noindent 
In this case, the rest of the atmosphere is thought to be calm and homogeneous.
At the layer output, the coherence function of the complex amplitude, 
$<\Psi_j\left({\bf x}\right)
\Psi_j^\ast\left({\bf x} + {\pmb \xi}\right)>$, leads to,

\begin{equation}
{\cal B}_j\left({\pmb \xi}\right) = <e^{i\left[\psi_j\left({\bf x}\right)-
\psi_j\left({\bf x} + {\pmb \xi}\right)\right]}>.  
\end{equation}

The quantity $\psi_j({\bf x})$ can be considered to be the sum of a large 
number of independent variables, and therefore, has Gaussian statistics. 
This equation is similar to Fourier
transform of the probability density function at unit frequency; therefore,

\begin{equation}
{\cal B}_j\left({\pmb \xi}\right) = e^{-\frac{1}{2}{\cal D}_{\psi_j}\left(
{\pmb \xi}\right)}, 
\end{equation}

\noindent
The term ${\cal D}_{\psi_j}\left({\pmb \xi}\right)$ is the 2-d structure 
function of the phase, $\psi_j({\bf x})$ that can be read as (Fried, 1966),

\begin{equation} 
{\cal D}_{\psi_j}\left({\pmb \xi}\right) = <| \psi_j\left({\bf x}
\right)-\psi_j\left({\bf x} + {\pmb \xi}\right)|^2>.
\end{equation}

\subsubsection{Computation of phase structure function}
 
Let the covariance of the phase, ${\cal B}_{\psi_j}\left( {\pmb \xi}\right)$
be defined as,

\begin{equation}
{\cal B}_{\psi_j}\left( {\pmb \xi}\right) 
= <\psi_j\left({\bf x}\right)\psi_j\left({\bf x} + {\pmb \xi}\right)>, 
\end{equation}

\noindent
and by replacing $\psi_j({\bf x})$, 

\begin{equation}
{\cal B}_{\psi_j}\left({\pmb \xi}\right)=k^2\int_{h_j}^{h_j+\delta h_j} dz
\int_{h_j-z}^{h_j+\delta h_j-z}{\cal B}_n\left({\pmb \xi}, \zeta\right)d\zeta,  
\end{equation}

\noindent 
where $\zeta = z^\prime -z$ and the 3-d refractive index covariance is,

\begin{equation}
{\cal B}_n\left({\pmb \xi}, \zeta\right) = <n({\bf x}, z)n\left({\bf x} 
+ {\pmb \xi}, z^\prime\right)> 
\end{equation}
 
\noindent 
Since the thickness of the layer, $\delta h_j$, is large compared to the 
correlation scale of the turbulence, the integration over $\zeta$  
from $-\infty \; to \; +\infty$, leads to,

\begin{equation}
{\cal B}_{\psi_j}\left({\pmb \xi}\right)=k^2\delta h_j\int {\cal B}_n\left(
{\pmb \xi}, \zeta\right)d\zeta.   
\end{equation}

\noindent 
The phase structure function is related to its covariance (equation 35);
therefore,

\begin{equation}
{\cal D}_{\psi_j}\left({\pmb \xi}\right) = 2k^2\delta h_j\int \left[{\cal B}_n
\left({\bf 0}, \zeta \right)-{\cal B}_n\left({\pmb \xi},\zeta\right)\right]
d\zeta.  
\end{equation}

The refractive index structure function defined in equation (37) is 
evaluated as, 
\begin{equation}
{\cal D}_n\left({\pmb \xi}, \zeta\right) = 
{\cal C}_n^2\left(\xi^2 + \zeta^2\right)^{1/3}, 
\end{equation}

\noindent
and using equation (35), equation (48) can be integrated to yield,

\begin{equation}
{\cal D}_{\psi_j}\left({\pmb \xi}\right) = 
2.91k^2{\cal C}_n^2\xi^{5/3}\delta h_j. 
\end{equation}

The covariance of the phase is deduced by substituting equation (50), 
in equation (42),

\begin{equation}
{\cal B}_{h_j}\left({\pmb \xi}\right) = 
e^{-\frac{1}{2}\left(2.91k^2{\cal C}_n^2\xi^{5/3}\delta h_j\right)}.  
\end{equation}

\noindent 
Using the Fresnel approximation, the covariance of the phase at the ground 
level due to a thin layer of turbulence at some height off the ground is given
by, 

\begin{equation}
{\cal B}_\circ\left({\pmb \xi}\right) = {\cal B}_{h_j}\left({\pmb \xi}\right).
\end{equation}
  
For high altitude layers the complex field will fluctuate both in phase and in
amplitude (scintillation), and therefore, the wave structure function, 
${\cal D}_{\psi_j}\left({\pmb \xi}\right)$, is not strictly true as at the 
ground level. The turbulent layer acts like a diffracting screen; however, 
correction in the case of astronomical observation remains small (Roddier, 
1981). 
 
The wave structure function after passing through $N$ layers can be expressed as
the sum of the $N$ wave structure functions associated with the individual 
layer. For each layer, the coherence function is multiplied by the term,
$e^{-\frac{1}{2}\left[2.91k^2{\cal C}_n^2(h_j)\xi^{5/3} \delta h_j\right]}$; 
therefore, the coherence function at ground level is given by,

\begin{eqnarray}
{\cal B}_\circ\left({\pmb \xi}\right) &=& \prod_{j=1}^N e^{-\frac{1}{2}
\left[2.91k^2{\cal C}_n^2(h_j)\xi^{5/3}\delta h_j\right]} \nonumber\\
&&= e^{-\frac{1}{2}
\left[2.91k^2\xi^{5/3} \sum_{j=1}^N {\cal C}_n^2(h_j)\delta h_j\right]}. 
\end{eqnarray}

\noindent
This expression may be generalized for a star at an angular distance  
$\gamma$ away from the zenith viewed through all over the turbulent atmosphere, 

\begin{equation}
{\cal B}_\circ\left({\pmb \xi}\right) = e^{-\frac{1}{2}\left[2.91k^2\xi^{5/3}
\sec\gamma\int {\cal C}_n^2(z)dz\right]}. 
\end{equation}

\subsubsection{Seeing limited images}

\label{subsubsec:seeing}
 
The term `seeing' is the total effect of distortion in the path of the star
light via different contributing layers of the atmosphere to the detector 
placed at the focus of the telescope. 
Let the MTF of the atmosphere and a telescope together be described as in figure
1. The long-exposure PSF is defined by the ensemble average, 
$<{\cal S}({\bf x})>$, which is independent of the direction. If the object 
emits incoherently, the average illumination, $<{\cal I}({\bf x})>$, of a 
resolved object, ${\cal O}({\bf x})$, obeys the convolution relationship, 

\begin{equation}
<{\cal I}({\bf x})> = {\cal O}({\bf x}) \star <{\cal S}({\bf x})>.  
\end{equation}

\noindent
Using 2-d FT, the above equation translates into,

\begin{equation}
<\widehat{{\cal I}}({\bf u})> = \widehat{{\cal O}}({\bf u})\cdot 
<\widehat{{\cal S}}({\bf u)}>, 
\end{equation}

\noindent 
where $<\widehat{{\cal S}}({\bf u})>$ denotes the transfer function for 
long-exposure images, ${\bf u}$, the spatial frequency vector with magnitude, 
$u$ and $\widehat{\cal O}({\bf u})$ the object spectrum. 
The argument of equation (56) is expressed as,

\begin{equation}
arg|\widehat{\cal I}{\bf (u)}| = \psi{\bf (u)} + \theta_1 - \theta_2, 
\end{equation}
 
\noindent 
where $\psi({\bf u})$ is the Fourier phase at ${\bf u}$, 
$arg| \; |$ stands for, `the phase of', and $\theta_j$ the apertures, 
corresponding to the seeing cells. The transfer function is the product of 
the atmosphere transfer function (wave coherence function), ${\cal B}({\bf u})$,
and the telescope transfer function, ${\cal T}({\bf u})$,

\begin{equation}
<\widehat{{\cal S}}({\bf u})> = {\cal B}({\bf u})\cdot{\cal T}({\bf u}). 
\end{equation}

\begin{figure}
\centerline {\psfig{figure=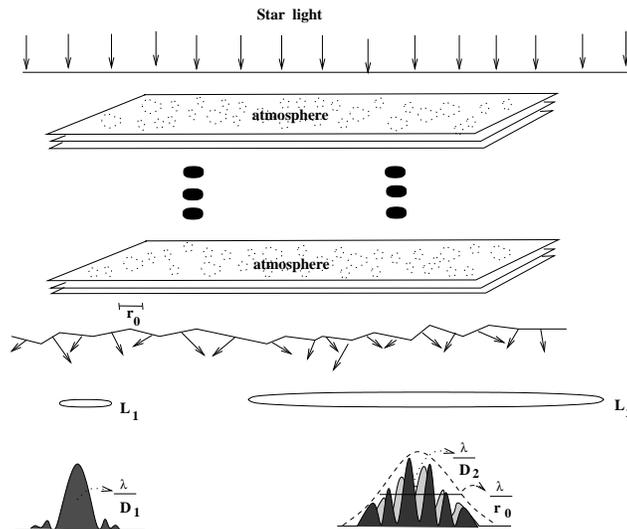,height=7cm}}
\caption {Plane-wave propagation through the multiple turbulent layers.
${\tt L}_1$ and ${\tt L}_2$ represent the small and large telescopes with
respective diameters ${\tt D}_1$ and ${\tt D}_2$.} 
\end{figure}

For a long-exposure through the atmosphere, the resolving power, ${\cal R}$, of 
any optical telescope can be expressed as,

\begin{equation}
{\cal R} = \int{\cal B}({\bf u})\cdot{\cal T}({\bf u})d{\bf u}. 
\end{equation}

\noindent
It is limited either by the telescope or by the atmosphere, depending on the
relative width of the two functions, ${\cal B}({\bf u})$ and 
${\cal T}({\bf u})$. 

\begin{eqnarray}
{\cal R} &=& \int{\cal T}({\bf u})d{\bf u} = \frac{\pi}{4}\left(\frac{D}
{\lambda}\right)^2 \, \, \, D \ll r_\circ. \\ 
&& = \int{\cal B}({\bf u})d{\bf u} = \frac{\pi}{4}\left(\frac
{r_\circ}{\lambda}\right)^2 \, \, \,  D \gg r_\circ.
\end{eqnarray}

\paragraph{Fried's parameter}

According to equation (54), ${\cal B}({\bf u})$, can be expressed as,
 
\begin{eqnarray}
{\cal B}({\bf u}) &=& {\cal B}_\circ(\lambda{\bf u}) \nonumber\\ 
&&= e^{-\frac{1}{2}\left[2.91k^2(\lambda u)^{5/3}\sec\gamma\int
{\cal C}_n^2(z) dz\right]}. 
\end{eqnarray}

\noindent
Therefore, equation (61) is translated into, 

\begin{equation}
{\cal R} = (6\pi/5)\left[\frac{1}{2}\left(2.91k^2
\lambda^{5/3}\sec\gamma\int{\cal C}_n^2(z) dz\right)\right]^{-6/5}
\Gamma(6/5).  
\end{equation}

Fried, (1966) had introduced the critical diameter $r_\circ$, for a telescope; 
therefore, placing $D = r_\circ$ in 
equation (60), equation (62) takes the following form, 

\begin{eqnarray}
{\cal B}({\bf u}) &=& e^{-3.44\left(\lambda u/r_\circ\right)^{5/3}}.\\
{\cal B}_\circ\left({\pmb \xi}\right) &=& 
e^{-3.44\left(\xi/r_\circ\right)^{5/3}}.
\end{eqnarray}

\noindent
The phase structure function (equation 43) across the telescope aperture
(Fried, 1966) becomes,

\begin{equation}
{\cal D}_{\psi}\left({\pmb \xi}\right) 
= 6.88\left(\frac{\xi}{r_\circ}\right)^{5/3}.
\end{equation}

\noindent 
By replacing the value of ${\cal B}_\circ\left({\pmb \xi}\right)$, in 
equation (54), an expression for $r_\circ$ in terms of the distribution of the 
turbulence in the atmosphere is found to be.
 
\begin{equation}
r_\circ = \left[0.423 k^2\sec\gamma\int{\cal C}_n^2(z)dz\right]^{-3/5}. 
\end{equation}

\noindent 
The Fried's parameter may be thought of as the diameter of telescope that would
produce the same diffraction-limited FWHM of a point source image as the 
atmospheric turbulence would with an infinite-sized mirror. 

\paragraph{Seeing at the telescope site}
 
The major sources of image degradation 
predominantly comes from thermal and aero-dynamic disturbances in the 
atmosphere surrounding the telescope and its enclosure. These sources
include: (i) convection in and around the building and the dome, obstructed 
location near the ground, off the surface of the telescope structure, 
(ii) thermal distortion of the primary and secondary mirrors when they are
warmer than the ambient air, (iii) dissipation of heat by the secondary mirror 
(Zago, 1995), (iv) rise in temperature at the primary mirror cell, 
and (v) at the focal point causing temperature gradient close to the 
detector. Degradation in image quality can occur due to opto-mechanical 
aberrations as well as mechanical vibrations of the optical system. 
 
Various corrective measures have been proposed to improve the seeing. These
measures include: (i) insulating the surface of the floors and walls, 
(ii) introducing an active cooling system to eliminate the heat produced by 
electric equipment on the telescope and elsewhere in the dome, and 
(iii) installing a ventilator to generate a downward air flow through the
slit to counteract the upward action of the bubbles (Racine, 1984, Ryan and 
Wood, 1995). Floor-chilling systems to dampen the natural convection have been
implemented which keeps the temperature of the primary mirror closer to the air 
volume (Zago, 1995). Saha and Chinnappan (1999) have found that the average
observed $r_\circ$ is higher during the later part of the night than the earlier
part. This change might indicate that the slowly cooling mirror creates thermal 
instabilities that decrease slowly during the night.

\section{SINGLE APERTURE DIFFRACTION-LIMITED IMAGING}
  
\label{sec:aperture}
 
Ever since the development of the SI technique (Labeyrie, 1970), it is widely 
employed both in the visible, as well as in the infrared (IR) bands at 
telescopes to decipher diffraction-limited informations. The following 
sub-sections deal with single aperture speckle imaging and related avenues, 
other techniques, AO imaging systems, dark speckle imaging, and high resolution
sensors.

\subsection{Speckle imaging}

\label{subsec:speckle}

If a point source is imaged through the telescope by using the pupil function 
consisting of two sub-apertures ($\theta_1, \theta_2$), corresponding to the two
seeing cells separated by a vector $\lambda {\bf u}$, a fringe pattern is 
produced with narrow spatial frequency bandwidth that moves within broad PSF
envelope; with increasing distance between the sub-apertures, the fringes move 
with an increasingly larger amplitude. The introduction of a third sub-aperture 
gives three pairs of sub-apertures and yields the appearance of 
three intersecting patterns of moving fringes. Covering the telescope aperture 
with $r_\circ$-sized sub-apertures synthesizes a filled aperture ${\it p}_j$ 
(each pair of them, ${\it p}_n, {\it p}_m$, is separated by a baseline) 
interferometer. The intensity at the focal-plane, ${\cal I}$, according 
to the diffraction theory (Born and Wolf, 1984), is determined by the 
expression,

\begin{equation}
{\cal I} = \sum_{n,m}<\Psi_n\Psi^\ast_m>.  
\end{equation}

\noindent
The term, $\Psi_n\Psi^\ast_m$, is multiplied by $e^{i\psi}$, where $\psi$ is 
the random instantaneous shift in the fringe pattern. Each sub-aperture is small
enough for the field to be coherent over its extent. Atmospheric turbulence does
not affect the contrast of the fringes produced but introduces phase delays. 
If the integration time is shorter than the evolution time of the phase 
inhomogeneities, the interference fringes are preserved but their phases are 
randomly distorted, which produces `speckles'. (The formation of speckles stems 
from the summation of coherent vibrations having random characteristics. It can 
be modeled as a 2-dimensional random walk with Fresnel's vector representation
of vibrations). Each speckle covers an area of 
the same order of magnitude as the Airy disc of the telescope. The number of 
correlation cells is proportional to the square of $D/r_\circ$ and the number 
of photons, $N_p$, per speckle is independent of its diameter. The lifetime of 
speckles, $\tau_\circ \sim r_\circ/\Delta\nu$, where $\Delta\nu$ is the 
velocity dispersion in the turbulent seeing layers across the line of sight. 

The structure of the speckle pattern changes randomly over a short interval of 
time. The sum of several such statistically uncorrelated patterns from a point 
source can result in a uniform patch of light a few arcseconds 
($^{\prime\prime}$) wide. Figures 2 and 3 depict the speckles of a 
binary star HR4689 and the results of summing 128 specklegrams, respectively.
Averaging $\widehat{\cal I}({\bf u})$ over many frames, the resultant 
for frequencies greater than $r_\circ/\lambda$, 
tends to zero because the phase-difference, $\theta_1 - \theta_2$, mod~$2\pi$,  
between the two apertures is distributed uniformly between $\pm\pi$, with zero 
mean. The Fourier component performs a random walk in the complex 
plane and averages to zero: 

\begin{equation}
<\widehat{\cal I}{\bf (u)}> = 0, \ \ \ \  u > r_\circ / \lambda. 
\end{equation}
 
\begin{figure}
\centerline {\psfig{figure=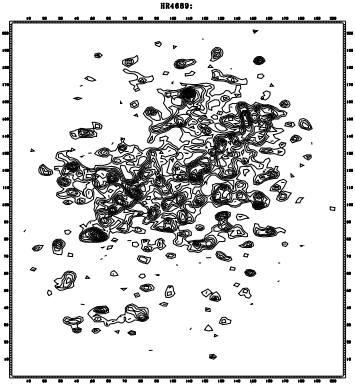,height=5.5cm,angle=270}}
\caption{Specklegram of a binary star, HR4689 obtained at the VBT, Kavalur, 
India.} 
\end{figure}

\begin{figure}
\centerline {\psfig{figure=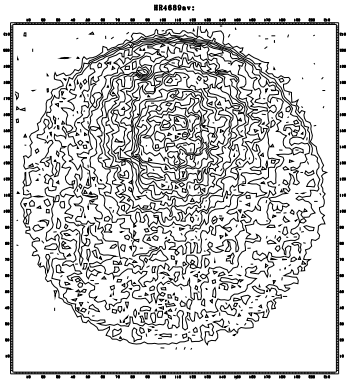,height=5.5cm,angle=270}}
\caption{The result of summing 128 specklegrams of HR4689 demonstrating the
destructions of finer details of the image by the atmospheric turbulence.}
\end{figure}

In general, a high quantum efficiency detector is needed to record magnified 
short-exposure images for such observation. To compensate for atmospherically 
induced dispersion at zenith angles larger than a few degrees, either a 
counter-rotating computer-controlled dispersion-correcting prism or a 
narrow-bandwidth filter is used. 

\subsubsection{Speckle interferometry (SI)}

\label{subsubsec:inter}

Speckle interferometry estimates a `power spectrum' which is the ensemble
average of the squared modulus of an ensemble of FT from a set of specklegrams,
${\cal I}_k({\bf x}), k = t_1, t_2, t_3, \ldots, t_M$. 
The intensity of the image, ${\cal I}({\bf x})$,  
in the case of quasi-monochromatic incoherent source can be expressed as,

\begin{equation} 
{\cal I}({\bf x}) = \int{\cal O}({\bf x}^\prime){\cal S}({\bf x} - 
{\bf x}^\prime) d{\bf x}^\prime. 
\end{equation} 

\noindent 
where ${\cal O}({\bf x}^\prime)$, is an object at a point anywhere in the 
field of view. 

The variability of the corrugated wavefront yields `speckle 
boiling' and is the source of speckle noise that arises from difference in 
registration between the evolving speckle pattern and the boundary of the PSF 
area in the focal-plane. These specklegrams have additive noise contamination, 
${\cal N}_j({\bf x})$, which includes all additive measurement of uncertainties.
This may be in the form of (i) photon statistics noise, and (ii) all distortions
from the idealized iso-planatic model represented by the convolution of 
${\cal O}({\bf x})$ with ${\cal S}({\bf x})$ that includes non-linear 
geometrical distortions. For each of the short-exposure instantaneous records, 
the imaging equation applies,

\begin{equation} 
{\cal I}{(\bf x)} = {\cal O}{(\bf x)}\star{\cal S}{(\bf x)} + {\cal N}({\bf x}),
\end{equation} 

\noindent
Denoting $\widehat{\cal N}({\bf u})$, for the noise spectrum. 
the Fourier space relationship between object and the image is

\begin{equation} 
\widehat{\cal I}({\bf u}) = \widehat{\cal O}({\bf u})\cdot\widehat{\cal S}
({\bf u}) + \widehat{\cal N}({\bf u}). 
\end{equation} 

\noindent 
Taking the modulus square of the expression and averaging over many frames,
the average image power spectrum is,

\begin{equation} 
<|\widehat{\cal I}({\bf u})|^{2}> = |\widehat{\cal O}({\bf u})|^{2} 
\cdot<|\widehat{\cal S}({\bf u})|^{2}> + <|\widehat{\cal N}({\bf u})|^2>. 
\end{equation} 

\noindent
Since $|\widehat{\cal S}({\bf u})|^{2}$ is a random function in which the 
detail is continuously changing, its ensemble average becomes smoother. 

By the Wiener-Khintchine theorem (Mendel and Wolf, 1995), the inverse FT of 
equation (73) gives the autocorrelation of the object, 
${\cal A}[{\cal O}{\bf (x)}]$. 

\begin{equation}
{\cal A}[{\cal O}{\bf (x)}] = {\cal F}^{-} [|\widehat{\cal O}{\bf (u)}|^2],
\end{equation}

\noindent 
In this technique, the atmospheric phase contribution is eliminated but
the averaged signal is non-zero, i.e., 

\begin{equation}
<\widehat{\cal I}^{\cal A}({\bf u})> \neq 0.  
\end{equation}

\noindent
The argument of the equation (72) is given by the expression,

\begin{eqnarray}
arg|\widehat{\cal I}{\bf (u)}|^2 &=& 
\psi({\bf u}) + \theta_1 - \theta_2 + \psi {\bf (-u)} - \theta_1 + 
\theta_2 \nonumber\\
&& = 0. 
\end{eqnarray}

The transfer function of ${\cal S}({\bf x})$, is generally estimated by 
calculating the power spectrum of the instantaneous intensity from an unresolved
star. Saha and Maitra (2001) developed an algorithm, where a Wiener parameter, 
$w_1$, is added to PSF power spectrum. The classic Wiener filter that resulted 
from electronic information theory where diffraction-limits do not mean much, is
meant to deal with signal dependent `colored' noise. In practice, this term is 
usually just a constant, a `noise control parameter' whose scale is estimated
from the noise power spectrum. In this case, it assumes that the noise is white 
and that one can estimate its scale in regions of the power spectrum where the 
signal is zero (outside the diffraction-limit for an imaging system). 

\begin{equation}
|\widehat{O}{\bf (u)}|^2 = \frac{<|\widehat{I}{\bf (u)}|^2>}
{[<|\widehat{S}{\bf (u)}|^2> + w_1]}.
\end{equation}

The SI technique in the case of the components in a 
group of stars retrieves the separation, position angle with 180$^\circ$ 
ambiguity, and the relative magnitude difference at low light levels. 
Figure 4 depicts the autocorrelation of a binary system, HR4689. Another 
algorithm called directed vector autocorrelation (DVA) method is found to be 
effective in eliminating the 180$^\circ$ ambiguity (Bagnuolo et al. 1992). 

\begin{figure} 
\centerline {\psfig{figure=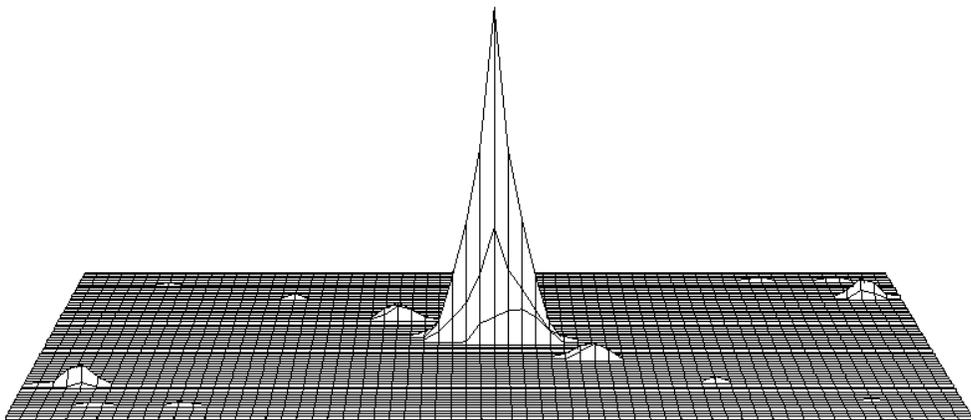,height=5.5cm,angle=270}}
\caption {Autocorrelation of a binary system, HR4689; the second star
in the binary is one of the two distinct bumps and symmetrically placed on 
either side of the main peak. The bumps at the edge of the figure are the 
artifacts.}
\end{figure} 

\subsubsection{Speckle holography}

If a reference point source is available within the iso-planatic patch 
($\sim7^{\prime\prime}$), it is used as a key to reconstruct the target 
in the same way as a reference coherent beam is employed in holographic 
reconstruction (Liu and Lohmann, 1973). Let the point source be 
represented by a Dirac impulse, $A_\circ\delta({\bf x})$ at the 
origin and ${\cal O}_1({\bf x})$ be the nearby object to be reconstructed. The 
intensity distribution in the field of view is 

\begin{equation}
{\cal O}({\bf x}) = A_\circ\delta({\bf x}) + {\cal O}_1({\bf x}). 
\end{equation}

\noindent
The squared modulus of its FT is derived as,

\begin{equation} 
|\widehat{\cal O}({\bf u})|^2 = A_\circ^2 + A_\circ\widehat{\cal O}_1({\bf u})
+ A_\circ\widehat {\cal O}_1^\ast({\bf u})
+ \widehat{\cal O}_1({\bf u})\widehat{\cal O}_1^\ast({\bf u}).   
\end{equation} 
 
\noindent
The inverse FT of this equation translates into, 

\begin{equation}
{\cal A}[{\cal O}{(\bf x)}] = A_\circ^2\delta{(\bf x)} + A_\circ
{\cal O}_1{(\bf x)} + A_\circ{\cal O}_1{(\bf - x)} + 
{\cal A}[{{\cal O}_1}{(\bf x)}],  
\end{equation}
 
\noindent
The first and the last terms in equation (80) are centered at the origin. If the
object is far enough from the reference source, ${\cal O}({\bf x})$, its mirror 
image, ${\cal O}{(\bf - x)}$, is therefore recovered apart from a 180$^\circ$ 
rotation ambiguity. 
 
\subsubsection{Differential speckle interferometry}
 
Differential speckle interferometry (DSI) is a method to observe the objects in
different spectral channels simultaneously and to compute the average 
cross-correlation of pairs of speckle images (Beckers, 1982). 
Let ${\cal O}_1{(\bf x)}$ and ${\cal O}_2{(\bf x)}$, be respectively the source 
brightness distributions at $\lambda_1$ and $\lambda_2$, and 
${\cal I}_1{(\bf x)}$ and ${\cal I}_2{(\bf x)}$, their 
associated instantaneous image intensity distributions. The relation between 
the object and the image in the Fourier space becomes,

\begin{eqnarray} 
\widehat{\cal I}_1({\bf u}) &=& 
\widehat{\cal O}_1({\bf u})\cdot\widehat{\cal S}_1({\bf u}), \\
\widehat{\cal I}_2({\bf u}) &=& 
\widehat{\cal O}_2({\bf u})\cdot\widehat{\cal S}_2({\bf u}), 
\end{eqnarray} 
 
\noindent 
where $\widehat{\cal S}_1{(\bf u)}$ and $\widehat{\cal S}_2{(\bf u)}$ are the 
related transfer functions. The average cross-spectrum is given by,

\begin{equation} 
<\widehat{\cal I}_1{(\bf u)}\widehat{\cal I}_2^\ast{(\bf u)}> = \widehat
{\cal O}_1{(\bf u)}\widehat{\cal O}_2^\ast{(\bf u)}\cdot<\widehat{\cal S}_1
{(\bf u)}\widehat{\cal S}_2^\ast({\bf u})>.
\end{equation} 

\noindent 
The transfer function,
$<\widehat{\cal S}_1{(\bf u)}\widehat{\cal S}_2^\ast{(\bf u)}>$, can be 
calibrated on the reference point source for which,  
$<\widehat{\cal O}_1{(\bf u)}> = <\widehat{\cal O}_2{(\bf u)}> = 1$. If the
two spectral windows are close enough ($\Delta\lambda /\lambda \ll r_\circ/D$), 
the instantaneous transfer function is assumed to be identical in both the
channels [${\cal S}_1 = {\cal S}_2 = {\cal S}$]. Therefore, equation (83) 
becomes,

\begin{eqnarray} 
\widehat{\cal O}_1({\bf u}) &=& \frac{<\widehat{\cal I}_1({\bf u})
\widehat{\cal I}_2^\ast({\bf u})>}{\widehat{\cal O}_2^\ast({\bf u})
<|\widehat{\cal S}({\bf u})|^2>} \nonumber\\
&& = \widehat{\cal O}_2({\bf u})
\frac{<\widehat{\cal I}_1({\bf u})\widehat{\cal I}_2^\ast({\bf u})>}
{<|\widehat{\cal I}_2({\bf u})|^2>}.
\end{eqnarray} 

The noise contributions from two different detectors are uncorrelated, and 
thereby their contributions cancel out. The DSI estimates the ratio,
$\widehat{\cal O}_1({\bf u})/\widehat{\cal O}_2({\bf u})$ and the differential 
image, ${\cal D}_{\cal I}({\bf x})$, is obtained by performing an inverse 
FT of this ratio,

\begin{equation} 
{\cal D}_{\cal I}({\bf x}) =  
{\cal F}^{-1}\left[\frac{<\widehat{\cal I}_1({\bf u})\widehat{\cal I}_2^\ast
({\bf u})>}{<|\widehat{\cal I}_2({\bf u})|^2>}\right], 
\end{equation} 

\noindent 
where ${\cal D}_{\cal I}({\bf x})$ represents an image of the object in the 
emission feature having the resolution of the object imaged in the continuum. 

\subsubsection{Speckles and shadow bands}

When any planetary body of a notable size passes in front of a star, the light 
coming from the latter is occulted. The profiles of stellar occultations by
the Moon show diffraction patterns as the star is being occulted, provided
the data is recorded at high time resolution. The method remains useful 
because of the extraordinary geometric precision it provides. The notable 
advantage of occultation of binary stars is that it can determine relative 
intensities and measure the separations comparable to those measured by long 
baseline interferometers. Speckle surveys have resolved known occulting
binaries down to a separation of about $<$~0.025$^{\prime\prime}$ (Mason, 1996).
Further, this method provides a means of determining the limiting magnitude 
difference of SI. The shortcomings of this technique may be noted due to its 
singular nature: the object may not occult again until one Saros cycle later 
(18.6 yr), and must be limited to a belt of the sky (10\% of the celestial 
sphere). 
 
\subsubsection{Speckle spectroscopy}
 
The application of the SI technique to speckle spectroscopic observations 
enables one to obtain spectral resolution with high spatial resolution 
of astronomical objects simultaneously. The intensity distribution, 
${\cal W}({\bf x})$, of an instantaneous objective prism speckle spectrogram 
is expressed as,

\begin{equation} 
{\cal W}({\bf x}) = \sum_m{\cal O}_m({\bf x}-{\bf x}_m)\star{\cal G}_m({\bf x})
\star {\cal S}({\bf x}), 
\end{equation} 
 
\noindent 
where ${\cal O}_m({\bf x}-{\bf x}_m)$ denotes the intensity of m$^{th}$ object 
pixel and ${\cal G}_m({\bf x})$ is the spectrum of the object pixel. 
In the narrow wavelength bands ($<$30~nm), the PSF, ${\cal S}({\bf x})$, is 
wavelength independent. The objective prism spectrum,
$\sum_m{\cal O}_m({\bf x}-{\bf x}_m)\star{\cal G}_m(\bf x)$, can be 
reconstructed from the speckle spectrograms.

In a speckle spectrograph, either a prism or a grism can be employed to 
disperse 1-d specklegrams (Grieger et al. 1988). An imaging spectrometer uses 
two synchronized detectors to record the dispersed speckle pattern and the 
specklegrams of the object (Baba, Kuwamura et al. 1994); a reflection grating 
acts as disperser. 

\subsubsection{Speckle polarimetry}
 
In general, radiation is polarized and the measurement of polarization 
parameters is important in understanding of the emission mechanisms. Processes 
such as electric and magnetic fields, scattering, chemical interactions, 
molecular structure, and mechanical stress cause changes in the polarization 
state of an optical beam. Applications relying on the study of these changes 
cover a vast area, among them are astrophysics and molecular biology.  
The importance of such observations in astronomy is to obtain 
information such as the size and shapes of dust envelopes around stars, the 
size and shape of the dust grains, and magnetic fields. Among other astronomical
objectives worth investigating are: (i) the wavelength dependence of the degree 
of polarization and the rotation of the position angle in stars with extended 
atmospheres, (ii) the wavelength dependence of the degree of polarization and 
position angle of light emitted by stars present in very young 
($\leq2\times10^6$ years) clusters and associations. 
  
The modified incident polarization caused by the reflection of a mirror  
is characterized by two parameters: (i) the ratio between the reflection 
coefficients of the electric vector components which are perpendicular and 
parallel to the plane of incidence, known as $s$ and $p$ components 
respectively, (ii) the relative phase-shift between these electric vibrations. 
The effect on the statistics of a speckle pattern is the degree of 
depolarization caused by the scattering at the surface. If the light is 
depolarized, the resulting speckle field is considered to be the sum of two
component speckle fields produced by scattered light polarized in two
orthogonal directions. The intensity at any point is the sum of the
intensities of the component speckle patterns (Goodman, 1975). These patterns
are partially correlated; therefore, a polarizer that transmits one of the
component speckle patterns is used in the speckle camera system (Falcke
et al. 1996). The advantage of using a speckle camera over a conventional 
imaging polarimeter is that it helps in monitoring the short-time variability 
of the atmospheric transmission. 

\subsubsection{Speckle imaging of extended objects}

Image recovery is relatively simple where the target is a point source. 
Nevertheless, interferometric observations can reveal the fundamental
processes on the Sun that take place on sub-arcsecond scales 
concerning convection and magnetic fields. The limitations come from (i) the 
rapid evolution of solar granulation that prevents the collection of long 
sequences of specklegrams for reconstruction, (ii) the lack of efficient 
detectors to record a large number of frames within the stipulated time before 
the structure changes. Another major problem of reconstructing images comes from
difficulty in estimating the PSF due to the lack of a reference point source. 
The spectral ratio technique (Von der L\"uhe, 1984), which is based on a 
comparison between long and short-exposure images, has been employed (Wilken et 
al. 1997) to derive Fried's parameter. Models of the speckle transfer 
function (Korff, 1973) and of average short-exposure MTF have also been
applied to compare the observed spectral ratios with theoretical values.  
High resolution solar images obtained during partial solar eclipse may help in 
estimating the seeing effect (Callados and V\`azquez, 1987). The limb of the 
moon eclipsing the sun provides a sharp edge as a reference object. The 
intensity profile falls off sharply at the limb. The departure of 
this fall off gives an indirect estimate of the atmospheric PSF. 
 
\subsection{Other techniques}

\label{subsec:other}

Several other methods, viz., pupil-plane techniques such as wavefront shearing 
interferometry, phase-closure methods, and phase-diversity techniques are also 
employed at single telescope in order to obtain diffraction-limited information.

\subsubsection{Shearing interferometry}

Shearing interferometers make use of the principle of self referencing, that is,
they combine the wavefront with a shifted version of itself to form 
interferences. Fringes are produced by two partially or totally superimposed 
pupil images created by introducing a beam splitter. At each point, interference
occurs from the combination of only two points on the wavefronts at a given 
baseline, and therefore, behaves as an array of Michelson-Fizeau 
interferometers. An important property of these interferometers is their ability
to work with partially coherent light, which offers better S/N ratio on bright 
sources, and is insensitive to calibration errors due to seeing fluctuations and
telescope aberrations. A rotational shear interferometer was used at the 
telescope to map the visibility of fringes produced by $\alpha$~Ori (Roddier 
and Roddier, 1988). In this technique, the 2-d FT is obtained by rotating one 
pupil image about the optical axis by a small angle with respect to the other. 
If the rotation axis coincides with the center of the pupil, the two images 
overlap. All the object Fourier components within a telescope's diffraction 
cutoff frequency are measured simultaneously. 

\subsubsection{Phase-closure methods}

The phase of the visibility may be deduced from a closure-phase that is 
insensitive to the atmospherically induced random phase errors, as well as to 
the permanent phase errors introduced by the imaging systems (Jennison, 1958) 
using three telescopes. The observed phases, $\psi_{ij}$, on the 
different baselines contain the phases of the source Fourier components 
$\psi_{0,ij}$ and also the error terms, $\theta_j, \theta_i$, introduced by 
errors at the individual antennas and by the atmospheric variations at each 
antenna. The observed fringes are represented by the following equations,

\begin{eqnarray}
\psi_{12} &=& \psi_{0,12} + \theta_2 - \theta_1. \\
\psi_{23} &=& \psi_{0,23} + \theta_3 - \theta_2. \\
\psi_{31} &=& \psi_{0,31} + \theta_1 - \theta_3. 
\end{eqnarray}
 
\noindent 
where the subscripts refer to the antennae at each end of a particular 
baseline. The closure phase, $\beta_{123}$ is the sum of phases of the source 
Fourier components and is derived as,

\begin{eqnarray} 
\beta_{123} &=& \psi_{12} + \psi_{23} + \psi_{31}, \\  
&&= \psi_{0,12} + \psi_{0,23} + \psi_{0,31}.
\end{eqnarray} 

\noindent 
This equation implies cancellations of the antennae phase errors.
Using the measured closure phases and amplitudes as observables, the object
phases are determined (mostly by least square techniques, viz., singular
value decomposition, conjugate gradient method). From the estimated object 
phases and the calibrated amplitudes, the degraded image is reconstructed.

Baldwin et al. (1986) reported the measurements of the closure-phases obtained 
at a high light level with a three hole aperture mask set in the pupil-plane of 
the telescope. The non-redundant aperture masking method, in which the 
short-exposure images are taken through a multi-aperture screen, has several 
advantages. These are: (i) an improvement of signal-to-noise (S/N) ratios for 
the individual visibility and closure-phase measurements, (ii) attainment of the
maximum possible angular resolution by using the longest baselines, and (iii) 
built-in delay to observe objects at low declinations. But the system is 
restricted to high light levels, because the instantaneous coverage of spatial 
frequencies is sparse and most of the available light is discarded. 

\subsubsection{Phase-diversity imaging}

Phase-diversity (Gonsalves, 1982, Paxman et al. 1992) is a post-collection
technique that uses a number of intensity distributions encoded by known
aberrations for restoring high spatial resolution detail while imaging in the
presence of atmospheric turbulence. The phase aberrations are estimated from two
simultaneously recorded images. Phase-diverse speckle is an extension of this 
technique, whereby a time sequence of short-exposure image pairs is detected at 
different positions in focus and out of focus near the focal-plane. Incident 
energy is split into two channels by a simple beam splitter: one is collected 
at a conventional focal-plane, the other is defocussed (by a known amount) and a
second detector array permits the instantaneous collection of the latter. It is 
less susceptible to the systematic errors caused by the optical hardware, and 
is found to be more appealing in astronomy (Baba, Tomita et al. 1994, Seldin 
and Paxman, 1994). 

\subsection{Adaptive-optics (AO)}

\label{subsec:optics}

Significant technological innovations over the past several years
have made it possible to correct perturbations in the wavefronts
in real time by incorporating a controllable phase distortion in
the light path, which is opposite to that introduced by the atmosphere 
(Babcock, 1953, Rousset et al. 1990). This technique has advantages over 
post-detection image restoration techniques that are limited by noise. 
Adaptive-optics (AO) systems are employed in other branches of physics as well. 
Liang et al. (1997) have constructed a camera equipped with adaptive-optics that
allows one to image a microscopic size of single cell in the living human 
retina. They have shown that a human eye with adaptive-optics correction can 
resolve fine gratings that are invisible to the unaided eye. AO systems are 
useful for spectroscopic observations, as well as for low light level imaging 
with future very large telescopes, and ground-based LBOIs. 

\subsubsection{Greenwood frequency}

Turbulence cells are blown by wind across the telescope aperture, hence, the 
wind velocity dictates the speed with which a corrective action must be taken. 
Greenwood (1977) derives the mean square residual wavefront error as a 
function of servoloop bandwidth for a first order controller, which is given by,

\begin{equation}
\sigma^2_{cl} = \left(\frac{{\it f}_G}{{\it f}_{3db}}\right)^{5/3} \, \, 
{rad}^2,
\end{equation}

\noindent 
where ${\it f}_{3db}$ is the closed loop bandwidth of the wavefront compensator 
and ${\it f}_G$ the Greenwood frequency that is defined by the relation,

\begin{equation}
{\it f}_G = \frac{0.426{\it v}}{r_\circ},
\end{equation}

\noindent 
where ${\it v}$ is the wind velocity in the turbulent layer of air. For imaging 
in near-IR to ultraviolet, the AO system bandwidths need to have a response time
of order several hundreds to 1~KHz. It is easier to achieve 
diffraction-limited information using AO systems at longer wavelengths since
$r_\circ$ is proportional to the six-fifths power of the wavelength,

\begin{equation}
r_\circ \propto \lambda^{6/5}.
\end{equation}

\noindent 
The above equation (94) implies that the width of seeing limited images,
$1.22\lambda/r_\circ \propto \lambda^{-1/5}$ varies with $\lambda$. 
The number of degrees of freedom, i.e., the number of actuators on the 
deformable mirror (DM) and the number of sub-aperture in the wavefront sensor, 
in an AO system should be determined by the following equation,

\begin{equation}
(D/r_\circ)^2 \propto \lambda^{-12/5}.
\end{equation}

\subsubsection{AO imaging system}

The required components for implementing an AO system are wavefront sensing, 
wavefront phase error computation and a flexible mirror whose surface is 
electronically controlled in real time to create a conjugate 
surface enabling compensation of the wavefront distortion (Roggemann et al. 
1997 and references therein). In order to remove the low frequency tilt error, 
generally the incoming collimated beam is fed by a tip-tilt mirror. After 
traveling further, it reflects off of a DM that eliminates high frequency 
wavefront errors. A beam-splitter devides the beam into two parts: one is 
directed to the wavefront sensor to measure the residual error in the wavefront 
and to provide information to the actuator control computer to compute the DM 
actuator voltages and the other is focused to form an image. 

Implementation of the dynamically controlled active optical components 
consisting of a tip-tilt mirror system in conjunction with closed-loop control 
electronics has several advantages: (i) conceptually the system is 
simple, (ii) fainter guide stars increase the sky coverage, 
and (iii) field of view is wider (Glindemann, 1997). These systems are limited
to two Zernike modes (x and y tilt), while a higher order system compensating 
many Zernike mode (Zernike polynomials are an orthogonal expansion over unit 
circle) is required to remove high frequency errors. 
  
A variety of DMs have been developed for the applications of (i) high energy 
laser focusing, (ii) laser cavity control, (iii) compensated imagery through 
atmospheric turbulence etc. Several wavefront sensors such as (i) lateral 
shearing interferometer, (ii) Shack-Hartman (SH) sensor, and (iii) the curvature
sensor are in use as well. The technical details of these DMs and sensors are 
enumerated in the recent literatures (Roggemann et al. 1997, Roddier, 1999, and 
Saha, 1999).

The phase reconstruction method finds the relationship between the measurements 
and the unknown values of the wavefronts and can be categorized as being either
zonal or modal, depending on whether the estimate is either a phase value in a 
local zone or a coefficient of an aperture function (Rousset, 1999). In the case
of curvature sensing, the computed sensor signals are multiplied by a control 
matrix to convert wavefront slopes to actuator control signals, the output of 
which are the increments to be applied to the control voltages on the DM. 
A conjugate shape is created using these data by controlling a DM.
 
The real time computation of the wavefront error, as well as correction of 
wavefront distortion involves digital manipulation of wavefront sensor data
in the wavefront sensor processor, the re-constructor and the low-pass filter,
and converting to analog drive signals for the DM actuators. 
The functions are (i) to compute sub-aperture gradients, phases at 
the corners of each sub-aperture, low-pass filter phases, and (ii) 
to provide actuator offsets to compensate the fixed optical system errors 
and real time actuator commands for wavefront corrections.
 
\subsubsection{Artificial source}

An AO system requires a reference source to monitor the atmospheric aberrations
for measuring the wavefront errors, as well as for mapping the phase on the 
entrance pupil. It is generally not possible to find a sufficiently bright
reference star close enough to a target star. In order to alleviate this 
limitation, many observatories are currently implementing the artificial laser 
guide star (Ragazzoni and Bonnacini, 1996, Racine et al. 1996, Lloyd-Hart
et al. 1998). However, the best results are still obtained with natural guide 
stars. 

An artificial guide star can be obtained using either the resonance scattering 
by sodium in the mesosphere at 90~km (Foy and Labeyrie, 1985) or Rayleigh 
scattering between 10 and 20~km altitude (Fugate et al. 1994). 
A pulsed laser (tuned to the sodium $D_2$ line to excite sodium 
atom) is used to produce a bright compact glow in the upper atmosphere. 
Concerning the flux backscattered by a laser shot, Thompson and 
Gardner (1988) stressed the importance of investigating two basic problems: 
(i) the cone effect which arises due to the parallax between the remote 
astronomical source and artificial source, and (ii) the angular anisoplanatic 
effects. These effects can be restored by imaging the various turbulent layers 
of the atmosphere onto different adaptive mirrors (Tallon et al. 1988). 
Scattering of the upward propagating laser beam is due to Rayleigh scattering,
mostly by $N_2$ molecules. Mie scattering by aerosol or cirrus clouds may be 
important at lower altitudes but are usually variable and transient.
  
\subsubsection{Multi-conjugate adaptive-optics}

Multi-conjugate AO system enables near-uniform compensation for the atmospheric 
turbulence over considerably larger field of view than can be corrected with 
normal AO system. This method employs an ensemble of guide stars that allows for
3-d tomography of the atmospheric turbulence and a number of altitude-conjugate
DMs to extend the compensated field of view. However, its performances depends 
on the quality of the wavefront sensing of the individual layers. Ragazzoni et 
al. (2000) have demonstrated this type of tomography. This new technique pushes 
the detection limit by $\sim$1.7~mag on unresolved objects with respect to
seeing limited images; it also minimizes the cone effect. This technique will be
useful for the extremely large telescopes of 100~m class, e.g., the 
OverWhelmingly Large (OWL) telescope (Dierickx and Gilmozzi, 1999). However, the
limitations are mainly related to the finite number of actuators in a DM, 
wavefront sensors, and guide stars.
 
\subsubsection{Adaptive secondary mirrors}

The usage of a adaptive secondary mirror (ASM) for corrections making relay 
optics obsolete that are required to conjugate a DM at a reimaged pupil, as well
as to minimize thermal emission is a new key innovation (Bruns et al. 1997). 
The other notable advantages are (i) enhanced photon throughput that 
measures the proportion of light which is transmitted through an optical 
set-up, (ii) introduction of negligible extra IR emissivity, (iii) causes no 
extra polarization, and (iv) non-addition of reflective losses (Lee et al. 
2000). Due to the interactuator spacing, the resonant frequency of such a 
mirror may be lower than the AO bandwidth. The ASM system uses a SH sensor with 
an array of small lenslets, which adds two extra refractive surfaces to the 
wavefront sensor optical beam (Lloyd-Hart, 2000). An $f/15$ AO secondary with 
336 actuators is in the final stages of testing and will be installed on the
6.5~m Telescope of MMT observatory, Mt. Hopkins, Arizona in 2002 (Wehinger, 
2001).
 
\subsubsection{High resolution coronagraphy}

The relevance of using coronagraphy in imaging or spectroscopy of faint 
structure near a bright object can be noted in terms of reducing the light 
coming from the central star, and filtering out of the light at low spatial 
frequency; the remaining light at the edge of the pupil corresponds to high 
frequencies. A coronagraph reduces off-axis light from an on-axis source with 
an occulting stop in the image-plane as well as with a matched Lyot stop in the 
next pupil plane. While using the former stop the size of the latter pupil 
should be chosen carefully to find the best trade-off between the throughput and
image suppression. The limitations come from the light diffracted by the 
telescope and instrument optics. Coronagraphy with dynamic range can be a 
powerful tool for direct imaging of extra-solar planets. Nakajima (1994) 
estimates that imaging with such a method with a low order AO system in a 6.5~m 
telescope could detect Jupiter-size extra-solar planets at separation 
$\sim$1.5$^{\prime\prime}$ with S/N of 3 in 10$^4$~s. Rouan et al. (2000) 
describes a four-quadrant phase-mask coronagraph where a detection at a contrast
of more than 8~mag difference between a star and a planet is feasible.

\subsection{Dark speckle method}

The dark speckle method uses the randomly moving dark zones between speckles 
$-$ `dark speckles'. It exploits the light cancellation effect in a random 
coherent field; highly destructive interferences that depict near black 
spots in the speckle pattern (Labeyrie, 1995) may occur occasionally.
The aim of this method is to detect faint objects around a star when the
difference of magnitude is significant. If a dark speckle is at the location of 
the companion in the image, the companion emits enough light to reveal itself. 

The required system consists of a telescope with an AO system, a coronagraph, a 
Wynne corrector, and a fast photon-counting camera with a low dark noise. 
If a pixel of the photon-counting camera is illuminated by the
star only (in the Airy rings area), because of the AO system, the
number of photons in each pixel, for a given interval (frame), is
statistically given by a Bose-Einstein distribution. The number of photons
per frame in the central peak of the image of a point source obeys a
classical Poisson distribution. For the pixels containing the image of the
companion, the number of photons, resulting from both the star and the
companion, is given by a different distribution (computed by mixing
Bose-Einstein and Poisson distributions). One noticeable property is that
the probability to get zero photons in a frame is very low for the pixels
containing the image of the companion, and much higher for the pixels
containing only the contribution from the star. Therefore, if the
`no photon in the frame' events for each pixel is counted, and for a very large 
number of frames, a `dark map' can be built that will show the pixels for which
the distribution of the number of photons is not Bose-Einstein type, therefore
revealing the location of a faint companion. The role of the
Wynne corrector is to give residual speckles the same size regardless the
wavelength. Otherwise, dark speckles at a given wavelength would be
overlapped by bright speckles at other wavelengths. With the current technology,
by means of the dark speckle technique at a 3.6~m telescope should allow 
detection of a companion with $\Delta m_k\approx$6-7~mag. Figures 5 and 6 
depict the coronagraphic images of the binary stars, HD192876 and HD222493,
respectively (Boccaletti et al. 2001); the data were obtained with ADONIS in the
K band (2.2~$\mu$m) on the European Southern Observatory's (ESO)  
3.6~m telescope. Due to the lack of a perfect 
detector (no read-out noise) at near-IR band, every pixel under the defined 
threshold (a few times the read-out noise) is accounted as a dark speckle. 

\begin{figure} 
\centerline {\psfig{figure=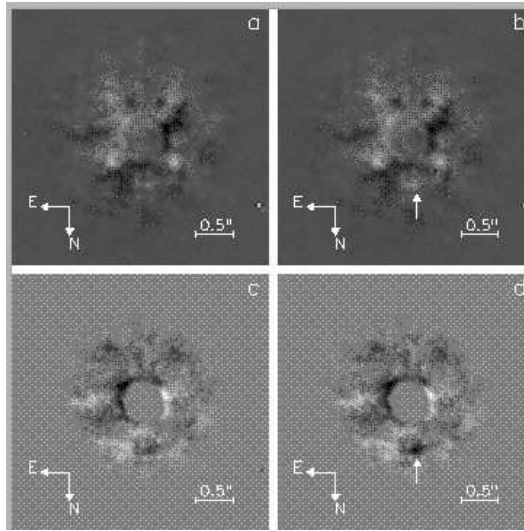,height=7cm}}
\caption {Coronagraphic images of the star HD192876 (Courtesy: A. Boccaletti).
An artificial companion is added to the data to assess the detection threshold 
($\Delta m_K$=6.0~mag, $\rho=0.65"$); (a) direct image : co-addition of 
400$\times$60~ms frames, (b) same as
(a) with a $\Delta m_K$=6.0~mag companion (SNR~1.8), (c) dark speckle analysis,
and (d) dark speckle analysis with the companion (SNR~4.8); the detection
threshold is about $\Delta m_K$=7.5~mag on that image, i.e an improvement of
1.5~mag compared to the direct image (Boccaletti et al. 2001).}   
\end{figure} 

\begin{figure} 
\centerline {\psfig{figure=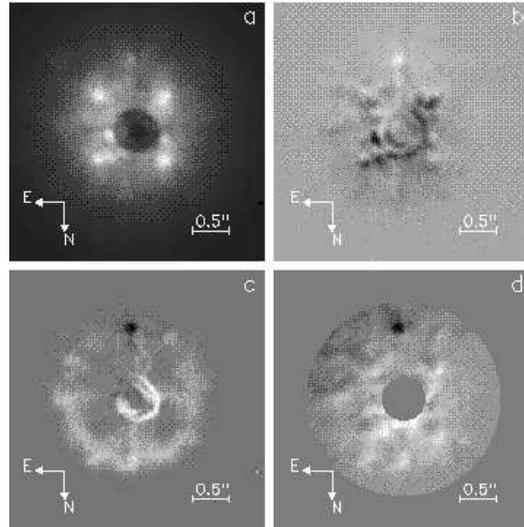,height=7cm}}
\caption {Coronagraphic images of the binary star HD222493 
($\Delta m_K$=3.8~mag, $\rho=0.89"$); (a) direct image: co-addition of 
600$\times$60~ms frames, 
(b) subtraction of the direct image with a reference star (SNR=14.6), (c) dark 
speckle analysis (constant threshold) and subtraction of a reference star, and 
(d) dark speckle analysis (radial threshold) and subtraction of a reference 
star (SNR=26.7) (Boccaletti et al. 2001: Courtesy: A. Boccaletti).}
\end{figure} 

Phase boiling, a relatively new technique that consists of adding a small amount
of white noise to the actuators in order to get a fast temporal decorrelation
of the speckles during long-exposure acquisition, may produce better results.
Aime (2000) has computed the S/N ratio for two different cases: short-exposure 
and long-exposure. According to him, even with an electron-noise limited 
detector like a CCD or a near-IR camera multi-object spectrometer (NICMOS), the 
latter can provide better results if the halo has its residual speckles smoothed
by fast residual `seeing' acting during the long-exposure than building a dark 
map from short-exposures in the photon-counting mode. Artificial very fast 
seeing can also be generated by applying fast random noise to the actuators, at 
amplitude levels comparable to the residual seeing left over by the AO system. 

The question is, what is easiest: dark speckle analysis or a `hyper-turbulated' 
long-exposure? Labeyrie (2000) made simulations supporting the Aime's (2000) 
results. Boccaletti (2001) has compared the dark speckle signal-to-noise ratio 
(SNR) with the long-exposure SNR (Angel, 1994). The speckle lifetime has to be 
of order 0.1~ms. Currently it is impossible to drive a DM at this frequency 
(10~kHz). With the 5~m Palomar telescope Boccaletti (2001) tried to smooth the 
speckle pattern by adding a straightforward 
random noise on the actuators (the DM is equipped with 241 actuators) at maximum
speed of 500~Hz. Effectively, the halo is smoothed, but its intensity is also
increased, so that the companion SNR is actually decreased. Blurring the speckle
pattern would probably require wavefront sensor telemetry; implementation 
of a hyper-turbulated long-exposure at the Palomar is still under study
(Boccaletti, 2001). 

\subsection{High resolution sensors}

\label{subsec:sensors}

All the techniques that are described above require a high quality sensor so as 
to enable one to obtain snap shots with a very high time resolution of the order
of (i) frame integration of 50~Hz, or (ii) photon recording rates of a 
few MHz. The performance relies on the characteristics of such sensors, e.g., 
(i) the spectral bandwidth, (ii) the quantum efficiency, (iii) the detector 
noise that includes dark current, read-out and amplifier noise, (iv) the time 
lag due to the read-out of the detector, and (v) the array size and the spatial 
resolution.

\subsubsection{Frame-transfer camera systems}
 
The frame-transfer intensified CCD (ICCD) camera employs micro-channel plate 
(MCP) as an intensifier. The photoelectron is accelerated into a channel of 
the MCP releasing secondaries and producing an output charge cloud of about 
$10^3 - 10^4$ electrons with 5~-~10 kilovolt (KV) potential. With further 
applied potential of $\sim$5~-~7~KV, these electrons are accelerated to impact 
a phosphor, thus producing an output pulse of $\sim10^5$ photons. These photons
are directed to the CCD by fiber optic coupling. The main disadvantage of 
such system is the poor gain statistics results in the introduction of a noise 
factor between 2 and 3.5. Recent development of a non-intensified CCD 
device which effectively reduces readout noise to less than one electron rms
has enabled substantial internal gain within the CCD before the signal reaches 
the output amplifier (Mackay et al. 2001). Such a detector, although the photon
counting performance appears moderate for the moment, shows promise 
for quantitative measurement of diffraction-limited stellar images. 
  
\subsubsection{Photon-counting detectors}
  
The marked advantage of a photon-counting system is that of reading the signal 
a posteriori to optimize the correlation time of short-exposures in order to 
overcome the loss of fringe visibility due to the speckle lifetime; 
the typical values for an object of m$_v$ = 12 over a field of 
2.5$^{\prime\prime}$ are less than 50~photons/ms through a narrow-band filter. 
The other notable features are, (i) capability of determining the position of a 
detected photon, (ii) ability to register individual photons with equal 
statistical weight and produces signal pulse (with dead time of a few ns), and 
(iii) low dark noise. 

The major short comings of the photon-counting system that is based on frame 
integration (Blazit, 1986) arise from the (i) calculations of the 
coordinates which are hardware-limited, and (ii) limited dynamic range of the 
detector. Non-detectability of a pair of photons closer than a minimum
separation by the detector yields a loss in high frequency information; this,
in turn, produces a hole in the center of the autocorrelation $-$ Centreur hole,
resulting in the degradation of the power spectra or bispectra (FT of triple 
correlation) of speckle images. 

Several 2-d photon-counting sensors that allow recording of the position and
time of arrival of each detected photons have been developed such as, 
(i) precision Analog Photon Address (PAPA; Papaliolios et al. 1985), (ii) 
resistive anode position detector (Clampin et al. 1988), (iii) multi anode 
micro-channel array (MAMA; Timothy, 1993), (iv) wedge-and-strip anodes, (v) 
delay-line anodes, (vi) silicon anode detector etc. Baring PAPA which is based 
on a high gain image intensifier and a set of photomultiplier tubes, these 
sensors detect the charge cloud from a high gain MCP. They provide spatial event
information by means of the position sensitive readout set-up; the encoding 
systems identify each event's location. The short-comings of the MCPs are 
notably due to its local dead-time which essentially restricts the conditions 
for use of these detectors for high spatial resolution applications. These 
constraints are also related with the luminous intensity and the pixel size. 

\subsubsection{Infrared sensors}

In the infrared band, no photon-counting is possible with the current 
technology. Nevertheless, a near-IR focal-plane array, NICMOS, has been
developed. It consists of 256$\times$256 integrating detectors organized in four
independent 128$\times$128 quadrants and is fabricated in HgCdTe 
grown on a saphire substrate that is very rugged and provides a good thermal 
contraction match to silicon multiplexer (Cooper et al. 1993).  
The typical NICMOS3 FPAs have read noise less than 35~e- with less than 
1~e-/sec detector dark current at 77~K and broadband quantum efficiency 
is better than 50\% in the range of 0.8 to 2.5~$\mu$m.  

\section{DILUTE-APERTURE INTERFEROMETRY}

\label{sec:dilute}
 
Modern technology has solved many of the problems that were pioneered by 
Michelson and Pease (1921). The light collected by an array of separated 
telescopes could be coherently combined to measure the Fourier components of the
brightness distribution of a star. While at an interesting stage of 
development, currently of more limited imaging capabilities, the 
following sub-sections elucidate the current state of the art of such arrays. 

\subsection{Aperture-synthesis interferometry}

\label{subsec:synthesis}

The potential of the aperture-synthesis interferometry in the optical domain is 
demonstrated by the spectacular images produced with aperture-masking of a 
single telescope (Tuthill et al. 2000). The principle behind this method may be 
considered as the operation of a conventional filled-aperture telescope composed
of $N$ elemental areas, in which there are $N(N~-~1)/2$ independent baselines, 
with $N~-~1$ unknown phase errors. This implies that by using many telescopes 
in an interferometric array, most of the phase information can be retrieved. 
The signal in the $n$th area due to a source of emission is expressed as,

\begin{equation}
V_n = {\it a}_n cos(\omega t + \psi_n),
\end{equation}

\noindent 
where ${\it a}_n$ is the amplitude of the signal and $\psi_n$ the relative
phase of the radiation.
If these signals are added together vectorily and time averaged, the 
intensity of the light ${\cal I}_n$ is derived as, 

\begin{eqnarray}
{\cal I}_n &\propto &\frac{1}{2}\sum^N_{j=1}\sum^N_{k=1}{\it a}_j{\it a}_k
cos(\psi_j - \psi_k) \nonumber\\
&&= \frac{1}{2}\sum^N_{j=1}{\it a}_j^2 + \sum_{j=1}^{N-1}\sum^N_{k=j+1}
{\it a}_j{\it a}_kcos(\psi_j - \psi_k).
\end{eqnarray}

\noindent 
The first term is proportional to the sum of the power received by the 
elementary areas. The resolving power is derived from the cross product. 
Each term can equally be measured with two elementary areas in 
positions $j$ and $k$. The term $\psi_j - \psi_k$ is expressed as,

\begin{equation}
\psi_j - \psi_k = \frac{2\pi}{\lambda}{\bf B}_{jk}\cdot {\bf s},
\end{equation}

\noindent 
where ${\bf B}_{jk}$ is the separation of the two elemental areas, ${\bf s}$
the unit vector defining to the source.

\subsubsection{Aperture-synthesis imaging}

Generally three or more telescopes are required for aperture-synthesis, two
can suffice if the object includes a point source usable as a phase reference.
Different spectral channels are employed for differential visibility 
measurements (continuum and spectral line) and use data from one part of the 
spectrum such as the continuum emission to calibrate another part (Mourard et 
al. 1989). The continuum channel is supposed to originate from the unresolved 
region of a star, e.g., photosphere and the spectral line centered on a part of 
the spectrum created in an extended region, such as circumstellar medium. The 
phase-referenced technique with a fringe-tracking
channel was also used at Mark~III interferometer to recover the phase 
information (Quirrenbach et al. 1996). Another convenient method is to measure 
the instantaneous phases of fringes from a bright point source that lies 
within the iso-planatic patch in order to correct the corrupted phases on the 
target. Shao and Colavita (1992) used fringe-phase information to determine 
precise relative positions of nearby stars. 

The direct measurements of the closure phase together with the measurements of
visibility amplitude allow one to reconstruct an image of any object using 
three or more independent telescopes. This technique has been successfully 
demonstrated by Baldwin et al. (1998) in the visible band at the Cambridge 
optical aperture-synthesis telescope (COAST). Each pair of telescopes in an 
array yields a measure of the amplitude of 
the spatial coherence function of the object at a spatial frequency 
${\bf B}/\lambda$. In order to make an image from an interferometer, one needs 
estimates of the complex visibilities over a large portion of the $(u, v)$ 
plane, both the amplitudes and phases. The $(u, v)$ coordinates corresponding 
to a snapshot projection of the baseline that are sampled from a star of
declination $\delta_\ast$, when its hour angle is $H$, is given by (Fomalont and
Wright, 1974):

\begin{eqnarray}
u &=& ({\rm B}'_{EW}\cos H-{\rm B}'_{NS}\sin\theta_l\sin H)/\lambda \\
v &=& ({\rm B}'_{EW}\sin\delta_\ast\sin H \nonumber\\
&&+{\rm B}'_{NS}\left(\sin\theta_l
\sin\delta_\ast\cos H+\cos\theta_l\cos\delta_\ast\right))/\lambda,
\end{eqnarray}

\noindent 
where ${\rm B}'_{EW}$ and ${\rm B}'_{NS}$ are orthogonal
East-West and North-South components of the baseline vector at the ground of an 
interferometer located at the terrestrial latitude $\theta_l$. 

\subsubsection{Astrometry}

The accurate determination of relative-positions of stars will provide crucial 
data for astrophysics. For example, precise parallax distances of Cepheids will 
help to establish a period/absolute magnitude relationship in order to calibrate
distances of galaxies, thus reducing the uncertainty on the value of $H_0$. 
An interferometer measures the angle projected onto baseline. Stellar fringes
must be observed at two or more baseline orientations to determine two angular
coordinates of an astronomical object. Hipparcos (1997) uses the phase-shift
measurement of the temporal evolution of the photometric level of two stars
seen drifting through a grid. 

\begin{figure} 
\centerline {\psfig{figure=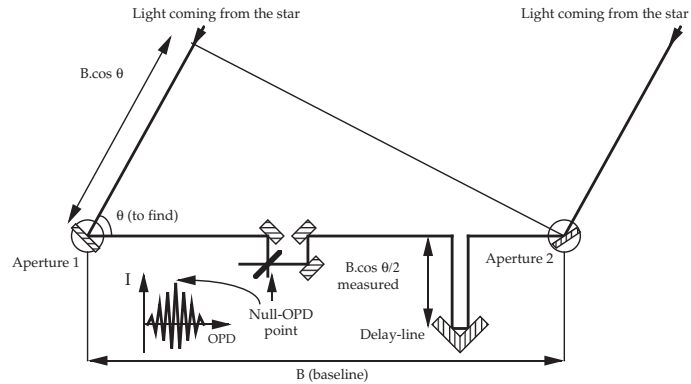,height=5cm}}
\caption {Principle of an interferometer for astrometry (Saha and Morel, 2000)}
\end{figure} 

For a two-aperture interferometer (see figure 7) the external optical delay 
$d$, while an object with an angle $\theta$ is observed in a broad spectral 
range (i.e. white light), is:

\begin{equation}
d=|{\bf B}| \times \cos\theta. 
\end{equation}

\noindent 
This delay can be determined from the position of the optical delay-line of the
instrument set up such that the central fringe of the interference pattern
appears in a narrow observation window. The position, as well as $|{\bf B}|$,
are measured by laser metrology. Hence, $\theta$ is deduced with a high
precision. For ground-based interferometers, the baseline is fixed to the Earth 
and will rotate with the Earth, while in space the interferometer must reorient 
the baseline to measure both angular coordinates.  For a space-borne 
interferometer, the issue is to find a reference for the angle measured. 
Usually, a grid of far objects like quasars are used as a reference frame. 
There are two modes of observation possible: the `wide-angle'
and the `narrow-angle' modes. In wide-angle mode, the large angle difference
between the reference and the studied object usually requires collector motions.
In narrow-angle, the two objects are in the field of view of the instrument,
therefore, no motions are required and the accuracy of the measurement is
improved. However, it is difficult to have always a correct reference star
within the field of view for any studied object. Narrow-angle astrometry is,
therefore, more suitable for wobble characterization. 

\subsubsection{Nulling interferometry}

Nulling interferometry was first proposed by Bracewell (1978) for applications 
in radio astronomy. Such an interferometer could be employed at the upcoming
large interferometers to observe faint structures close to non-obscured
central sources. This technique can also be used in space to
search for extra-solar Earth-like planets through their thermal emission and
to determine the atmospheric signature of life with spectroscopic analysis 
(Angel et al. 1986, Hinz et al. 1998). Here the light collected by two apertures
is combined to generate a deep destructive interference 
fringe at the stellar position, thus selectively nulling the star by many orders
of magnitude related to the surrounding off-axis environs, such as a planetary 
system. Basically, a $\pi$ phase-shift is introduced in one wavefront segment, 
so that when it interferes with another segment of the same wavefront and 
perfect cancellation is achieved. Therefore, the 
central fringe of the interference pattern is dark, allowing the fringe pattern 
from a faint object to appear. Unlike a coronagraph, where useful imaging is 
possible beyond several Airy radii from the on-axis stellar source, this method 
is expected to be effective within the core of a telescope's PSF, and so can be 
employed for stars at greater distances, where a larger sample is available 
(Serabyn, 2000). The quality of a nulling is defined by the `null depth' $N_d$:

\begin{equation}
N_d=(1-{\cal V}\cos\varphi_e)/2\approx(\pi\sigma_{\rm \Delta\psi}/\lambda)^2,
\end{equation}

\noindent 
where $\varphi_e$ is the phase error between the two recombined beams,
${\cal V}$ the fringe visibility modulus, and $\sigma_{\rm \Delta\psi}$ the 
standard deviation of the optical path difference (OPD) between the two beams. 

\begin{figure} 
\centerline {\psfig{figure=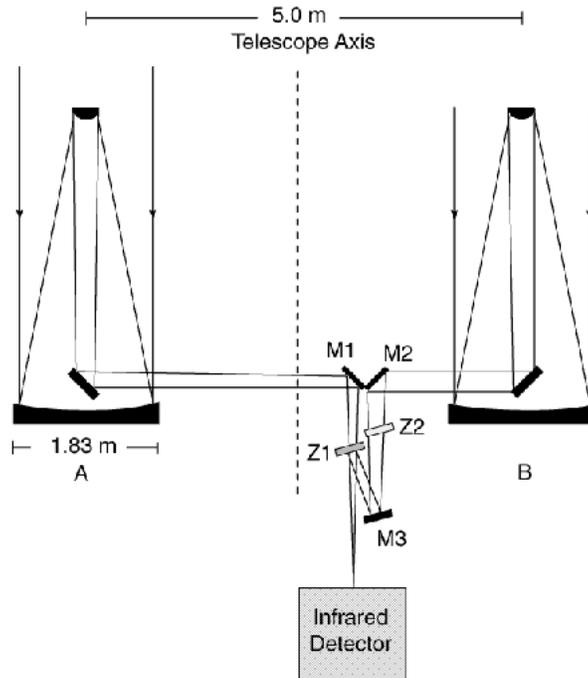,height=9cm}}
\caption {Schematic of the nulling interferometer at the MMT (Hinz et al. 
1998: Courtesy: P. M. Hinz). The wavefronts from the two co-mounted
telescopes {\tt A} and {\tt B} are translated via three mirrors 
{\tt M1, M2, M3} for superposition without relative rotation or tilt. The
zinc selenide beamsplitters {\tt Z1, Z2} are used to adjust the path length.} 
\end{figure} 
 
To create the $\pi$ phase-shift, a few important techniques, such as (i) 
using roof reflectors to achieve a reversal of sign of the electric vector
and (ii) introducing a precise thickness of glass whose index acts to retard 
all wavelengths by very nearly one-half wavelength may be used. Serabyn (2000) 
reported satisfactory nulling results obtained with fiber-coupled rotational 
shearing interferometer in visible wavelength. Hinz et al. (1998) have 
demonstrated the viability of nulling interferometry using two 1.8~m mirrors
of the original six-mirror MMT. They have detected the thermal image of the 
surrounding, circumstellar dust 
nebula around $\alpha$~Ori. Hinz et al. (2001) have measured the spatial 
extent of the mid-IR emission for a few Herbig Ae stars as well. Figure 8 
depicts the schematic of nulling interferometer at the MMT (Hinz et al. 1998).

\subsection{Fundamental limitations and technical challenges}

\label{subsec:technical}

Atmospheric seeing affects the measurements of fringe visibility by introducing 
phase aberrations across the wavefronts incident on the interferometer, the 
relative phase of the wavefronts at the apertures changes with time, and also 
varies the optical paths through the arms. Fringes are to be 
obtained in a time as short as atmospheric fluctuations ($\sim$0.01~s).
The optical interferometers require accurate alignments, high stability, full
control of any effect decreasing visibility. These limitations
come from the parameters, viz., (i) precise determination of visibility, (ii) 
sensitivity in measuring weak sources, (iii) accurate measurement of fringe
phases, and (iv) availability of range of baselines. 

\subsubsection{Signal-to-noise (S/N) ratio} 

The S/N ratio with which the visibility can be measured is a function of 
$N_p^\prime{\cal V}^2$. The fringe S/N ratio is given by the expression (Lawson,
1995),

\begin{equation}
{\rm SNR}\propto{N_p^\prime{\cal V}^2\over\sqrt{1+0.5 \times N_p^\prime
{\cal V}^2}}, 
\end{equation}

\noindent 
where $N_p^\prime$ is the number of photons detected per sub-aperture per 
integration time. 

The dependence on $N_p^\prime{\cal V}^2$ implies that interferometry becomes 
increasingly difficult for faint sources, particularly for those with complex 
structures. The $N_p^\prime{\cal V}^2$ limit can be addressed in various ways, 
viz., (i) using larger sub-apertures, (ii) slicing of the image at the entrance 
of the spectrograph, (iii) bootstrapping, and (iv) tracking fringes on a point 
source to increase integration time on the target.  

\subsubsection{Delay-lines}

\label{subsubsec:lines}

The real art of developing interferometers is to combine the beams in phase
with each other after they have traversed exactly the same optical path from
the source through each telescopes down to the beam combination point. The 
pathlengths of the two arms need to be equalized and maintained to a fraction 
of $c/{\Delta\nu}$. A correct determination of ${\bf B}$ that is unstable over 
time is also necessary; the situation becomes complicated in the presence of 
atmosphere. The paths are 
made equal by adjusting the position of mirrors in the optical delay-line that 
corrects the drift induced by the diurnal rotation of the tracked star. Of 
course, the difficulty comes from avoiding various aberrations and vignetting, 
particularly when light is fed through long and narrow pipes. The optical
delay in terms of telescope and source parameters translates into;

\begin{eqnarray}
d &=& {\rm B}'_{EW}\cos\delta_\ast sin H \nonumber\\ 
&& - {\rm B}'_{NS}(\sin\theta_l\cos H \cos\delta_\ast 
- cos\theta_l\sin\delta_\ast).
\end{eqnarray}

Until recently, the beam-recombining optical devices at both I2T and Grand 
interf\'erom\`etre \`a deux t\'elescopes (GI2T; Labeyrie et al. 1986) were 
kept on a computer controlled motor driven carriage parallel to the baseline in 
order to compensate the OPD. Of late, a delay-line that is movable is inserted 
in one of the arms of the I2T, by means of a cat's eye system; the other arm was
equipped with a fixed delay-line (Robbe et al. 1997). The present recombiner, 
recombineur pour grand interf\'erom\`etre (REGAIN) at GI2T also uses 
a delay-line featuring a cat's eye reflector with variable curvature mirror. 

A few interferometers, viz., Mark~III stellar interferometer (Shao et al. 
1988), US Navy prototype optical interferometer (NPOI; Armstrong et al. 1998),
Infrared Optical Telescope Array (IOTA; Carleton et al. 1994) 
use the vacuum delay-lines. Mark~III interferometer delay lines used laser
interferometers to measure the position of the delay line carts and nested 
servo loops for fine control of the OPD.  
Delay-lines of IOTA, have a 28~m travel `long' delay-line which 
is moved each time a new object is observed, when the OPD to 
compensate is very different. It does not move during fringe acquisition. IOTA
has a second `short' delay-line of about 2~m travel which tracks the
sidereal motion during fringe acquisition. Both the delay-lines use a dihedral
(two plane mirrors at 90$^\circ)$ mounted on a carriage. For the long
delay-line, the carriage is moved by a pulley-and-cable system powered by a
stepper motor, while for the short delay-line a linear motor system allows 
precise motion of the carriage (10~nm steps). For both delay-lines, measurement
of the carriage position is done by a laser metrology system using the 
Doppler-Fizeau effect of a laser beam sent to the carriage and bouncing back
(Morel, 2000).

Another method known as group-delay tracking, based on the integration of the 
moduli of all the computed FTs (Lawson, 1994, Lawson et al. 1998), is used with 
the interferometers SUSI, and COAST. The group-delay is proportional
to the rate of change of phase as a function of wavenumber, evaluated at the 
center of the band. This delay can be measured if the combined beams from an 
interferometer are dispersed in a spectrometer. The group-delay tracking yields 
a peak whose position is proportional to the OPD. 

\subsubsection{Beam recombination}

\label{subsubsec:recombination}

Fringes can be obtained either by utilizing the concept of merging speckles 
(Labeyrie, 1975) or by employing pupil-plane configuration (Tango and Twiss, 
1980). At the initial stage, the beam recombining optics that were employed at 
both I2T and GI2T consisted of (i) a recombining element for reconfiguring the 
pupil and fixing the fringe spacing, (ii) an image slicer (a series of 10 
wedges of different angles cemented on a field lens 
slices the image), (iii) the compensating system of the atmospheric dispersion, 
(iv) gratings (maximum spectral resolution 0.15~nm), and (v) detectors to 
record the fringes. The advantages of the dispersion mode are the capabilities 
(i) of allowing continuous observation of fringes across the spectral bandwidth,
(ii) of recording the fringes with longer integration time, and (iii) of 
selecting different spectral channels for differential visibility measurements. 
In the REGAIN combiner, each coud\'e beam coming from the telescopes meets a 
pupil stabilizer, a field rotator, a wedge prism and the beam combiner
(Rousselet-Perraut et al. 1996). Figure 9 depicts the process performed by an 
arm of the REGAIN table prior to recombination. 

\begin{figure} 
\centerline {\psfig{figure=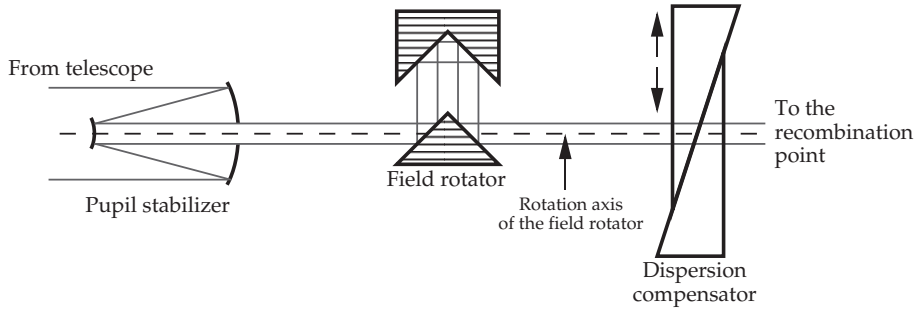,height=4cm}}
\caption {Optical processing of a beam from one arm of GI2T by the REGAIN
recombiner (Saha and Morel, 2000).}
\end{figure} 
 
Interferometry with multiple apertures, such as COAST uses multi-stage
four-way combiner. Light from the telescopes is first combined in pairs. These
pairs are recombined with other pairs. Each detector sees light from all of the
telescopes. Another method known as pair-wise recombination technique is 
employed at NPOI, where a beam combiner uses a different detector for each 
baseline. 

\paragraph{Fiber-linked recombination}

Fiber optics systems provide a perfect spatial filtering of the 
turbulence-induced corrugated wavefronts and find that the contribution 
variations are much reduced. The advantages of this may be envisaged in the 
form of (i) selecting the plane-wave part of a wavefront, (ii) splitting a 
guided wave into any desired intensity ratio, and (iii) combining two guided 
waves interferometrically. The significance of the fiber-linked unit 
for optical recombination (FLUOR), IOTA is that visibilities can be
calibrated with sub-1\% precision, which is important for many
astrophysical applications like stellar atmospheres studies, Cepheid pulsation
measurements for distance determinations, and detection of angular
anisotropies arising in disks around young stars.
 
The fiber delay-lines, using spools of fiber that can be stretched or relaxed 
to increase or decrease the optical pathlength are used in FLUOR. These fibers 
have been designed to propagate infrared light at near-IR (Mennesson et al.
(1999) in TEM mode, like a coaxial cable. Therefore, only plane-waves 
perpendicular to the axis of the fiber may propagate over long distance. 
Basically, this method results in a `spatial filtering', thus
smoothing the wavefront. The advantage of such a method for interferometry is 
a reduction of the uncertainty on the measured visibility, and the drawbacks are
the loss of optical coupling efficiency and larger photometric variations due to
the turbulence. Figure 10 depicts the schematic of the FLUOR recombiner.

\begin{figure}
\centerline {\psfig{figure=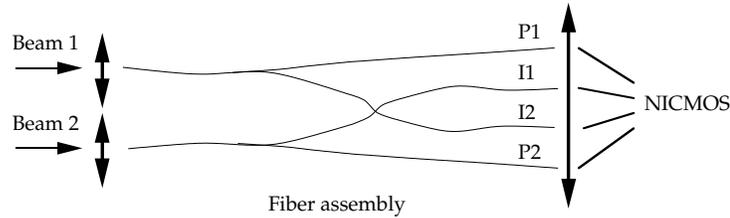,height=4cm}}
\caption {Schematic of the FLUOR recombiner. ${\tt P}1$ and ${\tt P}2$ are the 
photometric output fibers. ${\tt I}1$ and ${\tt I}2$ the interferometric 
output fibers. These outputs 
are imaged by a lens on a NICMOS infrared array detector (Saha and Morel, 2000).
}
\end{figure}

\paragraph{Integrated optics}

The integrated optics (IO), analogous to integrated chips in micro-electronics, 
potentially allows large tables of bulk optics to be replaced by miniature 
devices (Haguenauer et al. 2000). Such hardware provides easy access to spatial 
filtering and photometric calibration. The astronomical validation of this 
approach to interferometrically combine beams from separated telescopes at IOTA 
has been reported recently by Berger et al. (2001). Two different chips designed
for two-telescope beam combination in H~band were used by them. One is
manufactured using the ion exchange process: Na$^+$ ions from a glass substrate
are exchanged with Ag$^+$ ions in a molten salt through a dedicated mask 
and the other is manufactured using silica etching technique.   

\paragraph{Polarization}

For phased combination either in the image-plane or in the pupil-plane, the 
individual incoming beams from the arms of an interferometer should have 
identical pupil orientations, image orientations, and polarization 
characteristics (Traub, 2000). If the corresponding mirrors at each reflection 
(in the case of two telescope interferometer) are of same type, both the beams 
will experience the same phase-shifts; the respective $s$ and $p$ components 
combine independently in the focal-plane and produce identical fringe 
packets. If the sequence of reflections is different, the visibility 
${\cal V}_{pol}$, of an interferogram is,

\begin{equation}
{\cal V}_{pol} = |cos\frac{\psi_{sp}}{2}|,
\end{equation}
 
\noindent 
where $\psi_{sp}$ is the $s - p$ shift between the two beams. The loss of 
coherence due to misalignment of optical train, aging of coatings, and 
accumulation of dust can also be analyzed (Elias,  2001). 

At GI2T, the REGAIN uses the field rotators consisting of four plane mirrors for
each beam to compensate the polarization difference. In the COAST, starlight
passing through the central siderostat undergoes an additional two reflections
so that its $s$ and $p$ polarizations experience the same reflections as light
from the other siderostats (Baldwin et al. 1998). 

Fibers have a natural birefringence that introduce elliptic polarization at
the output, when linearly-polarized light is injected which causes a loss
of the  measured visibility. A solution for compensating the fiber
birefringence consists in winding the fibers into one or two loops
(Lef\`evre 1980). The supplementary birefringence introduced by this system
depends on the radius of the loops; it cancels the effects of the natural 
birefringence. With this system, the polarization plane can be rotated by 
twisting the fiber (by displacing the loops around the main fiber axis). Another
solution to minimize birefringence effects consists in using a Babinet 
compensator (a birefringent quartz crystal consisting of two thin prisms 
cemented together to form a thin parallel plate) at the input of each fiber. 

\paragraph{Dispersion}

A compensating system for correcting atmospherically induced dispersion is
essential at the recombiner. In the REGAIN, the different `chromatic 
dispersions' between the two beams are compensated by using two prisms which can
slide on their hypotenuse, forming therefore, a plate with adjustable thickness.
This thickness is modified every 4 minutes, following the variation of the 
altitude of the observed object (Rousselet-Perraut et al. 1996).  

In the case of fiber-linked recombiner, the dispersion of a fiber optics coupler
made by two fibers of 1 and 2 is expressed by the phase curvature:

\begin{equation}
\frac{{\it d}^2\psi}{{\it d}k^2} ,
\end{equation}

\noindent 
where $\psi$ is the phase of the spectrum of the interferogram. It is 
demonstrated by Coud\'e du Foresto et al. (1995) that 
the phase curvature can be given by:

\begin{equation}
\frac{{\it d}^2\psi}{{\it d}k^2} = -2\pi c\lambda^2(\nabla_2\Delta L 
+ L_1\Delta \nabla),
\end{equation}

\noindent 
where $\nabla_2$ is the dispersion coefficient of fiber 2, 
$\Delta L = L_2 - L_1$ the difference of length between the 
two fibers, and $\Delta \nabla = \nabla_2 - \nabla_1$ the
difference of dispersion. The dispersion coefficient depends on the
refractive indexes of both the core and the cladding of the fiber.
One problem of fiber dispersion is the `flattening' of the interferogram,
reducing the fringe contrast. To minimize the dispersion, the length of
each fiber must be calculated from the dispersion coefficients of each fiber.

\subsubsection{Calibration}

Due to the atmospheric turbulence affecting the wavefronts before 
recombination, measurements of ${\cal V}$ are biased by a random factor 
depending on the seeing quality. Instrumental flaws leading to optical 
aberrations and non-balanced flux between the two beams modify the measured 
visibility modulus as well. It is, therefore, important to calibrate each 
measure on an object by measuring ${\cal V}$ on an non-variable unresolved 
source (e.g., a farther star) in the neighborhood, preferably within 1$^\circ$
of the program star to minimize motions of telescopes and delay-lines, and at 
the same turbulence condition. The calibrator and the studied object 
observations should be interleaved for recording the fringes on both back and 
forth a few times during the observing run. Hence, one can interpolate the 
transfer function for each object-observation period. To reproduce the 
instrumental conditions, the calibrator must roughly be as bright as the object 
to calibrate; it should have ideally a spectral type and a magnitude similar to 
the studied object. The calibration of the resulting visibility is given by, 
${\cal V}_{cal}^2 = {\cal V}^2/{\cal V}^2_{ref}$ (Berio, Mourard et al. 1999). 

\subsubsection{Fringe-tracking}

\label{subsubsec:tracking}

Fringes are searched by adjusting the delay-line position; however, 
mechanical constraints on the instrument, errors on the pointing 
model, thermal drifts, various vibrations and atmospheric turbulence make the
null-OPD point changing. The error on the OPD must be less than the  
coherence length defined by:

\begin{equation}
l_c = c\cdot\tau_c = {\bar\lambda^2\over\Delta\lambda}, 
\end{equation}

\noindent 
where $\bar\lambda$ is the mean wavelength observed and $\Delta\lambda$ is the
spectral interval. This real-time control is called `fringe-tracking'. 

There exist three possible set-ups for fringe acquisition for visible spectrum.
In the first one, the OPD is temporally modulated by a sawtooth signal, using a 
fast and short-travel delaying device. The intensity of the recombined beams,
describes therefore, over the time a fringe pattern that is recorded by single 
pixel detectors. The second method consists of
imaging the dispersed recombined beam on a linear detector. In the third one, 
beams are dispersed prior to the recombination; recombination is done by
focusing them with a common lens like in Michelson stellar interferometer. In
IR, it is essential to use as little pixel as possible in order to reduce global
readout noise. The `white' fringes set-up is preferably used for IR 
observations.

\paragraph{Coherencing and cophasing}

Techniques to compensate the OPD drift between the two beams of a standard
optical interferometer may be classified into two categories: coherencing and 
cophasing. The aim of the former is to keep the OPD within the coherence area of
the fringes, while the latter, the OPD must remain much smaller than the 
wavelength: fast compensation of the OPD variations due to the differential 
`piston' mode of the turbulence is, therefore, done in order to `freeze' the 
fringes.

The coherencing with non-white fringes is compared with the active optics. In 
white light it is done, as for IOTA, by scanning the OPD while acquiring signals
to find the null-OPD point in the fringe pattern (Morel et al. 2000). This 
coherencing yields the OPD correction to apply to the delay-line, at a few Hz 
servo-loop rate. With a channeled spectrum, the OPD is proportional to the 
fringe frequency. Another method, viz., the real time
active fringe-tracking (RAFT) system has been applied to dispersed fringes 
on GI2T using 2-d FT (Koechlin et al. 1996). Both group-delay 
tracking and RAFT allow a slow servo-loop period (up to a few seconds) by 
multiplying the coherence length by the number of spectral channels used. 

Cophasing that may be compared with the AO system, is performed with white 
fringes, using the synchronous detection method; the OPD is quickly scanned over
a wavelength range. The signal acquired from the detector is then processed in 
order to yield the phase-shift and the visibility, which can be done by 
integrating signal (Shao and Staelin, 1977) over four $\lambda/4$ bins, named 
$A$, $B$, $C$ and $D$. Phase-shift and visibility modulus are then given by,

\begin{eqnarray}
\Delta\varphi &=& \arctan\left({B-D\over A-C}\right) \\
{\cal V} &=& {\pi\sqrt{(A-C)^2+(B-D)^2}\over\sqrt{2}(A+B+C+D)}. 
\end{eqnarray}

Fringe-tracking methods may be enhanced by introduction of {\it a priori} 
information, in order to allow observations at fainter ${\cal V}$ or fainter
magnitudes. Gorham (1998) has proposed to improve white light cophasing by
filtering data with a function computed to reduce the photon noise. The gain for
the tracking limit magnitude, at constant ${\cal V}$, is between 0.5 and
0.7. Methods introducing {\it a priori} information for GDT or RAFT have been
imagined as well (Padilla et al. 1998, Morel and Koechlin, 1998).

\paragraph{Bootstrapping}

\begin{figure} 
\centerline {\psfig{figure=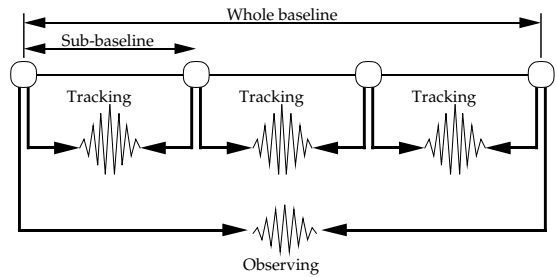,height=6cm}}
\caption {Principle of baseline bootstrapping. Apertures are represented by
circles (Saha and Morel, 2000).}
\end{figure} 

It is possible to stabilize the fringes using bootstrapping methods at the
interferometers. These methods can be categorized into two systems, namely, 
wavelength bootstrapping and baseline bootstrapping. It may be optimal to employ
the former that requires two recombiners, one for visibility measurement, the 
other one for fringe-tracking. For example, at long baselines, when the expected
fringe visibility is too low for tracking, it is possible to use a longer 
wavelength where the fringe contrast, for a white object, is higher. Meanwhile, 
fringes for computing the visibility are acquired at shorter wavelength than for
tracking. This method was applied at Mark~III interferometer (Quirrenbach et 
al. 1996). On the other hand, the latter is performed by tracking fringes over 
a connected series of short baselines to allow low visibility fringes to be 
measured on the longest baseline. In this method the photons are shared and it 
works well because of the closure-phase relationships. Here the baseline is 
divided into sub-baselines by adding apertures along the baseline. 
Fringe-tracking is performed on each sub-baseline, where the visibility is 
higher than with the entire baseline. Hence, fringes are tracked on the whole 
baseline as well. Figure 11 depicts the principle of baseline bootstrapping. 
NPOI employs such a system that enables one to
reach spatial frequencies beyond the first visibility null (Pauls et al. 1998). 
A related idea has been developed for the Keck interferometer, as well as for
the ESO's very large telescope interferometer (VLTI).

\paragraph{Role of AO systems}

An interferometer works well if the wavefronts from the individual
telescopes are coherent. The maximum useful aperture area is proportional
to $\lambda^{12/5}$. In order to improve the sensitivity of an interferometer 
where it is used to target faint sources, each telescope will have to be 
reasonably large. A large aperture produces more than 100 speckles in 
the image and the fringe pattern within each speckle is randomly phased. 
Enough photons are required to phase the telescopes into a coherent aperture, 
therefore, to enhance the instrumental visibility, an AO system should be 
incorporated. A typical image-width reduction of roughly 
a factor of 10 and a central intensity enhancement of a factor of 10$^{1.5}$ 
can be achieved on large telescopes (Traub, 2000).  
Though complete AO systems have not been implemented at any of
the interferometers to date, baring Keck-I which uses full-AO corrections,
a few of them, viz., GI2T, IOTA are using tip-tilt control system. This 
correction is sensed in the visible, using CCDs, and fringe detection is done 
in the near-IR. IOTA tip-tilt correction system uses a 32$\times$32 pixel CCD 
for each beam. The maximum rate is 200~Hz (Morel, 2000). A computer reads each
frame and computes the centroid. The value of the centroid position is sent
to a piezo mirror placed downstream the secondary mirror of each telescope.  

\subsection{Data processing}

\label{subsec:processing}

The optimal integration time required for measuring a visibility point is a 
trade-off between the number of photons to collect and the Earth rotation 
shifting the sampled point in the $(u, v)$ plane. Though with the large 
Binocular Telescope (LBT), the two 8.4~m mirrors will be installed on a common 
alt-az mounting, thus information in the $u, v$ plane can be continuously 
combined or coadded, most of
the interferometers use two apertures and are unable to recover the complex
visibility. Therefore, the information to extract from a batch of fringes is the
modulus of the visibility. Theoretically, using merely the Fourier transform
would give an optimal estimate of the visibility modulus, as demonstrated by
Walkup and Goodman (1973). However, white-light fringes obtained from
coherencing are flawed by the differential piston that modulates their 
frequency. Techniques used in radio-interferometry (where wavelengths are much 
longer), like fitting a sinewave through the fringe data, are therefore, not 
suitable. Perrin (1997) has proposed a method to remove the piston from fringes.
However, this method requires a high fringe SNR ratio and may only 
be applied when fringe SNR is important. 

\subsubsection{Recovery of visibility functions}

The model visibility amplitude ${\cal V}(s_\lambda)$ for a uniform source of 
diameter $\phi_{UD}$ is given by,

\begin{equation}
{\cal V}(s_\lambda) = \pm\frac{1}{{\sl F}_0}
\int_{-\phi_{UD}/2}^{\phi_{UD}/2}
{\sl B}(\theta) cos[2\pi s_\lambda \theta] d\theta,
\end{equation}

\noindent 
where ${\sl F}_0$ and ${\sl B}(\theta)$ are the respective total flux and the 
brightness distribution of the source, $\theta$ the position angle of the source
and $s_\lambda$ the baseline length in wavelengths. This equation reduces to,

\begin{equation}
{\cal V}(s_\lambda) = \frac{sin [\pi s_\lambda \phi_{UD}]}
{\pi s_\lambda \phi_{UD}}
\end{equation}

For the objects with circular symmetry, the visibility function is expressed
as,

\begin{equation}
{\cal V}(s_\lambda) = \biggl|\frac{2J_1(\pi s_\lambda \phi_{UD})}
{\pi s_\lambda \phi_{UD}}\biggr|,
\end{equation}

\noindent 
where $J_1(\pi s_\lambda \phi_{UD})$ is a Bessel function of the first kind.
However, a problem arises due to the limb-darkening of the stars. Observations 
of limb darkening measurements require one to collect data in the vicinity of 
and beyond the first zero or minimum of the visibility function. The radial 
intensity profile of a star may be given (Hestroffer, 1997) by:

\begin{equation}
{\cal I}(\mu)={\cal I}(0)\mu^{\alpha_l},
\end{equation}

\noindent 
where ${\cal I}(\mu)$ is the disk brightness at angle $\mu (= cos\theta)$,
$\theta$ the angle between the normal to the stellar surface and the 
direction to the observer, and 
$\alpha_l$ is the limb-darkening factor depending on the stellar atmosphere. 
The visibility function in this case may be derived as,

\begin{equation}
{\cal V}(s_\lambda) = \Gamma({\it n} + 1)
\frac{|2J_{\it n}(\pi s_\lambda \phi_{LD})|}
{(\pi s_\lambda \phi_{LD}/2)^{\it n}},
\end{equation}

\noindent 
where ${\it n} = (\alpha_l + 2)/2$ and $\phi_{LD}$ is the limb-darkened diameter
of the source. Many interferometers cannot measure low visibilities existing at 
high angular frequency (i.e, when $\sqrt{u^2 + v^2}$ is large), beyond the first
minimum of the visibility function. Reconstructions are, therefore, ambiguous 
and neither the diameter nor the limb-darkening factor may be accurately 
determined. Usually, $\alpha_l$ is an a priori information given by the stellar 
atmosphere model. The diameter is, therefore, deduced from $\alpha_l$ and the 
interferometric data. Figure 12 represents two visibility curves of the Mira 
type variable star R~Leo, obtained by the FLUOR/IOTA combination in the K band 
at two different epochs, which show both the change in equivalent UD disk 
diameter and (for the 1997 data) the diffusion by circumstellar material whose 
signature is a visibility curve substantially different from that of a uniform 
disk (Perrin et al. 1999). 

\begin{figure} 
\centerline {\psfig{figure=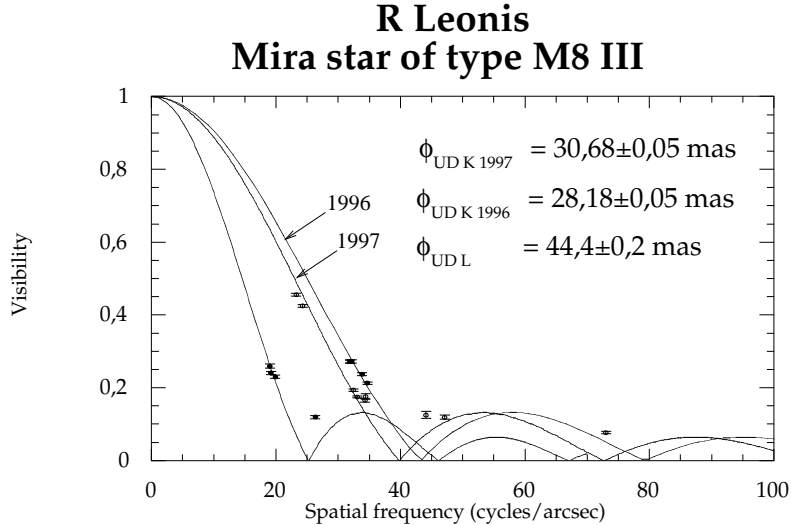,height=7cm}}
\caption {Visibility curves of the Mira variable R~Leo (Courtesy: V. 
Coud\'e du Foresto).}
\end{figure} 

For two unresolved sources, e.g., a binary system, the expression is 
formulated to,

\begin{equation}
{\cal O}({\bf x}) = {\sl B}_1\delta({\bf x} + {\bf x}_0) + 
{\sl B}_2\delta({\bf x} + {\bf x}_0 + {\bf x}_s), 
\end{equation}

\noindent 
where $|{\bf x}_s|$ is the angular separation between two sources and 
${\sl B}_1$, ${\sl B}_2$ the brightnesses of the source 1 and source 2, 
respectively. The visibility modulus corresponding to this function at ${\bf u}$
is, therefore:

\begin{equation}
|\widehat{\cal O}({\bf u})| = \sqrt{({\sl B}_1 - {\sl B}_2)^2 + 4{\sl B}_1
{\sl B}_2\cos^2(2\pi{\bf u} . {\bf x}_s)}.
\end{equation}

\noindent 
It is then useful to use a technique called `super-synthesis': the $(u, v)$
plane is swept during an observation lasting several hours, due to Earth
rotation. After a large variation of hour angle $H$, several visibility moduli 
are, therefore, measured at different $(u, v)$ points to determine the 
parameters (${\sl B}_1$, ${\sl B}_2$ and ${\bf x}_s$) of the system by fitting 
the function described by the equation (117).

The visibility function of the thin structures of the circumstellar shell may 
be computed by considering a coaxial uniform disk and a point source; therefore,
the function is written as,

\begin{equation}
{\cal V}(s_\lambda) = V_p + (1 - V_p) \biggl|\frac{2J_1(\pi s_\lambda\phi_{s})}
{\pi s_\lambda \phi_{s}}\biggr|,
\end{equation}

\noindent 
where $\phi_s$ is the diameter of the shell and $V_p$ the ratio of
power radiated by the star.

\subsubsection{Derivation of effective temperatures}

Combining photometry with the measurement of limb-darkened stellar diameter
yield the stellar emergent flux, ${\sl F}_e$ or surface brightness and is 
found from the relation,

\begin{equation}
{\sl F}_e = \frac{4{\sl F}_\nu}{\phi_{LD}^2},
\end{equation}

\noindent 
where ${\sl F}_\nu$ is the measured absolute monochromatic flux received from
the star at frequency $\nu$.

The stellar effective temperature $T_e$ is defined in terms of the emergent
flux by the Stefan-Boltzmann law. Integrating over all frequencies,

\begin{equation}
{\sl F_e} = \int {\sl F}_\nu d\nu = \sigma T_e^4 
\end{equation}

\noindent 
where $\sigma$ is Stefan constant.

\subsection{Ground-based optical/IR arrays}

\label{subsec:based}

Several ground-based LBOIs have been developed to obtain very high angular 
resolution informations of the stellar objects. However, three of them, viz., 
(i) I2T (Labeyrie, 1975), Plateau de Calern, France, (ii) Mark~III 
interferometer (Shao et al., 1988), Mt. Wilson, USA, and (iii) Infrared 
interferometer, SOIRD\'ET\'E (Gay and Mekarnia, 1988), Plateau de Calern, 
France are no longer in operation.  

\subsubsection{Direct detection interferometers} 

I. Labeyrie et al. (1986) have developed GI2T interferometer with two `boule' 
telescopes that run on North-South tracks with variable baseline of 12 to 65~m. This instrument at Plateau de Calern, France combines the features of the 
Michelson design and the radio interferometers; it operates in speckle mode. It 
consists of a pair of 1.52~m telescopes on altitude-altitude mounts. Each 
telescope is housed in a sphere (3.5~m diameter) made of concrete, which has 
three mirrors directing the horizontal afocal coud\'e beam to the recombiner 
optics. Light beams from both the telescopes are superposed at the foci in order
to produce Young's fringes. The driving system of the sphere consists of a pair 
of rings; each ring is motorized by 3 actuators acting in 3 orthogonal 
directions within 2 different tangential planes. The two rings alternately carry
the sphere, which in turn, produce continuous motion with a resolution of the 
order of 1~$\mu$m. The main drawback comes from the slow pointing of the 
telescopes; 4 or 5 stars can be tracked during the night. 

II. The SUSI, Sydney University, Australia has a very long baseline ranging from
5~m to 640~m (North-South) that are achieved with an array of 11 input stations 
equipped with a siderostat and relay optics, located to give a minimal baseline 
redundancy (Davis et al. 1999a); the intermediate baseline forms a geometric 
progression increasing in steps of $\sim$40\%. Starlight is steered by two 
siderostats of 20~cm diameter into the evacuated pipe system that carries the 
light to the atmospheric refraction corrector (at the central laboratory) 
consisting of the pairs of counter-rotating Risley prisms via a beam reducer. 
It proceeds towards either the optical path length compensator (OPLC) or is 
diverted towards the acquisition camera. On leaving OPLC, the beams from the two
arms of this interferometer may be switched over to one of the optical tables 
(blue or red) for recombination. 

III. The COAST, Cambridge, UK uses four independent telescopes consisting of 
50~cm siderostat flat feeding a fixed horizontal 40~cm Cassegrain telescope  
with a magnification of 16. These are arranged in a {\tt Y}-layout with one 
telescope on each arm, movable to a number of fixed stations and one telescope 
at the center of the {\tt Y} (Baldwin et al. 1998). Light from each siderostat 
passes through pipes containing air at ambient pressure into the beam combining 
laboratory (inside a tunnel). The four beams emerging from the path compensator 
are each split at a dichroic; the longer wavelength ($\lambda >$650~nm) of the 
visible band passes into the beam combining optics and the shorter ones are used
for acquisition and autoguiding. Each output beam passes through an iris
diaphragm and is focused by a long focus lens on to a fiber fed single-element 
avalanche photodiode detector for fringe detection.  

IV. The IOTA, situated at Mt. Hopkins, Arizona consists of three 45~cm collector
assembly located at various stations on the L-shaped baseline (5-38~m) that 
comprises a siderostat, an afocal Cassegrain telescope and an active relay 
mirror (Traub et al. 2000). In the recombining table, the optical differences 
are compensated by fixed and variable delays and the beams are recombined onto 
a beam splitter, producing two complementary interference signals (Carleton et 
al. 1994). A fast autoguiding system is used to correct the atmospheric
wavefront tilt errors. Two active delay-lines for three telescopes are provided;
a scanning piezo mirror is used to modulate the OPD between the two telescopes. 

V. The NPOI, located at Lowell Observatory, Arizona, is designed to 
measure positions with precision comparable to that of Hipparcos (1997). 
However, the aperture sizes limit the array to bright star astrometry. It is
developed as {\tt Y}-shaped baseline configuration that includes
sub-arrays for imaging, and for astrometry. For the astrometric mode, four
fixed siderostats (0.4~m diameter) are used with the baselines extendable from
19~m to 38~m. The astrometric sub-array has a laser metrology system to measure 
the motions of the siderostats with respect to one another and to the bedrock. 
While for the imaging mode, 6 transportable siderostats (0.12~m diameter) are 
used. Three siderostat positions are kept with equal space for each arm of the 
{\tt Y}. Coherence of imaging configuration is maintained by phase 
bootstrapping. The synchronous detection method applied to signals from several 
spectral channels, has been used as well (Benson et al. 1998). 

VI. The Palomar testbed interferometer (PTI), Palomar Observatory, California is
an IR phase-tracking interferometer that was developed as a test-bench for the 
Keck interferometer. The main thrust is to develop techniques and methodologies 
for doing narrow angle astrometry; therefore, it is designed to observe two 
stars simultaneously to measure the angle between them with high precision. It 
uses coherent fringe demodulation and active fringe-tracking systems with an 
array detector at 2.2~$\mu$m and active delay-lines with a range of $\pm$38~m. 
It is comprised of three 40~cm siderostats that are coupled to beam compressors.
These siderostats are used pairwise to provide baselines up to 110~m (Colavita 
et al. 1999). Both phase and group-delay measurements for narrow angle 
astrometry are being carried out (Lawson et al. 2000). Visibility is estimated 
from the fringe quadrature, either incoherently, or using source phase 
referencing to provide longer integration time (Colavita, 1999).

\subsubsection{Heterodyne interferometry} 

Heterodyne technique is generally used in radio-astronomy to reduce the 
high frequency signal to an intermediate one. Such a technique has 
the following advantages in the case of beam recombination in IR 
interferometry. They are: a larger coherence length, a simplification of the
transport of the signal from the collector to the recombiner (coaxial
cables instead of mirrors). Let ${\cal U}_{\it s}(t)$ and ${\cal U}_{\it l}(t)$,
be respectively the signals of a wave coming from a star and 
of an artificial source (laser), which are expressed as,

\begin{eqnarray}
{\cal U}_{\it s}(t) &=& {\it a}_{\it s0}.e^{-i[\omega_{\it s}.t - \psi]}, \\
{\cal U}_{\it l}(t) &=& a_{\it l0}.e^{-i\omega_{\it l}.t}.
\end{eqnarray}

\noindent
The laser is the phase reference. A detector like a photodiode,
illuminated by the sources (star + laser) yields an electrical signal
corresponding to the light intensity:

\begin{eqnarray}
{\cal I}(t) &=& |{\cal U}_{\it s}(t) + {\cal U}_{\it l}(t)|^2 \nonumber \\
&& =({\it a}_{\it s0}.e^{-i[\omega_{\it s}.t-\psi]}+{\it a}_{\it l0}.
e^{-i\omega_{\it l}.t}) \nonumber\\
&& \times({\it a}_{\it s0}.e^{i[\omega_{\it s}.t+\psi]}+{\it a}_{\it l0}.
e^{-i\omega_{\it l}.t}) \nonumber\\
&& ={\it a}_{\it s0}^2+{\it a}_{\it l0}^2+2.{\it a}_{\it l0}.{\it a}_{\it s0}.
\cos[(\omega_l-\omega_s)+\psi].
\end{eqnarray}

\noindent
If $\omega_{\it l}$ and $\omega_{\it s}$ are close, the frequency of ${\cal I}$ 
is low enough to fit in the bandwidth of the detector and its electronics (a few
GHz) and ${\cal I}$ carries the phase information from the radiation of the 
star. By correlating (multiplying) the signals ${\cal I}_1$ and ${\cal I}_2$ 
yielded by two apertures with heterodyning systems, one can extract a visibility
term. However, the lasers must have the same phases for the two apertures. 

The heterodyne interferometry at 8-11.5~$\mu$m spectral range was employed on 
the SOIRD\'ET\'E, Observatoire de Calern, France, an IR interferometer that 
consisted of a pair of 1~m telescopes with a 15~m East-West horizontal baseline 
(Gay and Mekarnia, 1988). It was an interesting project but ultimately a failure
and produced no significant scientific results. In this project,
beams were received in the central laboratory on a double cat's eye delay-line 
on a step by step movable carriage; natural OPD drift due to the Earth-rotation 
was used for acquiring fringes (Rabbia et al. 1990). Heterodyne 
interferometry is also used on the recently developed mid-IR spatial 
interferometer (ISI), Mt. Wilson, California (Townes et al. 1998), which is an 
outstanding success with many scientific results. It is well suited to study of 
circumstellar material around bright evolved stars. It features three movable 
telescopes; each telescope comprises a 1.65~m parabolic mirror and a 2~m 
flat mirror equipped with an automated guiding and tip-tilt control system at 
2~$\mu$m (Lipman et al. 1998). The star light from each aperture is first mixed
with a stable ${\rm CO}_2$ laser local oscillator, converting the signal to 
microwave frequencies, followed by pathlength matching and fringe detection in 
a correlator. On ISI, two ${\rm CO}_2$ lasers are used, the phase of one being 
controlled by the other. Here the interferometer noise is dominated by shotnoise
of the laser and thermal background is negligible for setting the sensitivity 
limit (Hale et al. 2000). ISI utilizes Earth rotation and periodic discrete 
changes of the baseline to obtain a wide range of effective baselines and map 
the visibility functions of the stellar objects.

\subsection{Projects under development and planned}

\label{subsec:under}

I. The Center for High Angular Resolution Astronomy (CHARA) array at Mt Wilson, 
California comprises six fixed 1~m telescopes arranged in a {\tt Y}-shaped 
configuration with a maximum baseline of $\sim$350~m that will operate 
at optical and IR wavelengths (McAlister et al. 1998) with a limiting resolution
of 0.2~mas. The key scientific goal of this interferometer is binary star 
astrometry, observations of stars with well-determined spectroscopic elements,
and determination of the metal abundance. 

II. The Mitaka optical-IR array, National Astronomical Observatory, Japan 
consists of several interferometers built one-by-one. The first of the series 
was MIRA-I (Machida et al. 1998) that has 25~cm siderostats and a 4~m baseline.
Its successor, MIRA-I.2 (Sato et al. 1998) has the same
baseline and slightly larger siderostats (30~cm). It features equipment
encountered on many operating interferometers: beam compressors (yielding 30~mm 
beams), delay-line operating in vacuum, tip-tilt correction system and laser
metrology. These instruments are specially designed for astrometry.

III. The large binocular telescope (LBT), on Mt. Graham, Arizona that consists 
of two 8.4~m primary mirrors (Hill, 2000) is under construction, The LBT 
mirrors are co-mounted on a fully steerable alt-az mounting, in which variable 
delay lines for the path equalization are not needed. At near-IR wavelengths,
a field of view of one arcminute or more is expected with unprecedented
spatial resolution of order of 8-9~mas at $\lambda \sim 1~\mu$m and a variable
baseline of 0-23~m (Wehinger, 2001). 

Another interferometer, the Magdalene Ridge observatory array, USA with three 
element (2$\times$2.4~m plus 0.8~m) has also received initial funding and is
under development. 

\subsubsection{Interferometers of heterogenous nature}

\label{subsubsec:heterogenous}

The Keck interferometer, Mauna Kea, USA (Colavita et al. 1998), and (ii) VLTI,
ESO, at Paranal, Chile (Derie et al. 2000) are of
a heterogenous nature. The problem comes from the recombination of two 
telescopes, one large and one small because the S/N ratio would be the one given
by the small telescopes. Nevertheless, the recent success of obtaining 
interferometric fringes from the starlight by the two large telescopes at both
Keck (http://www.jpl.nasa.gov/news) and VLTI (Glindemann and Paresce, 
2001) will have an enormous impact on developing future large optical arrays. 

The Keck interferometer comprises of two 10~m and four 1.8~m `outrigger' 
telescopes; for imaging the main telescopes would be used with outriggers to
fill in incomplete parts of the $u, v$ plane. It will combine 
phased pupils provided by adaptive-optics for the main telescopes and fast 
tip/tilt correction on the outriggers. Beam recombination will be carried out 
by 5 two-way combiners at 1.5-2.4~$\mu$m for fringe-tracking, astrometry, and 
imaging. A project for a 10~$\mu$m nulling-combiner for exo-zodiacal disk 
characterization is in the planning stages. A differential phase 
technique to aim at detecting faint sources near a bright object is also
under development (Akeson et al. 2000). 

While in the case of VLTI, beams are received from the movable 
telescopes in a central laboratory for recombination and are made to 
interfere after introducing suitable optical delay-lines. coud\'e beams from 
these apertures are sent through delay-lines operating in rooms at atmospheric 
pressure but at accurately controlled temperature. The beams reach an optical 
switch-yard to be directed to one of the four expected recombiners, e.g., (i) 
a single mode fiber recombiner (2.2~$\mu$m) that intends to debug the upstream 
sub-system of VLTI, (ii) a beam splitter based recombiner that operates at 
10~$\mu$m, (iii) a recombiner that operates between 1~$\mu$m - 2.5~$\mu$m, and
(iv) a recombiner for narrow angle astrometry.

\subsubsection{Interferometry with large arrays}

The next generation imaging interferometers with at least 15 or more elements 
should have the snapshot capability in an instantaneous mode. Beams from 
separated telescopes of such an interferometer must be recombined in the focal 
point as in the case of a Fizeau interferometer that is optically equivalent
to single large telescope masked with a multi-aperture screen so as to
reproduce exactly the ensemble of collecting telescopes (Traub, 1986).

Development of an optical very large array (OVLA), an array of 27 telescopes of 
1.5~m diameter was initiated more than a decade ago by Labeyrie, Lamaitre et 
al. (1986). Each telescope is housed in a fiberglass sphere (Dejonghe et al. 
1998) that is mounted on a six-legged robot to enable it to move on the ground 
while fringes are acquired, eliminating the need for optical delay-lines. A 
secondary mirror makes the beam afocal and compressed. A third steerable flat 
mirror sends this beam out through a slit located on the sphere to the central 
station where all other beams coming from other telescopes are combined into a 
single high resolution image. OVLA has also been considered for different 
possible aperture diameters including 12 to 25~m (Labeyrie, 1998). A new 
telescope structure has been imagined for this class of very large collectors: 
the `cage telescope', in which the sphere is replaced with an icosahedral truss 
steerable by a different mechanical system. Ridgway and Roddier (2000) have 
proposed a project of developing IR\_VLA with a total baseline of $\sim$1000~m, 
consisting of 27 telescopes with each aperture of 4~m. Another proposed 
project called `Large-Aperture Mirror Array' (LAMA) employs eighteen fixed 10~m 
liquid-mirror telescopes located within a circle of 60~m diameter which collect 
50\% of the light that falls within this area (Hickson, 2001).  

\subsubsection{Hyper-telescope imaging}

The `densified-pupil multi-aperture imaging interferometry' may provide direct 
images at their focal-plane, albeit it is a subtle technique. Conceptually 
it differs from the Fizeau interferometers which become inefficient when the 
sub-aperture spacing is large compared to their size. The reason is that most 
energy goes in a broad diffractive halo rather than in a narrow interference 
peak, which precludes obtaining usable snapshot images with kilometric or 
megametric arrays in space. 

Densifying the exit pupil, i.e., distorting it to increase the relative size of 
the sub-pupils, in such a way that the pattern of sub-aperture centers is 
preserved, concentrates the halo and intensifies the image (Labeyrie, 1996). 
Figure 13 depicts the concept of the hyper-telescope. Pedretti et al. (2000)
have derived the integrated intensities of the central peaks of the images on 
the star Capella that are obtained by taking two separate exposures of 100~s in 
the Fizeau and densified-pupil mode of the hyper-telescope. The comparison
of these values showed an intensity gain of 24$\pm$3$\times$ of the densified
with respect to the Fizeau configuration. Imaging arrays of huge size, possibly 
approaching a million kilometers to observe neutron stars, may become feasible 
in this way, called `hyper-telescopes'. They provide an image with full 
luminosity in a narrow field of $\approx\lambda/B_s$, where $B_s$ is the 
distance between the sub-apertures of the array. 

In practice for ground arrays, an elliptical track is one way of compensating
the Earth's rotation, but with delay lines, a periodic dilute-aperture can
also be built at the scale of 10 kilometers. In space, it would be a 
periodic hexagonal paving in the case of hyper-telescope version  
proposed by Labeyrie (2001) for space interferometer, Terrestrial Planet 
Finder (TPF). This version has a Fizeau focus followed by a small 
pupil densifier, and a coronagraph (Boccaletti et al. 2000). Unlike the 
ground-based elliptic ring where images are directly obtained at a 
recombination station located at a focus of the ellipse or periodic arrays on
the ground, space arrays can be globally pointed. 

The cophasing of the array may be done hierarchically (Pedretti and Labeyrie, 
1999) by cophasing triplets of beams (yielding a honeycomb pattern in the 
image-plane), then triplets of triplets, etc. Piston errors are measurable
from the triplet images. Another way of analyzing the piston errors (Labeyrie,
1999a) is an extension of the classical dispersed-fringes used since
Michelson: a set of monochromatic images recorded with a spectro-imager is 
organized as an $x, y, \lambda$ data cube, and its 3-dimensional FT is 
calculated to extract piston errors in pairs or triplets of apertures.

\begin{figure} 
\centerline {\psfig{figure=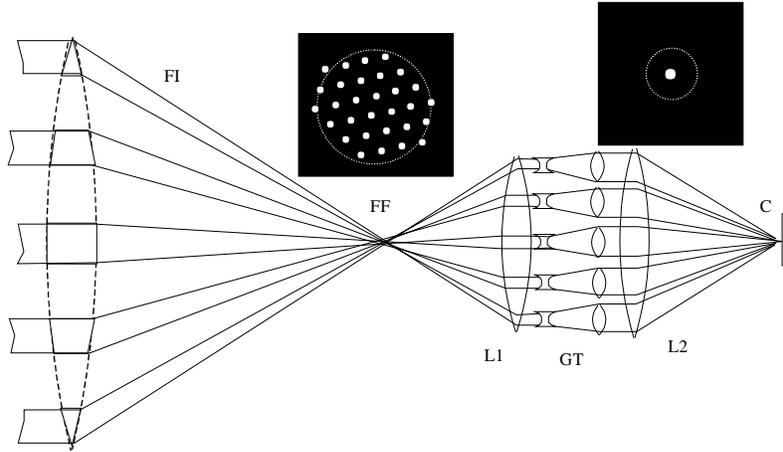,height=6cm}}
\caption {Principle of the hyper-telescope (Courtesy: A. Labeyrie). The focal
image ${\tt FF}$ provided by a Fizeau interferometer ${\tt FI}$ is re-imaged by 
lenses ${\tt L}1$ and ${\tt L}2$ on camera ${\tt C}$, through an array of 
miniature and inverted Galilean telescopes ${\tt GT}$. They densify the
exit beam, thus shrinking the diffractive envelope (dotted circle) of the 
focal pattern with respect to the interference peak. For an off-axis star, they 
also attenuate the local tilt of the flat wavefront transmitted from 
${\tt L}1$, while preserving the global tilt. The wavefront from an off-axis
star thus acquires stair steps while becoming densified at ${\tt L}2$. The
interference peak is displaced more than the envelope, but remains within it 
if the step is below one wavelength.
}
\end{figure} 

\subsection{Space-borne interferometry}

\label{subsec:borne}

The advantage of deploying LBI in space is that of observing can be done at any 
wavelength and for longer duration in the absence of atmospheric turbulence. The
difficulty comes from developing a technology featuring high precision 
positioning, as well as toughness required for space operation. A new 
generation of ultra-lightweight active mirrors (Burge et al. 2000, Angel, 2001) 
are essential to resolve the problems of size and weight. 

\subsubsection{Space technology 3}

This new generation of scientific spacecraft, ST3 (Gorham et al. 1999), 
NASA's mission, scheduled for 2003, consists of two
independent free-flying elements launched into an Earth-trailing heliocentric
orbit. One is a collector sending light from the observed object to the second
element featuring another collector, an optical delay-line and a beam
recombiner. The aim of ST3 is the demonstration and validation of technologies
that might be used for future space-borne interferometers.
Thus, the two elements of ST3 should be able to move up to 1~km from
each other, while being controlled by a laser metrology. However, the designed 
delay-line of ST3 can be up to 20~m of optical pathlength only. This instrument
will be used as an imaging interferometer for studying such objects as
Wolf-Rayet or Be stars (Linfield and Gorham, 1999).

\subsubsection{Space interferometry mission} 

Space Interferometry Mission (SIM) is also being designed and will be launched 
by NASA. The main goal of this interferometer will be to collect the new 
high-precision astrometry results, including the possibility of Jovian planet 
detection around stars up to 1 kilo-parsec distant and terrestrial planet 
detection around nearby stars (Unwin et al. 1998). The design consists of one 
free-flyer with a 10~m boom supporting 30~cm collectors. The expected angular 
accuracy is 1~$\mu$as in narrow-angle mode (with a 1$^\circ$ field of view) and 
4~$\mu$as in wide-angle mode. The sensitivity for astrometry is $m_v=20$ after 
a four-hour integration. This interferometer will work in the visible spectrum 
(0.4 to 0.9~$\mu$m). In order to get an accurate knowledge of  
{\bf B} for wide-angle astrometry without collector motions, it will feature two
auxiliary interferometers, aimed at reference stars (grid-locking). The 
schematic design of the SIM is depicted in figure 14.

Two new projects being pursued, DARWIN and TPF, are for an ozone search on 
extra-solar planets (Penny et al. 1998), and for detecting extra-solar planet 
directly (Beichman, 1998), respectively.

\begin{figure} 
\centerline {\psfig{figure=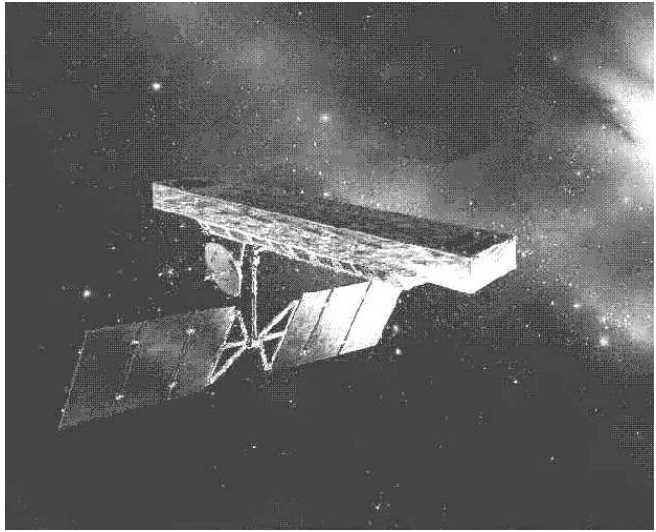,height=7cm}}
\caption {Schematic design of the SIM (Courtesy: NASA/JPL/Caltech).}
\end{figure} 

\section{IMAGE RECONSTRUCTION TECHNIQUES}
 
\label{sec:reconstruction} 

The diffraction-limited phase retrieval of a degraded image is indeed an art.
In other branches of physics too, e.g., electron microscopy, wavefront sensing,
and crystallography, one often wishes to recover phase. AO systems may also 
require image-processing algorithms since the real-time corrected image is 
often partial. Prior to using such algorithms, the basic operations to be
performed are dead pixel removal, debiasing, flat fielding, sky or background 
emission subtraction, and suppression of correlated noise. In what follows, the 
sub-sections VI-A to C describe methods to obtain components of the object FT 
and sub-section VI-D describes methods to reconstruct an image from
these components, which can usually be described as a `deconvolution'. 

\subsection{Shift-and-add algorithm} 

\label{subsec:add}

The Shift-and-add (SAA) technique (Lynds et al. 1976, Worden et al. 
1976) aligns and co-adds the recorded short-exposure images; the method is
analogous to the tip-tilt mirror of conventional AO systems. The position of the
brightest pixel, ${\bf x}_k$, is necessary to be located in each specklegram,
${\cal I}_k({\bf x})$, [${\cal I}_k({\bf x}_k) > {\cal I}_k({\bf x})$ for
all ${\bf x} \neq {\bf x}_k$], followed by shifting the specklegram (without
any rotation) to place this pixel at the center of the image space. The
SAA image, ${\cal I}_{sa}({\bf x})$, is obtained by averaging over
the set of the shifted specklegrams,

\begin{equation}
{\cal I}_{sa}(\bf x) = <{\cal I}_k({\bf x} + {\bf x}_k)>. 
\end{equation}

The large variations in the brightness of the brightest pixels are
observed in a set of speckle images; the contamination level
may not be proportional to its brightest pixel (Bates and McDonnell, 1986). 
The adjusted SAA image, ${\cal I}_{asa}({\bf x})$, can be defined as,

\begin{equation}
{\cal I}_{asa}({\bf x}) = <{\it w}[{\cal I}_k({\bf x}_k)]{\cal I}_k({\bf x} +
{\bf x}_k)>, 
\end{equation}
 
\noindent 
where ${\it w}[{\cal I}_k({\bf x}_k)]$ is the weighting in relation with
the brightness of the brightest pixel. The choice of the same quantity can be
made as ${\it w}\{{\cal I}_k({\bf x}_k)\} = {\cal I}_k({\bf x}_k)$.
An array of impulse is constructed by putting an impulse at each of the center 
of gravity with a weight proportional to the speckle intensity. This impulse 
array is considered to be an approximation of the instantaneous PSF and
is cross-correlated with the speckle frame. Disregarding the
peaks lower than the pre-set threshold, the m$^{th}$ speckle
mask, $mask_m({\bf x})$, is defined by,

\begin{equation}
mask_m({\bf x}) = \sum_{n=1}^M{\cal I}_m({\bf x}_{m,n})\delta
({\bf x} - {\bf x}_{m,n}).
\end{equation}

\noindent 
The $m^{th}$ masked speckled image, $m{\cal I}_m({\bf x})$, is expressed as,

\begin{equation}
m{\cal I}_m({\bf x}) = {\cal I}_m({\bf x}) \otimes mask_m({\bf x}).  
\end{equation}

The Lynds-Worden-Harvey image is obtained by averaging equation (127). 
For direct speckle imaging, the SAA image, ${\cal I}_{sa}({\bf x})$, 
is a contaminated one containing two complications - a convolution, 
${\cal S}_k({\bf x})$ and an additive residual, ${\cal C}({\bf x})$ - which
means,

\begin{equation}
{\cal I}_{sa}({\bf x}) = {\cal O}({\bf x}) \star {\cal S}({\bf x}) + 
{\cal C}({\bf x}), 
\end{equation}
 
\noindent 
where ${\cal S}({\bf x}) = \sum_{k=1}^k \delta({\bf x} - {\bf x^\prime}_k)d_k, 
{\bf x^\prime}_k$ being the constant position vectors and $d_k$, the positive 
constant. It is essential to calibrate ${\cal I}_{sa}({\bf x})$ with an 
unresolved point source and reduce it in the same way to produce 
${\cal S}({\bf x})$. The estimate for the object, ${\cal O}({\bf x})$, is 
evaluated from the inverse FT of the following equation,

\begin{equation}
\widehat{\cal O}({\bf u}) = \frac{\widehat{\cal I}_{sa}({\bf u})}{\widehat
{\cal I}_\circ({\bf u}) + \widehat{\cal N}({\bf u})}.  
\end{equation}

\noindent 
which is the first approximation of the object irradiance. 
This method is found to be insensitive to the telescope aberrations 
but sensitive to dominating photon noise. 

Another method called `selective image reconstruction' selects the few sharpest 
images that are recorded when the atmospheric distortion is naturally at 
minimum, from a large dataset of short-exposures (Dantowitz et al. 2000). 
Baldwin et al. (2001) have demonstrated the potential of such a technique. 

\subsection{Knox-Thomson method}
 
The Knox-Thomson (KT) method (Knox and Thomson, 1974) defines the correlation of
${\cal I}({\bf x})$ and ${\cal I}({\bf x})$ multiplied by a complex
exponential factor with spatial frequency vector ${\bf \Delta u}$. The 
approximate phase-closure is achieved by two vectors (see figure 15),
${\bf u}$ and ${\bf u} + {\bf \Delta u}$, assuming that the pupil phase is 
constant over ${\bf \Delta u}$. 
Let the general second-order moment be the cross spectrum, 
$<\widehat{\cal I}({\bf u}_1)\widehat{\cal I}^\ast({\bf u}_2)>$. It
takes significant values only if $|{\bf u}_1 - {\bf u}_2| < r_\circ/\lambda$; 
the typical value of $|{\bf \Delta u}|$ is $\sim$0.2 - 0.5~$r_\circ/\lambda$. 
Invoking equation (70), a 2-d irradiance distribution, ${\cal I}({\bf x})$ and 
its FT, $\widehat{\cal I}({\bf u})$, is defined by the equation,

\begin{equation}
\widehat{\cal I}{\bf (u)} = \int^{+\infty}_{-\infty} {\cal I}{\bf (x)}e^{-i2\pi 
{\bf u x}} d{\bf x}. 
\end{equation}

\noindent 
In image space, the correlations of ${\cal I}({\bf x})$, is derived as, 

\begin{equation}
{\cal I}({\bf x}_1, {\bf \Delta u}) = \int^{+\infty}_{-\infty} 
{\cal I}^\ast({\bf x})
{\cal I}({\bf x} + {\bf x}_1) e^{i2\pi{\bf \Delta u x}} d{\bf x}, 
\end{equation}
 
\noindent 
where ${\bf x}_1 = {\bf x}_{1x} + {\bf x}_{1y}$ are 2-d spatial 
co-ordinate vectors.
The KT correlation may be defined in Fourier space as products of, 
$\widehat{\cal I}({\bf u})$,

\begin{eqnarray}
\widehat{\cal I}({\bf u}_1, {\bf \Delta u}) &=& \widehat{\cal I}({\bf u}_1) 
\widehat{\cal I}^\ast ({\bf u}_1 + {\bf \Delta u}), \\
&&= \widehat{\cal O}({\bf u}_1)\widehat{\cal O}^\ast({\bf u}_1+{\bf \Delta u})
\nonumber\\
&&\widehat{\cal S}({\bf u}_1)\widehat{\cal S}^\ast({\bf u}_1+{\bf \Delta u}), 
\end{eqnarray}
 
\noindent 
where ${\bf u}_1 = {\bf u}_{1x} + {\bf u}_{1y}$, and ${\bf \Delta u}
= {\bf \Delta u}_x + {\bf \Delta u}_y$ are 2-d spatial frequency 
vectors. ${\bf \Delta u}$ is a small, constant offset spatial frequency. A 
number of sub-planes is used by taking different values of ${\bf \Delta u}$.
The argument of the equation (133) provides the phase-difference between the two
spatial frequencies separated by ${\bf \Delta u}$ and is expressed as, 

\begin{equation}
arg|\widehat{\cal I}^{KT}({\bf u}_1, {\bf \Delta u})|
= \psi({\bf u}_1) - \psi ({\bf u}_1 + {\bf \Delta u}).   
\end{equation}

\noindent 
Therefore, the equation (133) translates into,   

\begin{eqnarray}
\widehat{\cal I}{(\bf u}_1, {\bf \Delta u}) &=& |\widehat{\cal O}{(\bf u}_1)
||\widehat{\cal O}({\bf u}_1 + {\bf \Delta u})| \nonumber\\
&& |\widehat{\cal S}{(\bf u}_1)|| \widehat{\cal S}({\bf u}_1 + 
{\bf \Delta u})| \nonumber\\
&&e^{i[\theta^{KT}_{\cal O}({\bf u}_1, {\bf \Delta u})
+ \theta^{KT}_{\cal S}({\bf u}_1, {\bf \Delta u})]}.  
\end{eqnarray}

The object phase-spectrum of the equation (135) is encoded in the term, 
$e^{i\theta^{KT}_{\cal O}({\bf u}_1, {\bf \Delta u})} 
= e^{i[\psi_{\cal O}({\bf u}_1) - \psi_{\cal O}({\bf u}_1 + {\bf \Delta u})]}$.
In a single image realization, it is corrupted by the random phase-differences 
due to the atmosphere-telescope OTF, 
$e^{i\theta^{KT}_{\cal S}({\bf u}_1, {\bf \Delta u})} 
= e^{i[\psi_{\cal S}({\bf u}_1) - \psi_{\cal S}({\bf u}_1 
+ {\bf \Delta u})]}$. If equation (135) is averaged over a large number of 
frames, the feature $({\bf \Delta \psi_{\cal S}}) = 0$. 
$|\widehat{\cal O}({\bf u}_1+{\bf \Delta u})| \ \approx \   
|\widehat{\cal O}({\bf u}_1)|$, when ${\bf \Delta u}$ is small,
etc., and so,

\begin{eqnarray}
<\widehat{\cal I}{(\bf u}_1, {\bf \Delta u})> 
&=& |\widehat{\cal O}({\bf u}_1)| 
|\widehat{\cal O}({\bf u}_1 + {\bf \Delta u})| e^{i[\theta^{KT}_{\cal O}
({\bf u}_1, {\bf \Delta u})]} \nonumber\\
&& <\widehat{\cal S}({\bf u}_1)\widehat{\cal S}^\ast({\bf u}_1 + 
{\bf \Delta u})>,  
\end{eqnarray}
 
\noindent 
from which, together with equation (73), the object phase-spectrum, 
$\theta^{KT}_{\cal O}({\bf u}_1, {\bf \Delta u})$, can be determined.

\begin{figure} 
\centerline {\psfig{figure=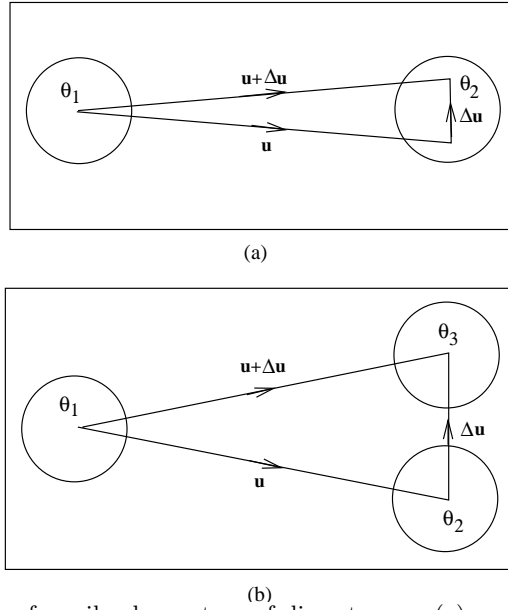,height=8cm}}
\caption {Diagrammatic representation of pupil sub-aperture of diameter, 
$r_\circ$, (a) approximate phase-closure achieved in KT method, (b) 
complete phase-closure is achieved in TC method.}
\end{figure} 

\subsection{Triple correlation technique}

The triple correlation (TC) 
is a generalization of closure phase technique (section IV.B.2.) that
is used in radio/optical interferometry where the number of closure phases is 
small compared to those available from bispectrum (Weigelt, 1977, Lohmann et 
al. 1983). It is insensitive to (i) the atmospherically induced random phase 
errors, (ii) the random motion of the image centroid, and (iii) the permanent 
phase errors introduced by telescope aberrations; any linear phase term in the 
object phase cancels out as well. The images are not required to be shifted to 
common centroid prior to computing the bispectrum. The other advantages are: 
(i) it provides information about the object phases with better S/N ratio from 
a limited number of frames, and (ii) it serves as the means of image recovery 
with diluted coherent arrays (Reinheimer and Weigelt, 
1987). The disadvantage of this technique is that it demands 
severe constraints on the computing facilities with 2-d data since the 
calculations are four-dimensional (4-d). 
  
A triple correlation is obtained by multiplying a shifted object,
${\cal I}({\bf x} + {\bf x}_1)$, with the original object, ${\cal I}({\bf x})$, 
followed by cross-correlating the result with the original one (for example, in
the case of a close binary star, the shift is equal to the angular separation 
between the stars, masking one of the two components of each double speckle). 
The calculation of the ensemble average TC is given by,

\begin{equation}
{\cal I}{\bf (x}_1, {\bf x}_2) = <\int^{+\infty}_{-\infty}{\cal I}
{\bf (x)}{\cal I}({\bf x} + {\bf x}_1){\cal I}({\bf x} + {\bf x}_2)d{\bf x}>,  
\end{equation}
 
\noindent 
where ${\bf x}_j = {\bf x}_{jx} + {\bf x}_{jy}$ are 2-d spatial 
co-ordinate vectors. The bispectrum is given by,

\begin{eqnarray}
\widehat{\cal I}({\bf u}_1, {\bf u}_2) &=& <\widehat{\cal I}({\bf u}_1) 
\widehat{\cal I}^\ast ({\bf u}_1 + {\bf u}_2) \widehat{\cal I}({\bf u}_2)>,\\
&&= \widehat{\cal O}({\bf u}_1)
\widehat{\cal O}^\ast({\bf u}_1+{\bf u}_2)\widehat{\cal O}({\bf u}_2)\nonumber\\
&&<\widehat{\cal S}({\bf u}_1)
\widehat{\cal S}^\ast({\bf u}_1+{\bf u}_2)\widehat{\cal S}({\bf u}_2)>, 
\end{eqnarray}
 
\noindent 
where ${\bf u}_j = {\bf u}_{jx} + {\bf u}_{jy}$, 
$\widehat{\cal I}({\bf u}_j) 
=\int{\cal I}({\bf x})e^{-i2\pi{\bf u}_j{\bf x}} d{\bf x}$, and

\noindent
$\widehat{\cal I}^\ast({\bf u}_1 + {\bf u}_2) 
= \int{\cal I}({\bf x}) e^{i2\pi({\bf u}_1 + {\bf u}_2).{\bf x}}d{\bf x}$. 
The object bispectrum is given by,

\begin{eqnarray}
\widehat{\cal I}_{\cal O}({\bf u}_1, {\bf u}_2) &=& \widehat{\cal O}({\bf u}_1)
\widehat{\cal O}^\ast({\bf u}_1 + {\bf u}_2)\widehat{\cal O}({\bf u}_2)
\nonumber\\ 
&&= \frac{<\widehat{\cal I}({\bf u}_1)\widehat{\cal I}^\ast({\bf u}_1 
+ {\bf u}_2) 
\widehat{\cal I}({\bf u}_2)>}{<\widehat{\cal S}({\bf u}_1)\widehat{\cal S}^\ast ({\bf u}_1 
+ {\bf u}_2)\widehat{\cal S}({\bf u}_2)>}.  
\end{eqnarray}

\noindent 
The modulus $|\widehat{\cal O}({\bf u})|$ of the 
object FT $\widehat{\cal O}({\bf u})$ can be evaluated from the 
object bispectrum $\widehat{\cal I}_{\cal O}({\bf u}_1, {\bf u}_2)$.
The argument of equation (138) gives the phase-difference and
is expressed as,

\begin{equation}
arg|\widehat{\cal I}^{TC}({\bf u}_1, {\bf u}_2)| = 
\psi({\bf u}_1) + \psi({\bf u}_2) 
- \psi({\bf u}_1 + {\bf u}_2).
\end{equation}

The object phase-spectrum is encoded in the term 
$e^{i\theta^{TC}_{\cal O}({\bf u}_1, {\bf u}_2)}$. 
It is corrupted in a single realization by the random phase-differences due to 
the atmosphere-telescope OTF, $e^{i\theta^{TC}_{\cal S}({\bf u}_1, {\bf u}_2)} 
= e^{i[\psi_{\cal S}({\bf u}_1) - \psi_{\cal S} ({\bf u}_1 + {\bf u}_2)
+ \psi_{\cal S}({\bf u}_2)]}$. If sufficient number of specklegrams are 
averaged, one can overcome this shortcoming. Let 
$\theta^{TC}_{\cal O}({\bf u}_1, {\bf u}_2)$ be the phase of the object 
bispectrum; then,

\begin{eqnarray}
\widehat{\cal O}({\bf u}) &=& |\widehat{\cal O}({\bf u})| e^{i\psi({\bf u})},\\
\widehat{\cal I}_{\cal O}({\bf u}_1,{\bf u}_2) &=& |\widehat{\cal I}_{\cal O}
({\bf u}_1, {\bf u}_2)| e^{i\theta^{TC}_{\cal O}({\bf u}_1, {\bf u}_2)}. 
\end{eqnarray}

\noindent 
Equations (142) and (143) may be inserted into equation (140), yielding
the relations,

\begin{eqnarray}
\widehat{\cal I}_{\cal O}({\bf u}_1, {\bf u}_2)  
&=& |\widehat{\cal O}({\bf u}_1)|  
|\widehat{\cal O}({\bf u}_2 )|
|\widehat {\cal O}({\bf u}_1+{\bf u}_2)| \nonumber\\ 
&& e^{i[\psi_{\cal O}({\bf u}_1) - \psi_{\cal O}({\bf u}_1 + {\bf u}_2)
+ \psi_{\cal O}({\bf u}_2)]}\rightarrow,  \\ 
\theta^{TC}_{\cal O}({\bf u}_1, {\bf u}_2) &=& \psi_{\cal O}({\bf u}_1) 
- \psi_{\cal O}({\bf u}_1 + {\bf u}_2) + \psi_{\cal O}({\bf u}_2). 
\end{eqnarray}

\noindent 
Equation (145) is a recursive one for evaluating the phase of the object 
FT at coordinate ${\bf u} = {\bf u}_1 + {\bf u}_2$.
The reconstruction of the object phase-spectrum from the phase of the
bispectrum is recursive in nature. The object phase-spectrum at $({\bf u}_1
+ {\bf u}_2)$ can be written as,

\begin{equation}
\psi_{\cal O}({\bf u}_1 + {\bf u}_2) 
= \psi_{\cal O}({\bf u}_1) + \psi_{\cal O}({\bf u}_2)
- \theta^{TC}_{\cal O} ({\bf u}_1, {\bf u}_2).   
\end{equation}

\noindent 
If the object spectrum at ${\bf u}_1$ and ${\bf u}_2$ is known,
the object phase-spectrum at $({\bf u}_1 + {\bf u}_2)$ can be computed.
The bispectrum phases are mod $2\pi$, therefore, the recursive reconstruction
in equation (139) may lead to $\pi$ phase mismatches between the computed
phase-spectrum values along different paths to the same point in frequency 
space. However, according to Northcott et al. (1988), phases from different
paths to the same point cannot be averaged to reduce noise under this
condition. A variation of the nature of computing argument of the term, 
$e^{i\psi_{\cal O}({\bf u}_1 + {\bf u}_2)}$, is needed to obtain the object 
phase-spectrum and the equation (146) translates into,

\begin{equation}
e^{i\psi_{\cal O}({\bf u}_1 + {\bf u}_2)} = 
e^{i[\psi_{\cal O}({\bf u}_1) + \psi_{\cal O}({\bf u}_2) - 
\theta^{TC}_{\cal O}({\bf u}_1,{\bf u}_2)]}. 
\end{equation}

The values obtained using the unit amplitude phasor recursive re-constructor
are insensitive to the $\pi$ phase ambiguities. Saha et al. (1999) have 
developed a code based on this re-constructor. The least-square formulation of 
the phase reconstruction (Glindemann et al. 1991) and the projection-slice 
theorem of tomography and the Radon transform (Northcott et al. 1988) have been
successfully applied in the development of phase reconstruction from the 
bispectrum. 

\subsection{Deconvolution algorithms}

\label{subsec:iterative}

Most deconvolution techniques, in which `a priori information' plays an 
essential role, can be simplified to the minimization /maximization of 
a criterion by using an iterative numerical method (Gerchberg and Saxton, 1972)
that bounces back and forth between the image-domain and Fourier-domain 
constraints until two images are found that produce the input image when 
convolved together (Ayers and Dainty, 1988). 

\subsubsection{Blind iterative deconvolution (BID) technique}

Let the degraded image, ${\cal I}({\bf x})$, be used as the operand. An initial 
estimate of the PSF, ${\cal S}({\bf x})$, has to be provided. The image is 
deconvolved from the guess PSF by Wiener filtering (see section IV.A.1), which 
is an operation of multiplying a suitable Wiener filter (constructed from  
$\widehat{\cal S}({\bf u})$, of the PSF) with $\widehat{\cal I}({\bf u})$. The 
technique of Wiener filtering damps the high frequencies and minimizes the mean 
square error between each estimate and the true spectrum. This filtered 
deconvolution takes the form,

\begin{equation}
\widehat{\cal O}({\bf u}) = \widehat{\cal I}({\bf u)}{\frac{\widehat{\cal O}_f
({\bf u})} {\widehat{\cal S}({\bf u})}}.
\end{equation}

\noindent 
The Wiener filter, $\widehat{\cal O}_f({\bf u})$, is derived as,

\begin{equation}
{\widehat{\cal O}_f({\bf u}) = {\frac {\widehat{\cal S}({\bf u})
\widehat{\cal S}^\ast({\bf u})}{|\widehat{\cal S}({\bf u})|^2 
+ |\widehat{\cal N}({\bf u})|^2}}}.
\end{equation}

\noindent 
The term, $\widehat{\cal N}({\bf u})$ can be replaced with a 
constant estimated as the rms fluctuation of the high frequency region in the 
spectrum where the object power is negligible. The Wiener filtering spectrum, 
$\widehat{\cal O}(\bf u)$, takes the form: 

\begin{equation}
\widehat{\cal O}({\bf u}) = \widehat{\cal I}({\bf u}){\frac{\widehat
{\cal S}^\ast({\bf u})}{\widehat{\cal S}({\bf u})\widehat{\cal S}^\ast({\bf u}) 
+ \widehat{\cal N}({\bf u})\widehat{\cal N}^\ast({\bf u})}}. 
\end{equation}

\noindent 
The result, $\widehat{\cal O}({\bf u})$, is transformed back to image space, 
the negatives in the image and the positives outside a prescribed domain 
(called object support) are set to zero. The average of negative intensities 
within the support are subtracted from all pixels. The process is repeated 
until the negative intensities decrease below the noise. A new estimate of the 
PSF is next obtained by Wiener filtering ${\cal I}({\bf x})$, with a filter 
constructed from the constrained object, ${\cal O}({\bf x})$; this completes one
iteration. This entire process is repeated until the derived values of 
${\cal O}({\bf x})$ and ${\cal S}({\bf x})$ converge to sensible solutions. 

BID has the ability to retrieve the diffraction-limited image of an 
object from a single specklegram without the reference star data (Saha and
Venkatakrishnan, 1997). Jefferies and 
Christou, (1993), have developed an algorithm which requires more than a single 
speckle frame for improving the convergence. Barnaby et al. (2000) used this 
algorithm, and parametric blind deconvolution (PBD) to 
post-process the data obtained with the AO system, and found that the secondary 
of 81~Cnc was 0.12~$m_v$ brighter than the primary at 0.85~$\mu$m. 

\subsubsection{Fienup algorithm}

Fienup (1978) algorithm reconstructs an object using only the modulus of 
its FT. At the $kth$ iteration, ${\cal G}_k{(\bf x)}$, an estimate of the object
FT, is compared with the measured one and made to conform with 
the modulus at all Fourier frequencies. The inverse transform of the result 
yields an image ${\cal G}_k^\prime{(\bf x)}$. This iteration is completed by 
forming a new estimate of the object that conforms to certain object-domain 
constraints, e.g., positivity and finite extent, such that,

\begin{eqnarray}
\widehat{\cal G}_k{(\bf u)} &=& |\widehat{\cal G}_k{(\bf u)}| 
e^{i\phi_k{(\bf u)}} = {\cal F}[{\cal G}_k{(\bf x)}], \\
\widehat{\cal G}_k^\prime{(\bf u)} &=& |\widehat{\cal I}{(\bf u)}| 
e^{i\phi_k{(\bf u)}}, \\
{\cal G}_k^\prime{(\bf x)} &=& {\cal F}^{-1}
[\widehat{\cal G}_k^\prime{(\bf u)}], \\
{\cal G}_{k+1}({\bf x}) &=& 
{\cal G}_k^{\prime} ({\bf x}), \, \, \, {\bf x} \notin \gamma, \nonumber\\
&& = 0, \, \, \, \, \, \, \, \, \, \,   {\bf x} \in \gamma,
\end{eqnarray}

\noindent 
where the region $\gamma$ is the set of all points at which 
${\cal G}_k^\prime{(\bf x)}$ violates the object-domain constraints, and
${\cal G}_k{(\bf x)}, \widehat{\cal G}_k^\prime{(\bf u)}$, and $\phi_k$ are 
estimates of ${\cal I}{(\bf x)}, \widehat{\cal I}{(\bf u)}$, and the phase 
$\psi$ of $\widehat{\cal I}{(\bf u)}$, respectively.

The above procedure may be accelerated if an estimate, ${\cal G}_{k+1}{(\bf x)}$
is formed as,

\begin{eqnarray} 
{\cal G}_{k+1}({\bf x}) &=& 
{\cal G}_k{(\bf x)}, \, \, \, \, \, \, \, \, \, \, \, \, \, \, \, \, \, \, \, \,
\, \, \, \, \, \, \, \, \, {\bf x} \notin \gamma, \nonumber \\ 
&& = {\cal G}_k{(\bf x)} - \beta {\cal G}_k^\prime{(\bf x)}, \, \, \, 
{\bf x} \in \gamma, 
\end{eqnarray}

\noindent 
where $\beta$ is a constant feedback parameter. 

\subsubsection{Other iterative algorithms}

I. The Richardson-Lucy (Richardson, 1972, Lucy, 1974) algorithm converges to 
the maximum likelihood
solution for Poisson statistics in the data which is appropriate for optical
data with noise from counting statistics. It forces the restored image to be
non-negative and conserves flux both globally and locally at each iteration.

II. The Magain, Courbin, Sohy, (MCS; Magain et al. 1998) algorithm is based on 
the principle that sampled data cannot be fully deconvolved without violating 
the sampling theorem (Shannon, 1949) that determines the maximal sampling 
interval allowed so that an entire function can be reconstructed from sampled 
data. The sampled image should be deconvolved by a narrower function instead of 
the total PSF so that the resolution of the deconvolved image is compatible 
with the adopted sampling. The positivity constraint unlike the traditional 
deconvolution methods, is not mandatory; accurate astrometric and photometric
informations of the astronomical objects can be obtained.

III. The myopic iterative step preserving algorithm, (MISTRAL; Conan et al. 
1998), is based on the Bayesian theorem that uses the probabilistic approach. It
incorporates a positivity constraint and some a priori knowledge of the object 
(an estimate of its local mean and a model for the power spectral density etc.).
It also allows one to account for the noise in the image, the imprecise 
knowledge of the PSF. MISTRAL has produced a few spectacular images after 
processing AO images. 

IV. A non-linear iterative image reconstruction algorithm, called Pixon method 
has been developed by Puetter and Yahil (1999) that provides statistically 
unbiased photometry and robust rejection of spurious sources. Unlike other
Bayesian methods, this technique does not assign explicit prior probabilities to
image models. It minimizes complexity by smoothing the image model locally. The 
model is then described using a few irregular-sized pixons containing similar
amounts of information, rather than many regular pixels containing
variable signal-to-noise data. Eke (2001) opines that it has the ability
to detect sources in low SNR data and to deconvolve a telescope beam in order to
recover the internal structure of a source.

\subsubsection{Aperture-synthesis mapping}

\label{subsubsec:mapping}

Reconstruction of complex images involves the knowledge of complex visibilities.
Interferometers with two apertures have limited possibilities for image
reconstructions due to absence of phase visibility recovering.
The phase of a visibility may be deduced from closure-phase that has been 
applied in optical interferometry (Baldwin et al. 1998). Process after data
acquisition consists of phase calibration and visibility phase reconstruction
from closure-phase terms by technique similar to bispectrum processing.
From complex visibilities acquired from a phased 
interferometric array , it is possible to reconstruct the image by actually 
interpolating the function in the $(u, v)$ plane. Let the output of a
synthesis array be a set of visibility functions ${\cal V}(u_j, v_j)$ that
are obtained by averaging the quasi-sinusoidal response of each interferometer 
pair and hence, the resultant brightness distribution be given by the equation:

\begin{equation}
{\cal I}^{\prime\prime}(x, y) = \sum_{j=1}^N{\cal V}(u_j, v_j){\it w}_j 
e^{[-2\pi i(u_jx + v_jy)]},
\end{equation}
 
\noindent 
where ${\cal I}^{\prime\prime}(x, y)$ is usually called the dirty image which
is the convolution of the true brightness distribution ${\cal I}^\prime(x, y)$
with the synthesized beam ${\cal S}^\prime(x, y)$, ${\it w}_j$ is a weight 
associated with $j^{th} N$. The map will show sufficient details when the 
$(u, v)$ coverage is good, but it is not the best representation of the sky and 
contains many artifacts notably the positive and negative side lobes around 
bright peaks. If the sampling of the $(u, v)$ plane is irregular and uneven, 
the dirty beam will have large sidelobes and confuse and obscure the structure 
of interest in the image.

Several non-linear methods have been introduced to restore the unmeasured
Fourier components in order to produce a more physical map. These methods
work on the constraint that the image must be everywhere positive or zero and
they make use of a priori knowledge about the extent of the source and 
statistics of the measurement processes. Since the restoration is not unique, 
the image restoration must select the best solution.

\paragraph{CLEAN}

The most commonly used approaches to deconvolve dirty image fall
into two groups: CLEAN algorithms (H\"ogbom, 1974) and maximum-entropy
methods (MEM, Ables, 1974). To some extent, CLEAN and MEM are complimentary 
in their application. Between them, CLEAN is the most routinely used algorithm,
particularly in radio astronomy, because it is both computationally efficient 
and intuitively easy to understand. This procedure is a non-linear technique 
that applies iterative beam removing method. It starts with a dirty image 
${\cal I}^\prime(x, y)$ made by the linear Fourier inversion procedure and 
attempts to decompose this image into a number of components, each of which 
contains part of the dirty beam ${\cal S}^\prime(x, y)$. One wishes to determine
the set of numbers ${\cal A}_i(x_i, y_i)$ such that

\begin{equation}
{\cal I}^\prime(x, y) = \sum_i {\cal A}_i{\cal S}^\prime(x-x_i, y-y_i) +
{\cal I}_R(x, y), 
\end{equation}

\noindent 
where ${\cal I}_R(x, y)$ is the residual brightness distribution after the 
decomposition. The solution is considered satisfactory if ${\cal I}_R(x, y)$
is of the order of the expected noise.
 
The algorithm searches the dirty image for the pixel with largest
absolute value ${\cal I}_{max}$ and subtracts a dirty beam pattern centered on
this point with amplitude $G_l{\cal I}_{max}$, where $G_l$ is called the loop
gain. The residual map is searched for the next largest value and the second 
beam stage is subtracted and so on. The iteration is stopped when the maximum
residual is consistent with the noise level. The iteration consists of a 
residual image that contains noise and low level source calculations plus a
set of amplitudes and positions of the components removed. These components
can be considered as an array of delta function and convolved with a clean
beams and added to produce a CLEANed image. The clean beam is usually chosen as 
a truncated elliptical Gaussian about the same size of the main lobe of the 
dirty beam. CLEAN results in a map with the same resolution as the original
dirty map without sidelobes.

\paragraph{Maximum entropy method (MEM)}

It is found that CLEAN generally performs well on small compact sources, 
while MEM does better on extended sources. MEM is employed in a variety of 
fields like medical imaging, crystallography, radio and X-ray astronomy  
as well. This procedure governs the estimation of probability distributions 
when limited information is available. In addition, it 
treats all the polarization component images simultaneously (unlike CLEAN which
deconvolves different polarization component images independently) and 
guarantees essential conditions on the image. It makes use of the highest 
spatial frequency information by finding the smoothest image consistent with
the interferometric data. While enforcing positivity and conserving the total 
flux in the frame, smoothness is estimated by the `entropy' $S$ that is of
the form,

\begin{equation}
S = - \sum_i {\it h}_j \ln \frac{{\it h}_j}{{\it m}_j},
\end{equation}

\noindent 
where ${\bf h} = [{\it h}_j]$ represents the image to be restored, and ${\bf m}
= [{\it m}_j]$ is known as prior image.

It can be shown that $S \leq 0$; the equality holds if ${\bf h} = {\bf m}$. The
value of $S$ is a measure of the similarity between ${\bf h}$ and ${\bf m}$ if 
the entropy $S$ is maximized without any data constraints. With data 
constraints, the maximum of $S$ will be less than its absolute maximum value
zero, meaning that ${\bf h}$ has been modified and deviated from its prior
model ${\bf m}$.

MEM solves the multi-dimensional constraints minimization problem. It uses
only those measured data and derives a brightness distribution which is the
most random, i.e., has the maximum entropy $S$ of any brightness distribution
consistent with the measured data. Maintaining an adequate fit to the data, it 
reconstructs the final image that fits the data within the noise level.

Monnier et al. (2001) have reconstructed the dust shells around two evolved 
stars, IK~Tau and VY~CMa using $(u, v)$ coverage from the contemporaneous 
observations at Keck-I and IOTA. Figure 16 depicts the MEM reconstructions of 
the dust shells around these stars. Their results clearly 
indicate that without adequate spatial resolution, it is improbable to cleanly 
separate out the contributions of the star from the hot inner radius of the 
shell (left panel in figure 16). They opined that image reconstructions 
from the interferometer data are not unique and yield results which depend 
heavily on the biases of a given reconstruction algorithm. 
By including the long baselines ($>$~20~m) data from the IOTA 
interferometer (right panel), the fraction of the flux arising from the central 
star can be included in the image reconstruction process by using the MEM 
prior. One can see for a dust shell such as IK~Tau, that
additional IOTA data are critical in accurately interpreting the physical
meaning of interferometer data. The thick dashed lines show the expected
dust shell inner radius from the data obtained at the ISI.

\begin{figure} 
\centerline {\psfig{figure=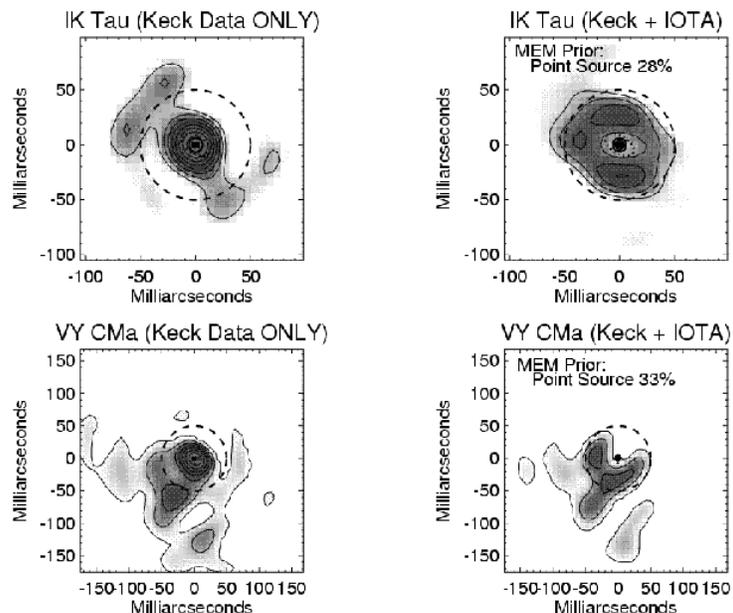,height=8cm,angle=90}}
\caption{MEM image reconstructions of the dust shell around IK~Tau and VY~CMa
(Courtesy: J. D. Monnier); the left panels show the reconstructions of them 
from data obtained with a single Keck telescopes using aperture masking 
(baselines up to 9~m at 2.2 $\mu$m, and the right panels show the dust shell 
reconstructions when the fractional amounts of star and dust shell emission is 
constrained to be consistent with both the Keck and IOTA data (Monnier et al.
2001).}
\end{figure} 

\paragraph{Self calibration}

In case of producing images with accurate visibility amplitude and poor or
absence of phases, self calibration can be used (Baldwin and Warner, 1976). 
Cornwell and Wilkinson, 1981) introduced a modification by explicitly solving
for the telescope-specific error as part of the reconstructing step. The
measured Fourier phases are fit using a combination of intrinsic phases plus
telescope phase errors. 

If the field of observation contains one 
dominating internal point source which can be used as an internal phase
reference, the visibility phase at other spatial frequencies is derived.
A hybrid map can be made with the measured amplitudes together with model
phase distributions. Since the measured amplitudes differ from the single point
source model, the hybrid map diffuses from the model map. Adding some new 
feature to the original model map, an improved model map is obtained in the 
next iteration. With clever selection of features to be added to the model in 
each iteration, the procedure converges.

\paragraph{Linear approach}

Another approach to the deconvolution problem is to formulate it as
a linear system of the form:

\begin{equation}
{\bf Ax} = {\bf b},
\end{equation}

\noindent
and then use algebraic techniques to solve this system.
The elements of {\bf A} contain samples of the dirty beam, the elements
of {\bf b} are samples of the dirty image, while {\bf x} contains components
of the reconstructed image. Without any additional constraints matrix {\bf A}
is singular: additional information has to be added to find a solution.
Assumptions that regularize the system include positivity and compact structure
of the source. An algebraic approach is not new in itself, but practical
applications of such techniques in astronomy have become feasible only recently,
e.g., for the non-negative least squares (NNLS, Briggs, 1995).

\paragraph{WIPE}

Lannes et al. (1994) present another technique, WIPE, based on a least squares 
approach, where support information is used to regularize the algorithm. Again, 
a linear system similar to that mentioned in the previous paragraph is solved,
but using a technique which iterates between ($u, v$) and image-planes.
Unlike CLEAN and MEM, WIPE suppresses excessive extrapolation of higher
spatial frequencies during the deconvolution.

\section{ASTRONOMICAL APPLICATIONS}

\label{sec:astrophysical}

Optical interferometry in astronomy is a boon for astrophysical studies. The
following sub-sections dwell on the astrophysical importance and the 
perspectives of interferometry. 

\subsection{Results from single aperture interferometry}

\label{subsec:from}

Single aperture interferometry has been contributing to study of the Sun and 
solar system, and of a variety of stellar astrophysical problems.

\subsubsection{Sun and solar system}

The existence of solar features with sizes of the order of 100~km or smaller 
was found by means of speckle imaging (Harvey, 1972, Harvey and Breckinridge, 
1973). From observations of photospheric granulation from disk center to limb 
at $\lambda = 550\pm5~nm$, Wilken et al. (1997) found a decrease of the 
relative rms-contrast of the center-to-limb of the granular intensity. A time 
series of high spatial resolution images reveal the highly dynamical evolution 
of sunspot fine structure, namely, umbral dots, penumbral filaments, facular 
points (Denker, 1998). Small-scale brightening in the vicinity of sunspots, 
were also observed in the wings of strong chromospheric absorption lines. These 
structures which are concomitant with strong magnetic fields show brightness 
variations close to the diffraction-limit of the Vacuum Tower Telescope 
($\sim$0.16$^{\prime\prime}$ at 550~nm), Observatorio del Teide (Tenerife). 
With the phase-diverse speckle method, Seldin et al. (1996) 
found that the photosphere at scales below 0.3$^{\prime\prime}$ is highly 
dynamic.

Speckle imaging has been successful in resolving the heavenly dance of 
Pluto-Charon system (Bonneau and Foy, 1980), as well as in determining shapes 
of asteroids (Drummond et al. 1988). Reconstructions of Jupiter with 
sub-arcsecond resolution have also been carried out by Saha et al. (1997).
 
\subsubsection{Stellar objects}

Studies of close binary stars play a fundamental role in measuring   
stellar masses, providing a benchmark for stellar evolution calculations; a 
long term benefit of interferometric imaging is a better calibration of the 
main-sequence mass-luminosity relationship. The 
parameter in obtaining masses involves combining the spectroscopic 
orbit with the astrometric orbit from the interferometric data. Continuous
observations are necessary to be carried out in order to derive accurate orbital
elements and masses, luminosities and distances. The radiative transfer 
concerning the effects of irradiation on the line formation in the expanding 
atmospheres of the component that is distorted mainly by physical effects, viz.,(i) rotation of the component, and (ii) the tidal effect can be modeled as well.
More than 8000 interferometric observations of stellar objects have been 
reported so far, out of which 75\% are of binary stars (Hartkopf et al. 1997).
The separation of most of the new components discovered are found to be less 
than 0.25$^{\prime\prime}$. From an inspection of the interferometric data,
Mason et al. (1999) have confirmed the binary nature of 848 objects, discovered
by the Hipparcos satellites; Prieur et al. (2001) reported high
angular resolution astrometric data of 43 binary stars that were also observed 
with same satellite. Torres et al. (1997) derived individual masses for 
$\theta^1$~Tau using the distance information from $\theta^2$~Tau; they found 
that the empirical mass-luminosity relation  
in good agreement with theoretical models. Gies et al. (1997) measured the 
radial velocity for the massive binary 15~Mon. With the speckle 
spectrograph, Baba, Kuwamura, Miura et al. (1994) have observed a binary star, 
$\phi$~And ($\rho$ = 0.53$^{\prime\prime}$) and found that the primary  
(a Be star) has H$\alpha$ in emission while the companion has H$\alpha$ in
absorption. The high angular polarization measurements of the pre-main 
sequence binary system Z~CMa at 2.2~$\mu$m by Fischer et al. (1998) reveal that
both the components are polarized; the secondary showed an unexpected by large 
degree of polarization. 

Studies of multiple stars are also an important aspect that can reveal  
mysteries. For instance, the R136a was thought to be a very 
massive star with a mass of $\sim$2500M$_\odot$ (Cassinelli et al. 
1981). Speckle imagery revealed that R136a was a dense cluster of stars 
(Weigelt and Baier, 1985, Pehlemann et al. 1992). R64 (Schertl et al. 1996), 
HD97950, and the central object of giant HII region starburst cluster NGC3603 
(Hofmann et al. 1995) have been observed as well. The star-like luminous blue 
variable (LBV), $\eta$~Carinae, an intriguing massive southern object, was found
to be a multiple object (Weigelt and Ebersberger, 1986). The polarimetric 
reconstructed images with a 0.11$^{\prime\prime}$ resolution in the H$\alpha$ 
line of $\eta$~Carina exhibit a compact structure elongated in consistence with 
the presence of a circumstellar equatorial disk (Falcke et al. 1996). 

Many supergiants have extended gaseous envelope which can be imaged in
their chromospheric lines. The diameter of a few such objects, $\alpha$~Ori and 
Mira, are found to be wavelength dependent (Bonneau and Labeyrie, 1973, 
Labeyrie et al. 1977, Weigelt et al. 1996). Recent studies have also confirmed 
the asymmetries on their surfaces; the presence of hotspots 
are reported as well (Wilson et al. 1992, Haniff et al. 1995, Bedding, 
Zijlstra et al. 1997, Tuthill et al. 1997, Monnier et al. 1999, Tuthill, 
Monnier et al. 1999). The surface imaging of long period variable stars 
(Tuthill, Haniff et al. 1999), VY~CMA reveals non-spherical circumstellar 
envelope (Wittkowski, Langer et al. 1998). Monnier et al. (1999) have found 
emission to be one-sided, inhomogeneous and asymmetric in IR and derived the 
line-of-sight optical depths of its circumstellar dust shell. The radiative
transfer modeling of the supergiant NML~Cyg reveals the multiple dust 
shell structures (Bl\"ocker et al. 2001). 
 
\begin{figure} 
\centerline {\psfig{figure=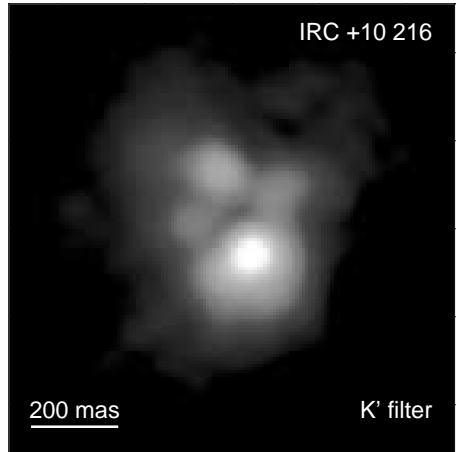,height=6cm}}
\caption {Speckle masking reconstruction of IRC+10216 (Courtesy: R. Osterbart);
the resolution of the object is found to be 76~mas for the K$^\prime$ band
(Osterbart et al. 1996).} 
\end{figure} 
 
High resolution imagery may depict the spatial distribution of circumstellar 
matter surrounding objects which eject mass, particularly in young 
compact planetary nebula (YPN) or newly formed stars in addition to T~Tauri
stars, late-type giants or supergiants. The large, older and evolved planetary 
nebula (PN) show a great variety of structure (Balick, 1987) that are (a) 
spherically symmetric (A39), (b) filamentary (NGC6543), (c) bipolar (NGC6302), 
and (d) peculiar (A35). The structure may form in the very early phases of the 
formation of the nebula itself which is very compact and unresolved. By making 
maps at many epochs, as well as by following the motion of specific structural 
features, one would be able to understand the dynamical processes 
at work. The structures could be different in different spectral lines e.g., 
ionization stratification in NGC6720 (Hawley and Miller, 1977), and hence maps 
can be made in various atomic and ionic emission lines too. Major results, such
as, (i) measuring angular diameters of several YPNs (Barlow et al. 1986, Wood 
et al. 1986). (ii) resolving five individual clouds around carbon star 
IRC+10216 (see figure 17) with a central peak surrounded by patchy circumstellar
 matter (Osterbart et al. 1996, Weigelt et al. 1998), (iii) exhibiting two 
lobes of the evolved object, the Red Rectangle (see figure 18) (Osterbart 
et al. 1996), (iv) revealing a spiral pinwheel in the dust around WR104 
(Tuthill, Monnier et al. 1999), and (v) depicting spherical dust shell around 
oxygen-rich AGB star AFGL~2290 (Gauger et al. 1999) are to be noted; images of
the young star, LkH$\alpha$~101 in which the structure of the inner accretion
disk is resolved have been reported as well (Tuthill et al. 2001).
Detailed information that is needed for the modeling of the 2-d 
radiative transfer concerning the symmetry $-$ spherical, axial or lack of 
clouds, plumes etc. of the objects $-$ can also be determined 
(Men'shchikov and Henning, 1997, Gauger et al. 1999). 
   
\begin{figure} 
\centerline {\psfig{figure=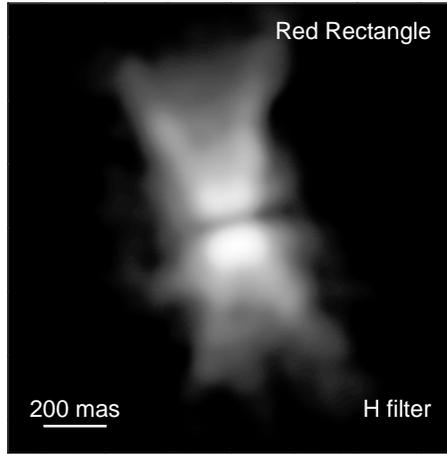,height=6cm}}
\caption {Speckle masking reconstruction of reflection nebula around the star 
HD44179, of Red Rectangle Red Rectangle (AFGL~915) exhibiting two lobes with 
$\rho$ =  $\sim$0.15$^{\prime\prime}$ (Courtesy: R. Osterbart); 
the dark lane between the lobes might be due to an obscuring dust disk 
and the central star is a close binary system (Osterbart et al. 1996).} 
\end{figure}

Both novae and supernovae (SN) have a complex nature of shells viz., multiple, 
secondary and asymmetric. High resolution mapping may depict the events near the
star and the interaction zones between gas clouds with different velocities. 
Soon after the explosion of the supernova SN1987A, various observers 
monitored routinely the expansion of the shell in different wavelengths 
(Nisenson et al. 1987, Papaliolios et al. 1989, Wood et al. 1989). The 
increasing diameter of the same in several spectral lines and the continuum was 
measured (Karovska et al. 1989). Nulsen et al. (1990) have derived
the velocity of the expansion as well and found that the size of this object was
strongly wavelength dependent at the early epoch $-$ pre-nebular phase 
indicating stratification in its envelope. A bright source at 
0.06$^{\prime\prime}$ away from the SN1987A with $\Delta m$ 2.7~mag 
at H$\alpha$ had also been detected. Recently, Nisenson and Papaliolios (1999) 
have detected a faint second spot, $\Delta m_v$~4.2~mag, on the opposite side of
SN1987A with $\rho$ = 160~mas.

\begin{figure} 
\centerline {\psfig{figure=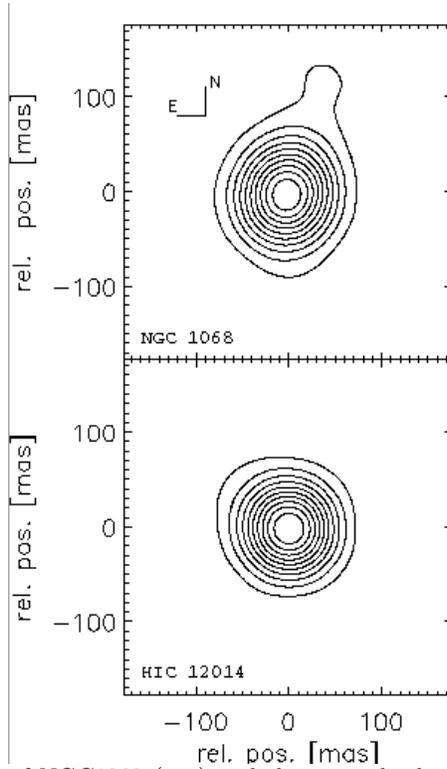,height=10cm}}
\caption {Speckle masking reconstruction of NGC1068 (top) and the
unresolved star HIC12014 (bottom). The contours are from 6\% to 100\% of peak
intensity (Wittkowski, Balega et al. 1998; Courtesy: M. Wittkowski).}    
\end{figure} 
 
Another important field of observational astronomy is the study of the physical 
processes, viz., temperature, density and velocity of gas in the active regions 
of active galactic nuclei (AGN); optical imaging by emission lines on 
sub-arcsecond scales can reveal the structure of the narrow-line region. The 
scale of narrow-line regions is well resolved by the diffraction-limit of a 
moderate-sized telescope. The time variability of AGNs ranging from minutes to 
decades can also be studied. The NGC~1068 is an archetype type 2 Seyfert galaxy.
Observations of this object corroborated with theoretical modeling like 
radiative transfer calculations have made significant contributions on its 
structure. Ebstein et al. (1989) found a bipolar structure of this object in the
[OIII] emission line. Near-IR observations at the Keck~I telescope trace a very 
compact central core and extended emission with a size of the order of 10~pc on 
either side of an unresolved nucleus (Weinberger
et al. 1999). Wittkowski, Balega et al. (1998) have resolved central 2.2~$\mu$m 
core by bispectrum speckle interferometry at the diffraction-limit of the 
Special Astrophysical Observatory (SAO) 6~m telescope, with a FWHM size of 
$\sim$2~pc for an assumed Gaussian 
intensity distribution. Figure 19 depicts the reconstructed image of the AGN, 
NGC1068. Subsequent observations by Wittkowski et al. (1999) indicate that this 
compact core is asymmetric with a position angle of $\sim$20$^\circ$ and an 
additional more extended structure in N-S direction out to $\sim$25~pc. 

Quasars (QSO) may be gravitationally lensed by stellar objects such as, 
stars, galaxies, clusters of galaxies etc., located along the line of sight.
The aim of the high angular imagery of these QSOs is to find their structure 
and components; their number and structure as a probe of the distribution of 
the mass in the Universe. The capability of resolving these objects 
in the range of 0.2$^{\prime\prime}$ to 0.6$^{\prime\prime}$ would allow the 
discovery of more lensing events. The gravitational image of the 
multiple QSO~PG1115+08 was resolved by Foy et al. (1985); one of the bright 
components, discovered to be double (Hege et al. 1981), was found to be 
elongated that might be, according to them, due to a fifth component of the QSO.

\subsubsection{Glimpses of AO observations}

Most of the results that obtained from the ground-based telescopes equipped with
AO systems are in the near-IR band; while results at 
visible wave lengths continue to be sparse. The contributions are in the 
form of studying (i) planetary meteorology (Poulet and Sicardy,
1996, Marco et al. 1997, Roddier et al. 1997); images of Neptune's ring arcs 
are obtained (Sicardy et al. 1999) that are interpreted as gravitational 
effects by one or more moons, (ii) nucleus of M31 (Davidge, Rigaut et al. 
1997), (iii) young stars and multiple star systems (Bouvier et al. 1997), 
(iv) galactic center (Davidge, Simons et al. 1997), (v) Seyfert galaxies, QSO 
host galaxies (Hutchings et al. 1998, 1999), and (vi) circumstellar 
environment (Roddier et al. 1996). The images of the objects such as, (a) the
nuclear region of NGC3690 in the interacting galaxy Arp~299 (Lai et al. 
1999), (b) the starburst/AGNs, NGC863, NGC7469, NGC1365, NGC1068, 
(c) the core of the globular cluster M13 (Lloyd-Hart et al. 1998).
and (d) R136 (Brandl et al. (1996) etc. are obtained from the moderate-sized
telescopes. The highest ever angular resolution AO images of the radio galaxy
3C294 in the near-IR bands have been obtained at Keck Observatory (Quirrenbach 
et al. 2001).

AO systems can also employed for studying young stars, multiple stars, 
natal disks, and related inward flows, jets and related outward flows,
proto-planetary disks, brown dwarfs and planets. Roddier et al. (1996) have 
detected a binary system consisting of a K7-MO star with an M4 companion that 
rotates clockwise; they suggest that the system might be 
surrounded by a warm unresolved disk. The massive star Sanduleak-66$^\circ$41 
in the LMC was resolved into 12 components by Heydari and Beuzit (1994). 
Success in resolving companions to nearby dwarfs has been reported (Beuzit et 
al. 2001, Kenworthy et al. 2001).
Macintosh et al. (2001) measured the position of the brown dwarf companion to 
TWA5 and resolved the primary into an 0.055$^{\prime\prime}$ double.
 
The improved resolution of crowded fields like globular clusters
would allow derivation of luminosity functions and spectral type,
to analyze proper motions in their central area. Simon et al. (1999)
have detected 292 stars in the dense Trapezium star cluster of the Orion nebula 
and resolved pairs to the diffraction-limit of a 2.2~m telescope. Optical and
near-IR observations of the close Herbig Ae/Be binary star NX~Pup (Brandner et 
al. 1995), associated with the cometary globular cluster I,  
Sch\"oller et al. (1996) estimated the mass and age of both the 
components and suggest that circumstellar matter around the former could be 
described by a viscous accretion disk. Line and continuum fluxes, and
equivalent widths are also derived for the massive stars in the Arches 
cluster (Blum et al. 2001).

\begin{figure} 
\centerline {\psfig{figure=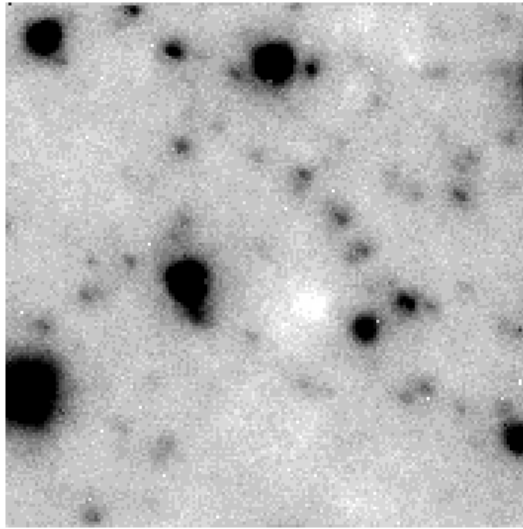,height=7cm}}
\caption {The ADONIS K$^\prime$ image of the Sgr window in the bulge of the
Milky Way (Bedding, Minniti et al. 1997). The image is 
8$^{\prime\prime}$$\times$8$^{\prime\prime}$ (Courtesy: T. Bedding).}
\end{figure} 

Stellar populations in galaxies in near-IR region provides the peak of the 
spectral energy distribution for old populations. Bedding, Minniti et al. 
(1997) have observed the Sgr~A window at the Galactic center of 
the Milky Way. They have produced an IR luminosity function and 
color-magnitude diagram for 70 stars down to $m_v\simeq$19.5 mag. Figure 20 
depicts the ADONIS K$^\prime$ image of the Sgr~A window.

Images have been obtained of the star forming region Messier 16 (Currie et al. 
1996), the reflection nebula NGC2023 revealing small-scale structure in the 
associated molecular cloud, close to the exciting star, in Orion (Rouan et al. 
1997). Close et al. (1997) mapped near-IR polarimetric observations of 
the reflection nebula R~Mon resolving a faint source, 
0.69$^{\prime\prime}$ away from R~Mon and identified it as a T~Tauri star. 
Monnier et al. (1999) found a variety of dust condensations that include a 
large scattering plume, a bow shaped dust feature around the red supergiant 
VY~CMa; a bright knot of emission 1$^{\prime\prime}$ away from the star is also 
reported. They argued in favor of the presence of chaotic and violent dust
formation processes around the star. Imaging of proto-planetary nebula (PPN), 
Frosty Leo and the Red Rectangle by Roddier et al. (1995) revealed a binary 
star at the origin of these PPNs. 
 
Imaging of the extragalactic objects particularly the central area of active 
galaxies where cold molecular gas and star formation occur is an important 
program. From the images of nucleus of NGC1068, Rouan et al. (1998), found 
several components that include: (i) an unresolved conspicuous core, (ii) an 
elongated structure, and (iii) large and small-scale spiral structures. Lai et 
al. (1998) have recorded images of Markarian 231, a galaxy 160~Mpc away 
demonstrating the limits of achievements in terms of morphological structures 
of distant objects. 

Aretxaga et al. (1998) reported the unambiguous detection of the host galaxy of
a normal radio-quiet QSO at high-redshift in K-band; detection of emission line 
gas within the host galaxies of high redshift QSOs has been reported as well 
(Hutchings et al. 2001). Observations by Ledoux et al. (1998) of broad 
absorption line quasar APM~08279+5255 at z=3.87 show the object consists of a 
double source ($\rho$ = 0.35$^{\prime\prime}$ $\pm$~0.02$^{\prime\prime}$; 
intensity ratio = 1.21~$\pm$~0.25 in H band). They proposed for a gravitational 
lensing hypothesis which came from the uniformity of the quasar spectrum as a 
function of the spatial position. 
Search for molecular gas in high redshift normal galaxies in the foreground of 
the gravitationally lensed quasar Q1208+1011 has also been made (Sams et al. 
1996). AO imaging of a few low and intermediate redshift quasars has been 
reported recently (M\'arquez et al. 2001).

High resolution stellar coronagraphy is of paramount importance in (i) detecting
low mass companions, e.g., both white and brown dwarfs, 
dust shells around asymptotic giant branch (AGB) and post-AGB stars,  
(ii) observing nebulosities leading to the formation of a planetary system, 
ejected envelops, accretion disk, and (iii) understanding of 
structure (torus, disk, jets, star forming regions), and dynamical 
process in the environment of AGNs and QSOs. By means of coronagraphic 
techniques the environs of a few interesting objects have been explored. They
include: (i) a very low mass companion to the astrometric binary Gliese~105~A 
(Golimowski et al. 1995), (ii) a warp of the circumstellar disk around the star 
$\beta$~Pic (Mouillet et al. 1997), (iii) highly 
asymmetric features in AG Carina's circumstellar environment (Nota et al. 
1992), (iv) bipolar nebula around the LBV R127 (Clampin et al. 1993), and 
(v) the remnant envelope of star formation around pre-main sequence stars
(Nakajima and Golimowski, 1995). 

\subsection{Impact of LBOIs in astrophysics}

The main objective of LBOIs is to measure the diameters, distances, masses
and luminosities of stars, to detect the morphological details, such as 
granulations, oblateness of giant stars, and the image features, 
i.e., spots and flares on their surfaces. Eclipsing binaries are also good 
candidates; for they provide information on circumstellar envelopes such 
as, the diameter of inner envelope, color, symmetry, radial profile etc. As
stated earlier (in section VII.A.2), good spectroscopic and interferometric
measurements are required to derive precise stellar masses since they 
depend on $sin^3 i$. A small variation on the inclination $i$ implies a 
large variation on the radial velocities. Most of the orbital calculations that 
are carried out with speckle observations are not precise to provide masses 
better than 10\% (Pourbaix, 2000). 

The results obtained so far with LBOIs are from the area of stellar angular 
diameters with implications for emergent fluxes, effective temperatures, 
luminosities and structure of the stellar atmosphere, dust and gas envelopes, 
binary star orbits with impact on cluster distances and stellar masses, relative
sizes of emission-line stars and emission region, stellar rotation, limb 
darkening, astrometry etc. (Saha, 1999, Saha and Morel, 2000, Quirrenbach, 2001 
and references therein). The angular diameter for more than 50 stars has been 
measured (DiBenedetto and Rabbia, 1987, Mozurkewich et al. 1991, Dyck et al. 
1993, Nordgren et al. 2000, Perin et al. 1999, Kervella et al. 2001, van Belle 
et al. 2001) with accuracy better than 1\% in some cases. 

Interesting results that have been obtained using I2T and Mark~III 
interferometers are: (i) measuring diameters, effective temperatures of 
giant stars (Faucherre et al. 1983, DiBenedetto and Rabbia, 1987), 
(ii) resolving the gas envelope of the Be star $\gamma$~Cas in the 
H$\alpha$ line (Thom et al. 1986), and structure of circumstellar shells 
(Bester et al. 1991), and (iii) determining orbits for spectroscopic, and 
eclipsing binaries (Armstrong et al. 1992, Shao and Colavita, 1994). 

The GI2T is being used regularly to observe the Be stars, LBVs, spectroscopic 
and eclipsing binaries, wavelength dependent objects, diameters of bright stars,
and circumstellar envelopes. However, the scientific programs 
are restricted by the low limiting visible magnitude down to 5 (seeing and 
visibility dependent). The first successful result that
is reported on resolving the rotating envelope of hot star, $\gamma$~Cas
came out of this instrument in 1989. Mourard et al. (1989) observed this star
with a spectral resolution of 1.5~\AA centered on H$\alpha$. As many as 
$\sim$300,000 short-exposure images were recorded by a photon-counting camera,
CP40 (Blazit, 1986). They have digitally processed each image, which contained 
$\sim$100 photons, using the correlation technique. The results were co-added to
reduce the effect of atmospheric seeing and photon noise, according to the 
principle of speckle interferometry (Labeyrie, 1970). With the central star as 
a reference, they have 
determined the relative phase of the shell visibility and showed clearly the 
rotation of the envelope. This result demonstrates the potential of observations
that combines spectral and spatial resolution. Through subsequent observations 
on later dates, Stee et al. (1995, 1998) derived the radiative transfer model.
Using the data obtained since 1988 with this instrument, Berio, Stee et al. 
(1999) have found the evidence of a slowly prograde of rotating density pattern 
in the said star's equatorial disk. Indeed $\gamma$~Cas has been a favorite 
target to the GI2T, with further systematic monitoring of multiple emission 
lines, the formation, structure, and dynamics of other Be stars can be 
addressed. The other noted results obtained in recent times with this instrument
include the mean angular diameter and accurate distance estimate of $\delta$~Cep
(Mourard et al. 1997), subtle structures in circumstellar environment such as 
jets in the binary system $\beta$~Lyr (Harmanec et al. 1996), clumpiness in the 
wind of P~Cyg (Vakili et al. 1997), and detection of prograde one-armed 
oscillation in the equatorial disk of the Be star $\zeta$~Tau (Vakili et al.
1998). With the SUSI instrument, Davis et al. (1998, 1999b) have determined the 
diameter of $\delta$~CMa with an accuracy of $\pm$1.8\%. 

From the data obtained at the COAST, aperture-synthesis maps of the double-lined
spectroscopic binary $\alpha$~Aur (Baldwin et al. 1996) depict
the milli-arcsecond orbital motion of the system over a 15 day interval. Images 
of $\alpha$~Ori reveals the presence of a circularly symmetric data with an 
unusual flat-topped and limb darkening profile (Burns et al. 1997). Young et 
al. (2000) have found a strong variation in the apparent symmetry of the
brightness distribution as function of wavelength. Variations of the cycle of 
pulsation of the Mira variable R~Leo have been measured (Burns et al. 1998).

With IOTA, the angular diameters and effective temperatures have been measured
for carbon stars (Dyck et al. 1996), Mira variables (van Belle et al. 1996, 
1997), cool giant stars (Perrin et al. 1998, 1999, Cepheids (Kervella et al. 
2001), and dust shell of CI~Cam (Traub et al. 1998).  
Millan-Gabet et al. (2001) have resolved circumstellar structure of Herbig
Ae/Be stars in near-IR. Figure 21 depicts the examples of H-band 
visibility data and models for 2 sources. The lower right panel in figure 21
illustrates the observed lack of visibility variation with baseline
rotation, consistent with circumstellar emission from dust which is
distributed either in a spherical envelope or in a disk viewed almost face-on.

\begin{figure} 
\centerline {\psfig{figure=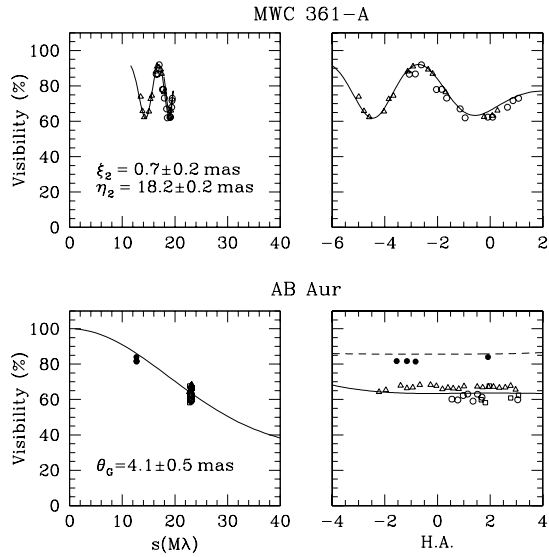,height=7.5cm}}
\caption {Examples of H-band visibility data and models for two Herbig Ae/Be 
stars (Courtesy: R. Millan-Gabet). The data and models are plotted as a function
of baseline length (left panels), and hour angle (right panels), which 
determines the baseline position angles; different symbols correspond to data 
obtained on different nights (Millan-Gabet et al. 2001). Solid symbols and 
dashed lines are for 21~m baseline data and models, and open symbols and solid 
lines are for 38~m baseline data and models. The upper panels show the data and 
best fit model for the binary detection in MWC~361-A, and displays the companion
offsets in right ascension ($\xi_2$) and declination ($\eta_2$). The
lower panels show the data and best fit Gaussian model for AB~Aur, and
displays the angular FWHM ($\theta_G$).}
\end{figure} 

Figure 22 summarizes the existing set of measurements of near-IR
sizes of said Herbig sources (credit: R. Millan-Gabet and J. D. Monnier),
using the data from Danchi et al. (2001), Millan-Gabet et al. (2001), and 
Tuthill et al. (2001). The agreement observed in most cases has
motivated in part a revision of disk physics in models of Herbig Ae/Be
systems (Natta et al. 2001). In the new models, the gas in the inner
disk is optically thin so that dust at the inner edge is irradiated
frontally, and expands forming a `puffed-up' inner wall. Due to the
extra heating, compared to the irradiation of a flat disk traditionally
considered, this model results in larger near-IR sources, essentially 
corresponding to the `puffed up' inner wall, which appears as a bright ring to 
the interferometer. Monnier et al. (2001) have reported the results of 
decomposing the dust and stellar signatures from the evolved stars (figure 16). 

\begin{figure} 
\centerline {\psfig{figure=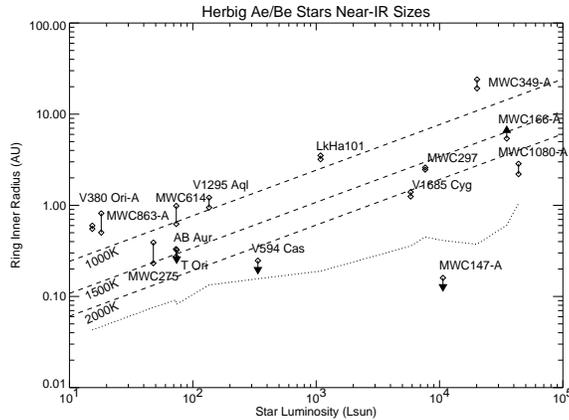,height=5.5cm,angle=90}}
\caption {Measurements of near-IR sizes of Herbig Ae/Be sources 
(Courtesy: R. Millan-Gabet and J. D. Monnier). 
The characteristic sizes plotted have been derived
from the visibility data using a simple model consisting of a central
point source plus a uniform ring. The sizes (ring 
inner radii) are plotted as a function of the luminosity of the central stars, 
so that they may be compared with the location of optically thin dust at typical
sublimation temperatures (dashed lines).} 
\end{figure} 

With PTI, Malbet et al. (1998) have resolved the young stellar 
object (YSO), FU~Ori in the near-IR with a projected resolution better than 
2~AU. Measurements of diameters and effective temperatures of cool stars and
Cepheids have been reported (van Belle et al. 1999, Lane et al. 2000). The 
visual orbit for the spectroscopic binary $\iota$~Peg with 
interferometric visibility data has also been derived (Boden et al. 1999).
Direct observations of an oblate photosphere of a main sequence star,
$\alpha$~Aql (van Belle et al. 2001) have been carried out. Ciardi et al.
(2001) have resolved the stellar disk of $\alpha$~Lyr in the near-IR. 
Employing NPOI, Hummel et al. (1998) have determined the orbital parameters of 
two spectroscopic binaries, $\zeta$~UMa (Mizar A), $\eta$~Peg (Matar) and 
derived masses and luminosities. The limb-darkened diameters have also been
measured of late-type giant stars (Hajian et al. 1998, Pauls et al. 1998, 
Wittkowski et al., 2001), and Cepheid variables (Nordgren et al. 2000). 

The observing programs of ISI have been aimed at determining the spatial
structure and temporal evolution of dust shell around long period variables.
From the data obtained with this instrument at 11.15~$\mu$m, Danchi et al. 
(1994) showed that the radius of dust formation depends on the spectral type
of the stars. Lopez et al. (1997) have found a strong dependence on the 
pulsation phase. Observations have been made of NML~Cyg, $\alpha$~Sco, as well 
as of changes in the dust shell around Mira and IK~Tau, (Hale et al. 1997, 
Lopez et al. 1997), mid-IR molecular absorption features of ammonia and silane 
of IRC+10216 and VY~CMA (Monnier et al. 2000).

Recent observations with the two of the VLT telescopes have measured the angular
diameter of the blue dwarf $\alpha$~Eri, which was found to be 1.92~mas 
(Glindemann and Paresce, 2001). Subsequently they (i) derived  
the diameters of a few red dwarf stars, and (ii) determined the variable 
diameters of a few pulsating Cepheid stars, as well as the measurement of the 
core of $\eta$~Carinae. 

\subsection{Perspectives of interferometry}

\label{subsec:perspec}

The future of high resolution optical astronomy lies with the new generation 
arrays but its implementation is a challenging task. Numerous technical 
challenges for developing such a system will require careful attention. 
Nevertheless, steady progress has enabled scientists to expand their knowledge 
of astrophysical processes. With improved technology, the interferometric 
arrays of large telescopes may provide snap-shot images at their recombined 
focus and yield improved images, spectra of quasar host galaxies, and 
astrometric detection and characterization of extra-solar planets. The expected 
limiting magnitude of a hyper-telescope imaging technique is found to be 
8.3~$m_v$ by numerical solution if 10-cm apertures are used and 20~$m_v$ for 
10-m apertures (Pedretti and Labeyrie, 1999). The limit is expected to increase 
with the CARLINA array (Labeyrie, 1999b), a 100-element hyper-telescope with a 
diameter of 200~m, shaped like Arecibo radio telescope in Puerto Rico. 

High precision astrometry also helps in establishing the cosmic distance scale; measurements of proper motion can confirm stars as members of cluster 
(known distance) that may elucidate the dynamics of the galaxy. The quest for 
extra-solar planets (Wallace et al. 1998) is a challenge for narrow angle 
astrometry. Very valuable astrometric results from space have already been 
obtained by the Hipparcos satellite (Perryman, 1998). Hipparcos used phase-shift
measurements of the temporal evolution of the photometric level of two stars
seen drifting through a grid. The successor of Hipparcos, Gaia (Lindengren and
Perryman, 1996), will probably use the same technique with improvements,
yielding more accurate results on a larger number of objects. However, only 
space-borne interferometers will achieve very high precision angular 
measurements. 

\subsubsection{Characterization of extra-solar planets}

As many as seventy-six Jovian-size planets orbiting stars have been identified 
by the Doppler-Fizeau effect (Mayor and Queloz, 1995, Butler and Marcy, 1996, 
http://exoplanets.org). For one of them, an atmosphere containing sodium
(observed in the sodium resonance doublet at 589.3~nm) has been detected in 
absorption as the planet transits its parent star HD209458 (Charbonneau et al.,
2001). It may also be possible to detect smaller planets by 
measuring the stellar photocenter motion due to the wobble. Such photocenter 
measurements will require diffraction-limited imaging even for the best possible
candidates. Interferometry can be used to measure the `wobble' in the 
position of a star caused by the transverse component of a companion's motion.
A planet orbiting around a star causes a revolution of the star around 
the center of gravity defined by the two masses. Like galaxy velocity, this 
periodical short motion has a radial counterpart measurable from the ground by 
spectrometry. 

The aim of the space interferometers like DARWIN and TPF is the discovery and 
characterization of terrestrial planets around
nearby stars (closer than 15~pc) by direct detection (i.e., involving the
detection of photons from the planet and not from the star as it is done with
Doppler-Fizeau effect detection or wobble detection). The difficulties  
for achieving Earth-like planet detection come from (i) minimizing scattered
light from the parent star and (ii) the presence of exo-zodiacal light 
(infrared emission from the dust surrounding the parent star). Interferometric 
nulling technique will be useful to address the first issue. 

\subsubsection{Astrobiology}

The knowledge of the chemical composition of any planetary atmosphere gives 
hints about the likelihood to find carbon-based life. Lovelock (1965) has 
suggested that the simultaneous presence on Earth of a highly oxidized gas, like
${\rm O}_2$, and highly reduced gases, like ${\rm CH}_4$ and ${\rm N}_2{\rm O}$ 
is the result of the biochemical activity. However, finding spectral signatures 
of these gases on an extra-solar planet would be very difficult. An alternative 
life indicator would be ozone (${\rm O}_3$), detectable as an absorption feature
at 9.6~$\mu$m. On Earth, ozone is photochemically produced from
${\rm O}_2$ and, as a component of the stratosphere, is not masked by other
gases. Finding ozone would, therefore, indicate a significant quantity of ${\rm
O}_2$ that should have been  produced by photosynthesis (L\'eger et al.
1993). Moreover, for a star-like the Sun, detecting ozone can be done 1000 times
faster than detecting ${\rm O}_2$ at 0.76~$\mu$m: estimates made by Angel and
Woolf (1997) show that the requirements for planet detection in the visible
with an 8~m telescope are not detectable with current technology. 

\subsubsection{Long term perspective}

Space-borne interferometry projects for years spanning from 2020 to 2050 already
exist; such projects must be regarded as drafts for future instruments.  
For the post-TPF era, NASA has imagined an enhanced version featuring
four 25~m telescopes and a $R\geq 1000$ spectrometer. This interferometer would
be able to detect an extra-solar planet lines of gases directly produced by
biochemical activity. The next step proposed by NASA is an array of 25
telescopes, 40~m diameter each, that would yield 25$\times$25 pixel images of 
an Earth-like planet at 10~pc, revealing its
geography and eventually oceans or chlorophyll zones.

A comparable project has been proposed by Labeyrie (1999a). It consists
of 150 telescopes, 3~m diameter each, forming an interferometer with a 150~km
maximum baseline. Such an instrument equipped with a highly efficient
coronagraph would give a 40$\times$40 pixel image of an Earth-like planet at 
3~pc. Figure 23 depicts a simulated image of an Earth-like planet detection 
(Labeyrie, 1999b). Developing LBI for lunar operation consisting of 20 to 27 
off-axis parabolic segments carried by robotic hexapodes that are movable during
observing run has also been suggested by Arnold et al. (1996). 

\begin{figure} 
\centerline {\psfig{figure=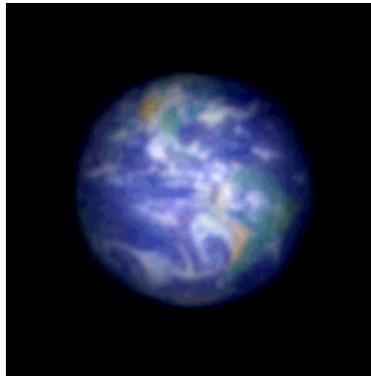,height=5cm}}
\caption {Simulated image of an Earth-like Planet detection (Courtesy: A. 
Labeyrie). An image of the Earth picture was convolved with the spread function 
from 150 point apertures, arrayed in 3 circles; it was multiplied by the 
hyper-telescope envelope, i.e. the spread function of the sub-pupils as well, 
and then pushed somewhat the contrast to attenuate the diffractive halo 
(Labeyrie, 2000).} 
\end{figure} 

\section{CONCLUSIONS}
 
Earth-bound astronomical observations are strongly affected by atmospheric 
turbulence that set severe limits on the angular resolution, which 
in optical domain, is rarely better than $\sim$~0.5$^{\prime\prime}$. A basic 
understanding of interference phenomenon is of paramount importance to other 
branches of physics and engineering as well. In recent years, a wide
variety of applications of speckle patterns has been found in many areas. 
Though the statistical properties of the speckle pattern is complicated, a 
detailed analysis of this pattern is useful in information processing. 

Image-processing is a very important subject. A second-order moment (power 
spectrum) analysis provides only the modulus of the object FT, whereas a 
third-order moment (bispectrum) analysis yields the phase allowing the object to
be fully reconstructed. A more recent attempt to go beyond the third order, 
e.g., fourth-order moment (trispectrum), illustrates its utility in finding 
optimal quadratic statistics through the weak gravitational lensing effect (Hu, 
2001). This algorithm provides a far more sensitive test than the bispectrum for
some possible sources of non-Gaussianity (Kunz et al., 2001), however, its
implementation in optical imaging is a difficult computational task. 
Deconvolution method is an important topic as well. It applies to imaging in 
general which covers methods spanning from simple linear 
deconvolution algorithms to complex non-linear algorithms. 
 
In astronomy, the field of research that has probably benefited the most from 
high angular resolution techniques using single telescopes, and will still 
benefit in the future, is undoubtedly the origin and evolution of stellar 
systems. This evolution starts with star formation, including multiplicity, 
and ends with the mass loss process which recycles heavier elements into the
interstellar medium. Large-scale star formation provides coupling between 
small-scale and large-scale processes. Stellar chemical evolution or 
nucleo-synthesis that is a result of star formation activity further influences 
evolutionary process. High resolution observations are fruitful for the 
detection of proto-planetary disks and possibly planets (either astrometrically,
through their influence on the disk, or even directly). The technique is also 
being applied to studies of starburst and Seyfert galaxies, AGNs, and 
quasars. Studies of the morphology of stellar 
atmospheres, the circumstellar environment of nova or supernova, YPN, long 
period variables (LPV), rapid variability of AGNs etc. are also essential. 

In spite of the limited capability of retrieving fully diffraction-limited 
images of the objects, AO systems are now offered to users of large telescopes. 
AO observations have contributed to the study of the solar system, and added
to the results of space borne instruments, for examples, monitoring of
the volcanic activity on Io or of the cloud cover on Neptune, the detection
of Neptune's dark satellites and arcs, and the ongoing discovery of
companions to asteroids etc.; they are now greatly contributing to the study of
the Sun itself as well (Antoshkin et al. 2000, Rimmele, 2000). Combination of 
AO systems with speckle imaging may enhance the results (Dayton et al. 1998). By
the end of the next decade (post 2010), observations using the AO system 
on a new generation telescope like OWL, will revolutionize the mapping of 
ultra-faint objects like blazars that exhibit the most rapid and the largest 
amplitude variations of all AGN (Ulrich et al. 1997), extra-solar planets etc.;
certain aspects of galactic evolution like chemical evolution in the Virgo 
cluster of galaxies can be studied as well. 
 
A host of basic problems needs a very high angular resolution for their 
solution. Among others, an important fundamental problem that can be addressed 
with the interferometry and AO is the origin and evolution of galaxies. 
The upcoming large facilities with phased arrays of multiple 8-10~m 
sub-apertures will provide larger collecting areas and higher spatial 
resolution simultaneously than the current interferometers. These instruments 
fitted with complete AO systems would be able to provide imaging and 
morphological informations on the faint extragalactic sources such as, galactic 
centers in the young universe, deep fields, and host galaxies. Measurement of 
such objects may be made feasible by the instruments with a fairly complete 
$u, v$ coverage and large field of view. The derivations of motions and 
parallaxes of galactic centers seem to be feasible with phase reference 
techniques. Origin of faint structures close to non-obscured central sources 
can also be studied in detail with interferometric polarization measurement.   

The capabilities of the proposed large arrays offer a revolution in the study of
compact astronomical sources from the solar neighborhood to cosmological 
distances. The aim ranges from detecting other planetary systems to imaging the 
black hole driven central engines of quasars and active galaxies; gamma-ray 
bursters may be the other candidate. Another important scientific objective is
the recording of spectra to derive velocity and to determine black hole masses 
as a function of redshift. At the beginning of the
present millennium, several such arrays will be in operation both on the ground 
as well as in space. Space-borne interferometers are currently planned to detect
planets either astrometrically (SIM) or directly (TPF). Projects like DARWIN and
other ambitious imaging instruments may also come to function. 

\section*{ACKNOWLEDGMENTS} 

The author expresses his gratitude to A. Labeyrie, S. T. Ridgway, and P. A.
Wehinger for comments on the review, and indebtedness to V. Coud\'e du Foresto, 
P. R. Lawson, and S. Morel for valuable communications. Thanks are also 
due to T. R.  Bedding, A. Boccaletti, P. M. Hinz, R. Millan-Gabet, J. D. 
Monnier, R. Osterbart, and M. Wittkowski for providing the images, figures etc.,
and granting permission for their reproduction, as well as to H. Bradt and P. 
Hickson for reading the manuscript. The services rendered by V. 
Chinnappan, K. R. Subramanian, and B. A. Varghese are gratefully acknowledged.


\begin{references}
 
\bibitem{}
Ables J. G., 1974, Astron. Astrophys. Suppl., {\bf 15}, 383. 
\bibitem{}
Aime C., 2000, J. Opt. A: Pure \& Appl. Opt., {\bf 2}, 411.
\bibitem{}
Akeson R., M. Swain, and M. Colavita, 2000, SPIE, {\bf 4006}, 321. 
\bibitem{}
Anderson J. A., 1920, Astrophys. J., {\bf 51}, 263.
\bibitem{}
Angel J. R. P., 1994, Nature, {\bf 368}, 203.
\bibitem{}
Angel J. R. P., 2001, Nature, {\bf 409}, 427.
\bibitem{}
Angel J. R., and N. J. Woolf, 1997, Astrophy. J., {\bf 475}, 373.
\bibitem{}
Antoshkin L. V. et al., 2000, SPIE, {\bf 4007}, 232.
\bibitem{}
Aretxaga I., D. Mignant, J. Melnick, R. Terlevich, and B. Boyle, 1998,
Mon. Not. R. Astron. Soc., {\bf 298}, L13.
\bibitem{}
Armstrong J. T., C. A. Hummel, and D. Mozurkewich, 1992, Proc. ESO-NOAO conf. 
eds., J. M. Beckers \& F. Merkle, ESO, Germany, 673.
\bibitem{}
Armstrong J. et al., 1998, Astrophys. J., {\bf 496}, 550.
\bibitem{}
Arnold L., A. Labeyrie, and D. Mourard, 1996, Adv. Space Res. {\bf 18}, 1148.
\bibitem{}
Ayers G. R., and J. C. Dainty, 1988, Opt. Lett., {\bf 13}, 457.
\bibitem{}
Baba N., S. Kuwamura, N. Miura, and Y. Norimoto, 1994, Astrophys. J., 
{\bf 431}, L111.
\bibitem{}
Baba N., S. Kuwamura, and Y. Norimoto, 1994, Appl. Opt., {\bf 33}, 6662.
\bibitem{}
Baba N., H. Tomita, and N. Miura, 1994, Appl. Opt., {\bf 33}, 4428.
\bibitem{}
Babcock H. W., 1953, Pub. Astron. Soc. Pac, {\bf 65}, 229.
\bibitem{}
Babcock H. W., 1990, Science,{\bf 249}, 253.
\bibitem{}
Bagnuolo Jr. W. G., B. D. Mason, D. J. Barry, W. I. Hartkopf, and H. A. 
McAlister, 1992, Astron. J., {\bf 103}, 1399.
\bibitem{}
Baldwin J. et al., 1996, Astron. Astrophys., {\bf 306}, L13.
\bibitem{}
Baldwin J., R. Boysen, C. Haniff, P. Lawson, C. Mackay, J. 
Rogers, D. St-Jacques, P. Warner, D. Wilson, and J. Young, 1998, 
SPIE., {\bf 3350}, 736.
\bibitem{}
Baldwin J. E., C. A. Haniff, C. D. Mackay, and P. J. Warner, 1986, Nature, 
{\bf 320}, 595.
\bibitem{}
Baldwin J., R. Tubbs, G. Cox, C. Mackay, R. Wilson, and M. Andersen, 2001,
Astron. Astrophys., {\bf 368}, L1.
\bibitem{}
Baldwin J. E., and P. J. Warner, 1978, Mon. Not. R. Astron. Soc., {\bf 182}, 
411.
\bibitem{}
Balick B., 1987, Astron. J., {\bf 94}, 671.
\bibitem{}
Barlow M. J., B. L. Morgan, C. Standley, and H. Vine, 1986, Mon. Not. R. 
Astron. Soc., {\bf 223}, 151.
\bibitem{}
Barnaby D., E. Spillar, J. Christou, and J. Drummond, 2000, Astron. J.,
{\bf 119}, 378.
\bibitem{}
Bates R., and M. McDonnell, 1986, `Image Restoration and Reconstruction',
Oxford Eng. Sc., Clarendon Press.
\bibitem{}
Beckers J. M., 1982, Opt. Acta., {\bf 29}, 361.
\bibitem{}
Bedding T. R., D. Minniti, F. Courbin, and B. Sams, 1997, Astron. Astrophys.,
{\bf 326}, 936.
\bibitem{}
Bedding T., A. Zijlstra, O. Von der L\"uhe, J. Robertson, R. Marson, 
J. Barton, and B. Carter, 1997, Mon. Not. R. Astron. Soc., {\bf 286}, 957.
\bibitem{}
Beichman C., 1998, SPIE, {\bf 3350}, 719.
\bibitem{}
Benson J., D. Mozurkewich, and S. Jefferies, 1998, SPIE, {\bf 3350}, 493. 
\bibitem{}
Berger J., P. Haguenauer, P. Kern, K. Perraut, F. Malbet, I. Schanen, M. Severi,
R. Millan-Gabet, and W. Traub, 2001, Astron. Astrophys., {\bf 376}, L31. 
\bibitem{}
Berio P., D. Mourard, D. Bonneau, O. Chesneau, P. Stee, N. Thureau, and F. 
Vakili, 1999, J. Opt. Soc. Am. A., {\bf 16}, 872.
\bibitem{}
Berio P., P. Stee, F. Vakili, D. Mourard, D. Bonneau, O. Chesneau, N. Thureau, 
D. Le Mignant, and R. Hirata, 1999, Astron. Astrophys., {\bf 345}, 203.
\bibitem{}
Bester M., W. C. Danchi, C. G. Degiacomi, and C. H. Townes, 1991, Astrophys. 
J., {\bf 367}, L27.
\bibitem{}
Beuzit J. et al., 2001, astro-ph/0106277.
\bibitem{}
Blazit A., 1986, SPIE, {\bf 702}, 259.
\bibitem{}
Bl\"ocker T., Y. Balega, K. -H. Hofmann, and G, Weigelt, 2001, Astron. 
Astrophys. (to appear).
\bibitem{}
Blum R., D. Schaerer, A. Pasquali, M. Heydari-Malayeri, P. Conti, and W. 
Schmutz, 2001, Astron. J., (to appear).
\bibitem{}
Boccaletti A., 2001, Private communication.
\bibitem{}
Boccaletti A., C. Moutou, D. Mouillet, A. Lagrange, and J. Augereau, 2001, 
Astron.  Astrophys., {\bf 367}, 371. 
\bibitem{}
Boccaletti A., P. Riaud, C. Moutou, and A. Labeyrie, 2000, Icarus, {\bf 145},
628.
\bibitem{}
Boden A. et al., 1999, Astrophys. J., {\bf 515}, 356.
\bibitem{}
Bonneau D., and R. Foy, 1980, Astron. Astrophys., {\bf 92}, L1.
\bibitem{}
Bonneau D., and A. Labeyrie, 1973, Astrophys. J, {\bf 181}, L1.
\bibitem{}
Born M., and E. Wolf, 1984, Principles of Optics, Pergamon Press.
\bibitem{}
Bouvier J., F. Rigaut, and D. Nadeau, 1997, Astron. Astrophys., {\bf 323}, 139.
\bibitem{}
Bracewell R. N., 1978, Nature, {\bf 274}, 780.
\bibitem{}
Brandl B., B. J. Sams, F. Bertoldi, A. Eckart, R. Genzel, S. Drapatz, R. 
Hofmann, M. Lowe, and A. Quirrenbach, 1996, Astrophys. J., {\bf 466}, 254.
\bibitem{}
Brandner W., J. Bouvier, E. Grebel, E. Tessier, D. de Winter, and J. L. Beuzit, 
1995, Astron. Astrophys., {\bf 298}, 816.
\bibitem{}
Briggs D. S., 1995, Ph. D. Thesis, New Mexico Institute of Mining and 
Technology.
\bibitem{}
Bruns D., T. Barnett, and D. Sandler, 1997, SPIE., {\bf 2871}, 890.
\bibitem{}
Burge J. H., B. Cuerdem, and J. R. P. Angel, 2000, SPIE, {\bf 4013}, 640.
\bibitem{}
Burns D. et al., 1997, Mon. Not. R. Astron. Soc., {\bf 290}, L11.
\bibitem{}
Burns D. et al., 1998, Mon. Not. R. Astron. Soc., {\bf 297}, 467.
\bibitem{}
Butler R., and G. Marcy, 1996, Astrophys. J, {\bf 464}, L153.
\bibitem{}
Cadot O., Y. Couder, A. Daerr, S. Douady, and A. Tsinocber, 1997, Phys. Rev.,
{\bf 56}, 427. 
\bibitem{}
Callados M., and M. V\`azquez, 1987, Astron. Astrophys., {\bf 180}, 223.
\bibitem{}
Carleton N. et al., 1994, SPIE, {\bf 2200}, 152.
\bibitem{}
Cassinelli J., J. Mathis, and B. Savage, 1981, Science, {\bf 212}, 1497.
\bibitem{}
Charbonneau D., T. Brown, R. Noyes, and R. Gilfiland, 2001, Astrophys. J.,
(to appear).
\bibitem{}
Ciardi D., G. van Belle, R. Akeson, R. Thompson, E. A. Lada, and S. Howell,
2001, Astrophys. J. (to appear).
\bibitem{}
Clampin M., J. Croker, F. Paresce, and M. Rafal, 1988, Rev. Sci. Instru., 
{\bf 59}, 1269.
\bibitem{}
Clampin M., A. Nota, D. A. Golimowski, C. Leitherer, and A. Ferrari, 1993,
Astrophys. J., {\bf 410}, L35.
\bibitem{}
Close L., F. Roddier, J. Hora, J. Graves, M. Northcott, 
C. Roddier, W. Hoffman, A. Doyal, G. Fazio, and L. Deutsch, 1997, 
Astrophys. J., {\bf 489}, 210.
\bibitem{}
Colavita M., 1999, Pub. Astron. Soc. Pac., {\bf 111}, 111.
\bibitem{}
Colavita M. et al., 1998, SPIE, {\bf 3350}, 776.
\bibitem{}
Colavita M. et al., 1999, Astron. J., {\bf 117}, 505.
\bibitem{}
Conan J. -M., L. M. Mugnier, T. Fusco, V. Michau, and G. Rousset, 1998, Appl. 
Opt., {\bf 37}, 4614.
\bibitem{}
Conan R., A. Ziad, J. Borgnino, F. Martin, and A. Tokovinin, 2000, SPIE, 
{\bf 4006}, 963.
\bibitem{}
Cooper D., D. Bui, R. Bailey, L. Kozlowski, and K. Vural, 1993, SPIE, 
{\bf 1946}, 170.
\bibitem{}
Cornwell T. J., and P. N. Wilkinson, 1981, Mon. Not. R. Astron. Soc., {\bf 196},
1067.
\bibitem{}
Coud\'e du Foresto V., G. Perrin, and M. Boccas, 1995, Astron. Astrophys., 
{\bf 293}, 278.
\bibitem{}
Coulman C. E., 1974, Solar Phys., {\bf 34}, 491.
\bibitem{}
Coulman C. E., 1985, Annu. Rev. Astron. Astrophys., {\bf 23}, 19.
\bibitem{}
Currie D., K. Kissel, E. Shaya, P. Avizonis, D. Dowling, and D. Bonnacini, 1996,
The Messenger, no. {\bf 86}, 31.
\bibitem{}
Danchi W. C., M. Bester, C. Degiacomi, I. Greenhill, and C. Townes, 1994, 
Astron. J., {\bf 107}, 1469.
\bibitem{}
Danchi W. C., P. G. Tuthill, and J. D. Monnier, 2001, Astrophys. J. (to appear).
\bibitem{}
Dantowitz R., S. Teare, and M. Kozubal, 2000, Astron. J., {\bf 119}, 2455.
\bibitem{}
Davidge T. J., F. Rigaut, R. Doyon, and D. Crampton, 1997, Astron. J., 
{\bf 113}, 2094.
\bibitem{}
Davidge T. J., D. A. Simons, F. Rigaut, R. Doyon, E. E. Becklin, and D. 
Crampton, 1997, Astron. J., {\bf 114}, 2586.
\bibitem{}
Davis J., and W. J. Tango, 1996, Pub. Astron. Soc. Pac., {\bf 108}, 456.
\bibitem{}
Davis J., W. Tango, A. Booth, and J. O'Byrne, 1998, SPIE, {\bf 3350}, 726.
\bibitem{}
Davis J., W. Tango, A. Booth, T. ten Brummelaar, R. Minard, and S. 
Owens, 1999a, Mon. Not. R. Astron. Soc., {\bf 303}, 773.
\bibitem{}
Davis J., W. J. Tango, A. J. Booth, E. D. Thorvaldson, and J. Giovannis, 1999b
Mon. Not. R. Astron. Soc., {\bf 303}, 783.
\bibitem{}
Dayton D., S. Sandven, J. Gonglewski, S. Rogers, S. McDermott, and S. Browne,
1998, SPIE, {\bf 3353}, 139.
\bibitem{}
Dejonghe J., L. Arnold, O. Lardi\`ere, J. -P Berger, C. Cazal\'e,
S. Dutertre, D. Kohler, and D. Vernet, 1998, SPIE, {\bf 3352}, 603.
\bibitem{}
Denker C., 1998, Solar Phys., {\bf 81}, 108.
\bibitem{}
Derie F., M. Ferrai, E. Brunetto, M. Duchateau, R. Amestica, and P. Aniol,
2000, SPIE, {\bf 4006}, 99.
\bibitem{}
DiBenedetto G. P., and Y. Rabbia, 1987, Astron. Astrophys., {\bf 188}, 114.
\bibitem{}
Diericks P., and R. Gilmozzi, 1999, Proc. `Extremely Large Telescopes', eds.,
T. Andersen, A. Ardeberg, and R. Gilmozzi, 43. 
\bibitem{}
Drummond J., A. Eckart, and E. Hege, 1988, Icarus, {\bf 73}, 1.
\bibitem{}
Dyck H., J. Benson, and S. Ridgway, 1993, Pub. Astron. Soc. Pac, {\bf 105}, 610.
\bibitem{}
Dyck H., G. van Belle, and J. Benson, 1996, Astron. J., {\bf 112}, 294.
\bibitem{}
Ebstein S., N. P. Carleton, and C. Papaliolios, 1989, Astrophys. J, {\bf 336}, 
103.
\bibitem{}
Eke V., 2001, Mon. Not. R. Astron. Soc., {\bf 320}, 106. 
\bibitem{}
Elias N. M., 2001, Astrophys. J., {\bf 549}, 647.
\bibitem{}
Falcke H., K. Davidson, K. -H. Hofmann, and G. Weigelt, 1996, Astron. Astrophys.
{\bf 306}, L17.
\bibitem{}
Faucherre M., D. Bonneau, L. Koechlin, and F. Vakili, 1983, Astron. Astrophys.
{\bf 120}, 263.
\bibitem{}
Fienup J. R., 1978, Opt. Lett., {\bf 3}, 27.
\bibitem{}
Fischer O., B. Stecklum, and C. Leinert, 1998, Astron. Astrophys., {\bf 334}, 
969.
\bibitem{}
Fizeau H., 1868, C. R. Acad. Sci. Paris, {\bf 66}, 934.
\bibitem{}
Fomalont E. B., and M. C. H. Wright, 1974, in Galactic and Extra-galactic
Radio Astronomy, eds., G. L. Verschuur, and K. I. Kellerman, 256. 
\bibitem{}
Foy R., D. Bonneau, and A. Blazit, 1985, Astron. Astrophys., {\bf 149}, L13.
\bibitem{}
Foy R., and A. Labeyrie, 1985, Astron. Astrophys., {\bf 152}, L29.
\bibitem{}
Fried D. L., 1966, J. Opt. Soc. Am., {\bf 56}, 1372.
\bibitem{}
Fugate R. et al., 1994, J. Opt. Soc. Am. A., {\bf 11}, 310.
\bibitem{}
Gauger A., Y. Y. Balega, P. Irrgang, R. Osterbart, and G. Weigelt, 1999,
Astron. Astrophys., {\bf 346}, 505.
\bibitem{}
Gay J., and D. Mekarnia, 1988, Proc. ESO-NOAO conf. ed., F. Merkle, ESO, 
FRG, 811. 
\bibitem{}
Gerchberg R. W., and W. O. Saxton, 1972, Optik, {\bf 35}, 237.
\bibitem{}
Gies R., B. Mason, W. Bagnuolo, M. Haula, W. I. Hartkopf,
H. McAlister, M. Thaller, W. McKibben, and L. Penny, 1997, Astrophys. J.,
{\bf 475}, L49.
\bibitem{}
Glindemann A., 1997, Pub. Astron. Soc. Pac, {\bf 109}, 682.
\bibitem{}
Glindemann A., R. G. Lane, and J. C. Dainty, 1991, Proc. `Digital Signal 
Processing', eds., V. Cappellini \& A. G. Constantinides, 59.
\bibitem{}
Glindemann A., and F. Paresce, 2001, http://www.eso.org/outreach. 
\bibitem{}
Golimowski D. A., T. Nakajima, S. R. Kulkarni, and B. R. Oppenheimer, 1995,
Astrophys. J, {\bf 444}, L101.
\bibitem{}
Gonsalves S. A., 1982, Opt. Eng., {\bf 21}, 829.
\bibitem{}
Goodman J. W., 1975, `Laser Speckle and Related Phenomena', ed., J. C. Dainty,
Springer-Verlag, Berlin, 9.
\bibitem{}
Goodman J. W., 1985, `Statistical Optics', Wiley, N. Y.
\bibitem{}
Gorham P. W., 1998, SPIE, {\bf 3350}, 116.
\bibitem{}
Gorham P., W. Folkner, and G. Blackwood, 1999, ASP conf.,  
{\bf 194}, eds. S. Unwin, and R. Stachnik, ISBN: 1-58381-020-X, 359.
\bibitem{}
Greenwood D. P., 1977, J. Opt. Soc. Am., {\bf 67}, 390.
\bibitem{}
Grieger F., F. Fleischman, and G. Weigelt, 1988, Proc. ESO-NOAO conf.  
ed. F. Merkle, ESO, FRG, 225.
\bibitem{}
Haguenauer P., M. Sevei, I. Schanen-Duport, K. Rousselet-Perraut, J. Berger,
Y. Duch\'ene, M. Lacolle, P. Kern, F. Melbet, and P. Benech, 2000, SPIE,
{\bf 4006}, 1107.
\bibitem{}
Hajian A. et al., 1998, Astrophys. J., {\bf 496}, 484.
\bibitem{}
Hale D., M. Bester, W. Danchi, W. Fitelson, S. Hoss, E. Lipman, J. Monnier,
P. Tuthill, and C. Townes, 2000, Astrohys, J., {\bf 537}, 998.
\bibitem{}
Hale D. et al., 1997, Astrophys. J., {\bf 490}, 826.
\bibitem{}
Hanbury Brown R., 1974, `The Intensity Interferometry, its Applications
to Astronomy', Taylor \& Francis, London.
\bibitem{}
Hanbury Brown R., and R. Twiss, 1958, Proc. Roy. Soc. A, {\bf 248}, 222.
\bibitem{}
Hanbury Brown R., R. C. Jennison, and M. K. Das Gupta, 1952, Nature, {\bf 170}, 
1061.
\bibitem{}
Haniff C., M. Scholz, and P. Tuthill, 1995, Mon. Not. R. Astron. Soc., 
{\bf 276}, 640.
\bibitem{}
Harmanec P. et al., 1996, Astron. Astrophys., {\bf 312}, 879.
\bibitem{}
Hartkopf W. I., H. A. McAlister, and B. D. Mason, 1997, CHARA Contrib. No. 4,
`Third Catalog of Interferometric Measurements of Binary Stars', W.I.
\bibitem{}
Hartley M., B. Mcinnes, and F. Smith, 1981, Q. J. Astr. Soc., {\bf 22}, 
272.
\bibitem{}
Harvey J. W., 1972, Nature, {\bf 235}, 90.
\bibitem{}
Harvey J. W., and J. B. Breckinridge, 1973, Astrophys. J., {\bf 182}, L137.
\bibitem{}
Hawley S. A., and J. S. Miller, 1977, Astrophys. J, {\bf 212}, 94.
\bibitem{}
Hege E., E. Hubbard, P. Strittmatter, and S. Worden, 1981,
Astrophys. J., {\bf 248}, 1.
\bibitem{}
Hestroffer D., 1997, Astron. Astrophys., {\bf 327}, 199.
\bibitem{}
Heydari M., and J. Beuzit, 1994, Astron. Astrophys., {\bf 287}, L17.
\bibitem{}
Hickson P., 2001, Private communication.
\bibitem{}
Hill J. M., 2000, SPIE, {\bf 4004}, 36.
\bibitem{}
Hinz P., R. Angel, W. Hoffmann, D. Mccarthy, P. Mcguire, M. Cheselka, J. Hora,
and N. Woolf, 1998, Nature, {\bf 395}, 251.
\bibitem{}
Hinz P., W. Hoffmann, and J. Hora, 2001, Astrophys. J. Lett. (to appear).
\bibitem{}
The Hipparcos Catalogue, 1997, ESA, SP-1200.
\bibitem{}
Hofmann K. -H., W. Seggewiss, and G. Weigelt, 1995, Astron. Astrophys., {\bf 300}, 403.
\bibitem{}
H\"ogbom J., 1974, Astron. Astrophys. Suppl., {\bf 15}, 417.
\bibitem{}
Hu W., 2001, astro-ph/0105117
\bibitem{}
Hummel C., D. Mozurkevich, J. Armstrong, A. Hajian, N. Elias, and D. Hutter,
1998, Astron. J., {\bf 116}, 2536.
\bibitem{}
Hutchings J., D. Crampton, S. Morris, D. Durand, and E. Steinbring, 
1999, Astron. J., {\bf 117}, 1109.
\bibitem{}
Hutchings J., D. Crampton, S. Morris, and E. Steinbring, 1998, 
Pub. Astron. Soc. Pac, {\bf 110}, 374.
\bibitem{}
Hutchings J., S. Morris, and D. Crampton, 2001, Astron. J., {\bf 121}, 80.
\bibitem{}
Ishimaru A., 1978, `Wave Propagation and Scattering in Random Media', Academic
Press, N. Y.
\bibitem{}
Jefferies S., and J. Christou, 1993, Astrophys. J, {\bf 415}, 862.
\bibitem{}
Jennison R. C., 1958, Mon. Not. R. Astron. Soc., {\bf 118}, 276.
\bibitem{}
Karovska M., L. Koechlin, P. Nisenson, C. Papaliolios, and C. Standley, 1989, 
Astrophys. J, {\bf 340}, 435.
\bibitem{}
Kenworthy M. et al., 2001, Astrophys. J. Lett. (to appear).
\bibitem{}
Kervella P., V. Coud\'e du Foresto, G. Perrin, M. Sch\"oller, W. Traub, and M. 
Lacasse, 2001, Astron. Astrophys., {\bf 367}, 876.
\bibitem{}
Knox K., and B. Thompson, 1974, Astrophys. J, {\bf 193}, L45.
\bibitem{}
Koechlin L., P. R. Lawson, D. Mourard, A. Blazit, D. Bonneau, F. Morand, P. 
Stee, I. Tallon-Bosc, and F. Vakili, 1996, Appl. Opt., {\bf 35}, 3002.
\bibitem{}
Kolmogorov A., 1941a, in `Turbulence', eds., S. K. Friedlander \& 
L. Topper, 1961, Wiley-Interscience, N. Y., 151.
\bibitem{}
Kolmogorov A., 1941b, in `Turbulence', eds., S. K. Friedlander \& 
L. Topper, 1961, Wiley-Interscience, N. Y., 156.
\bibitem{}
Kolmogorov A., 1941c, in `Turbulence', eds., S. K. Friedlander \& 
L. Topper, 1961, Wiley-Interscience, N. Y., 159.
\bibitem{}
Korff D., 1973, J. Opt. Soc. Am., {\bf 63}, 971.
\bibitem{}
Kunz M., A. Banday, P. Castro, P. Ferreira, and K. G\'orski, 2001, Astrophys. J.
Lett., (to appear).
\bibitem{}
Labeyrie A., 1970, Astron. Astrophys., {\bf 6}, 85.
\bibitem{}
Labeyrie A., 1975, Astrophys. J, {\bf 196}, L71.
\bibitem{}
Labeyrie A., 1995, Astron. Astrophys., {\bf 298}, 544.
\bibitem{}
Labeyrie A., 1996, Astron. Astrophys. Suppl., {\bf 118}, 517.
\bibitem{}
Labeyrie A., 1998, SPIE, {\bf 3350}, 960. 
\bibitem{}
Labeyrie A., 1999a, ASP Conf., {\bf 194}, eds. S. Unwin, and R. Stachnik, 350.
\bibitem{}
Labeyrie A., 1999b, Science, {\bf 285}, 1864.
\bibitem{}
Labeyrie A., 2000, Private communication. 
\bibitem{}
Labeyrie A., 2001, Private communication. 
\bibitem{}
Labeyrie A., L. Koechlin, D. Bonneau, A. Blazit, and R. Foy, 1977, Astrophys. 
J., {\bf 218}, L75.
\bibitem{}
Labeyrie A., G. Lamaitre, and L. Koechlin, 1986, SPIE, {\bf 628}, 323.
\bibitem{}
Labeyrie A., G. Schumacher, M. Dugu\'e, C. Thom, P. Bourlon, F. Foy,
D. Bonneau, and R. Foy, 1986, Astron. Astrophys., {\bf 162}, 359.
\bibitem{}
Lai O., D. Rouan, F. Rigaut, R. Arsenault, and E. Gendron, 1998, Astron.
Astrophys., {\bf 334}, 783.
\bibitem{}
Lai O., D. Rouan, F. Rigaut, F. Doyon, and F. Lacombe, 1999, Astron. Astrophys.,
{\bf 351}, 834.
\bibitem{}
Lane B., M. Kuchner, A. Boden, M. Crooch-Eakman, and S. R. Kulkarni, 2000,
Nature, {\bf 407}, 485.
\bibitem{}
Lannes A., E. Anterrieu, and K. Bouyoucef, 1994, J. Mod. Opt., {\bf 41}, 1537.
\bibitem{}
Lawson P. R., 1994, Pub. Astron. Soc. Pac., {\bf 106}, 917.
\bibitem{}
Lawson P. R., 1995, J. Opt. Soc. Am A., {\bf 12}, 306.
\bibitem{}
Lawson P., J. Baldwin, P. Warner, R. Boysen, C. Haniff, J. Rogers, D.
Saint-Jacques, D. Wilson, and J. Young, 1998, SPIE, {\bf 3350}, 753. 
\bibitem{}
Lawson P., M. Colavita, P. Dumont, and B. Lane, 2000, SPIE, {\bf 4006}, 397.
\bibitem{}
Ledoux C., B. Th\'eodore, P. Petitjean, M. N. Bremer, G. F. Lewis, R. A. Ibata,
M. J. Irwin, and E. J. Totten, 1998, Astron. Astrophys., {\bf 339}, L77. 
\bibitem{}
Lee J., B. Bigelow, D. Walker, A. Doel, and R. Bingham, 2000, Pub. Astron. Soc. 
Pac, {\bf 112}, 97.
\bibitem{}
Lef\`evre H. C., 1980, Electron. Lett., {\bf 16}, 778.
\bibitem{}
L\'eger A., M. Pirre, and F. J. Marceau, 1993, Astron. Astrophys., {\bf 277},
309.
\bibitem{}
Liang J., D. R. Williams, and D. T. Miller, 1997, J. Opt. Soc. Am. A, {\bf 14}, 
2884.
\bibitem{}
Lindengren L., and M. A. C. Perryman, 1996, Astron. Astrophys. Suppl., 
{\bf 116}, 579.
\bibitem{}
Linfield R., and P. Gorham, 1999, ASP Conf., 
{\bf 194}, eds. S. Unwin, and R. Stachnik, ISBN: 1-58381-020-X, 224.
\bibitem{}
Lloyd-Hart M., 2000, Pub. Astron. Soc. Pac, {\bf 112}, 264.
\bibitem{}
Lloyd-Hart M., J. R. Angel, T. Groesbeck, T. Martinez, B. Jacobsen,
B. McLeod, D. McCarthy, E. Hooper, E. Hege, and D. Sandler, 1998, Astrophys. J.,
{\bf 493}, 950.
\bibitem{}
Lipman E., M. Bester, W. Danchi, and C. Townes, 1998, SPIE, {\bf 3350}, 933. 
\bibitem{}
Liu Y. C., and A. W. Lohmann, 1973, Opt. Comm., {\bf 8}, 372.
\bibitem{}
Lohmann A., G. Weigelt, and B. Wirnitzer, 1983, Appl. Opt., {\bf 22}, 4028.
\bibitem{}
Lopez B. et al., 1997, Astrophys. J., {\bf 488}, 807.
\bibitem{}
Lopez B., 1991, `Last Mission at La Silla, April 19 $-$ May 8, on the Measure
of the Wavefront Evolution Velocity', ESO Internal Report.
\bibitem{}
Lovelock J. E., 1965, Nature, {\bf 207}, 568.
\bibitem{}
Lucy L., 1974, Astron. J., {\bf 79}, 745.
\bibitem{}
Lynds C., S. Worden, and J. Harvey, 1976, Astrophys. J, {\bf 207}, 174.
\bibitem{}
Machida Y. et al., 1998, SPIE, {\bf 3350}, 202.
\bibitem{}
Macintosh B., C. Max, B. Zuckerman, E. Becklin, D. Kaisler, P. Lowrence, A. 
Weinberger, J. Christou, G. Schneider, and S. Acton, 2001, astro-ph/0106479.
\bibitem{}
Mackay C. D., R. N. Tubbs, R. Bell, D. Burt, P. Jerram, and I. Moody, 2001,
SPIE, {\bf 4306}, (in press).
\bibitem{}
Magain P., F. Courbin, and S. Sohy, 1998, Astrophys. J., {\bf 494}, 472.
\bibitem{}
Malbet F. et al., 1998, Astrophys. J., {\bf 507}, L149.
\bibitem{}
Marco O., T. Encrenaz, and E. Gendron, 1997, Planet Sp. Sci., {\bf 46}, 547.
\bibitem{}
M\'arquez I., P. Petitjean, B. Th\'eodore, M. Bremer, G. Monnet, and J.
Beuzit, 2001, Astron. Astrophys., {\bf 371}, 97.
\bibitem{}
Masciadri E., J. Vernin, and P. Bougeault, 1999, Astron. Astrophys. Suppl., 
{\bf 137}, 203.
\bibitem{}
Mason B. D., 1996, Astron. J., {\bf 112}, 2260.
\bibitem{}
Mason B., C. Martin, W. I. Hartkopf, D. Barry, M. Germain, G. 
Douglass, C. Worley, G. Wycoff, T. ten Brummelaar, and O. Franz, 1999, 
Astron. J., {\bf 117}, 1890.
\bibitem{}
Mayor M., and D. Queloz, 1995, Nature, {\bf 378}, 355.
\bibitem{}
McAlister H., W. Bagnuolo, T. ten Brummelaar, W. I. Hartkopf, M. Shure,
L. Sturmann, N. Turner, and S. Ridgway, 1998, SPIE, {\bf 3350}, 947.
\bibitem{}
Mendel L., and E. Wolf, 1995, `Optical Coherence and Quantum Optics', 
Cambridge University Press, Cambridge.
\bibitem{}
Mennesson B., J. -M. Mariotti, V. Coud\'e du Foresto, G. Perrin, S. Ridgway,
C. Ruilier, W. Traub, M. Lacasse, and G. Maz\'e, 1999, Astron. Astrophys.,
{\bf 346}, 181.
\bibitem{}
Men'shchikov A., and T. Henning, 1997, Astron. Astrophys., {\bf 318}, 879.
\bibitem{}
Michelson A. A., 1891, Nature, {\bf 45}, 160.
\bibitem{}
Michelson A., and F. Pease, 1921, Astrophys. J, {\bf 53}, 249.
\bibitem{}
Millan-Gabet R., P. Schloerb, and W. Traub, 2001, Astrophys. J., {\bf 546},
358.
\bibitem{}
Monnier J., W. Danchi, D. Hale, P. Tuthill, and C. Townes,
2000, Astrophys. J, {\bf 543}, 868.
\bibitem{}
Monnier J., et al., 2001, AAS Meeting, {\bf 198}, 63.02.
\bibitem{}
Monnier J., P. Tuthill, B. Lopez, P. Cruzal\'ebes, W. Danchi, C. 
Haniff, 1999, Astrophys. J, {\bf 512}, 351.
\bibitem{}
Morel S., 2000, Private communication.
\bibitem{}
Morel S., and L. Koechlin, 1998, SPIE, {\bf 3350}, 257.
\bibitem{}
Morel S., W. Traub, J. Bregman, R. Mah, and C. Wilson, 2000, SPIE, {\bf 4006},
506.
\bibitem{}
Mouillet D., J. D. Larwood, J. C. Papaloizou, and A. M. Lagrange, 1997, Mon. 
Not. R. Astron. Soc., {\bf 292}, 896.
\bibitem{}
Mourard D., D. Bonneau, L. Koechlin, A. Labeyrie, F. Morand, P. Stee,
I. Tallon-Bosc, and F. Vakili, 1997, Astron. Astrophys., {\bf 317}, 789.
\bibitem{}
Mourard D., I. Bosc, A. Labeyrie, L. Koechlin, and S. Saha, 1989, Nature, 
{\bf 342}, 520.
\bibitem{}
Mozurkewich D., K. Johnston, R. Simon, D. Hutter,
M. Colavita, M. Shao, and X. Pan, 1991, Astron. J., {\bf 101}, 2207.
\bibitem{}
Nakajima T., 1994, Astrophys J., {\bf 425}, 348.
\bibitem{}
Nakajima T., and D. Golimowski, 1995, Astron. J., {\bf 109}, 1181.
\bibitem{}
Natta A., T. Prusti, R. Neri, D. Wooden, and V. Grinin, 2001, Astron. 
Astrophys., {\bf 371}, 186.
\bibitem{}
Nelkin M., 2000, Am. J. Phys., {\bf 68}, 310.
\bibitem{}
Nisenson P., and C. Papaliolios, 1999, Astrophys. J, {\bf 518}, L29.
\bibitem{}
Nisenson P., C. Papaliolios, M. Karovska, and R. Noyes, 1987, Astrophys. J, 
{\bf 320}, L15.
\bibitem{}
Nordgren T., J. Armstrong, M. Gierman, R. Hindsley, A. Hajian, J. Sudol, and
C. Hummel, 2000, Astrophys. J., {\bf 543}, 972.
\bibitem{}
Northcott M. J., G. R. Ayers, and J. C. Dainty, 1988, J. Opt. Soc. Am. A, 
{\bf 5}, 986.
\bibitem{}
Nota A., C. Leitherer, M. Clampin, P. Greenfield, and D. A. Golimowski, 1992, 
Astron. J., {\bf 398}, 621.
\bibitem{}
Nulsen P., P. Wood, P. Gillingham, M. Bessel, M. Dopita,
and C. McCowage, 1990, Astrophys. J, {\bf 358}, 266.
\bibitem{}
Osterbart R., Y. Balega, G. Weigelt, and N. Langer, 1996, Proc. IAU symp. 180,
eds., H. Habing \& G. Lamers, Kluwer Academic Pub. Netherlands, 362.
\bibitem{}
Padilla C., V. Karlov, L. Matson, K. Soosaar, and T. ten Brummelaar, 1998,
SPIE, {\bf 3350}, 1045.
\bibitem{}
Papaliolios C., M. Karovska, L. Koechlin, P. Nisenson, C. Standley, and S. 
Heathcote, 1989, Nature, {\bf 338}, 565.
\bibitem{}
Papaliolios C., P. Nisenson, and S. Ebstein, 1985, Appl. Opt., {\bf 24}, 287.
\bibitem{}
Pauls T., D. Mozurkewich, J. Armstrong, C. Hummel, J. Benson,
and A. Hajian, 1998, SPIE, {\bf 3350}, 467.
\bibitem{}
Paxman R., T. Schulz, and J. Fienup, 1992, J. Opt. Soc. Am., {\bf 9}, 1072.
\bibitem{}
Paxman R., J. Seldin, M. L\"ofdahl, G. Scharmer, and C. Keller, 1996, Astrophys.
J., {\bf 466}, 1087.
\bibitem{}
Pedretti E., and A. Labeyrie, 1999, Astron. Astrophys. Suppl., {\bf 137}, 543.
\bibitem{}
Pedretti E., A. Labeyrie, L. Arnold, N. Thureau, O. Lardi\'ere, A. Boccaletti,
and P. Riaud, 2000, Astron. Astrophys. Suppl., {\bf 147}, 285.
\bibitem{}
Pehlemann E., K. -H Hofmann, and G. Weigelt, 1992, Astron. Astrophys., 
{\bf 256}, 701.
\bibitem{}
Penny A., A. L\'eger, J. Mariotti, C. Schalinski, C. Eiora, R. Laurance, M.
Fridlund, 1998, SPIE, {\bf 3350}, 666.
\bibitem{}
Perrin G., 1997, Astron. Astrophys. Suppl., {\bf 121}, 553.
\bibitem{}
Perrin G., V. Coud\'e du Foresto, S. Ridgway, J. -M. Marrioti, W. Traub,
N. Carleton, M. Lacasse, 1998, Astron. Astrophys., {\bf 331}, 619.
\bibitem{}
Perrin G., V. Coud\'e du Foresto, S. Ridgway, B. Mennesson, C. Ruilier, J. -M. 
Marrioti, W. Traub, and M. Lacasse, 1999, Astron. Astrophys., {\bf 345}, 221.
\bibitem{}
Perryman M. A. C., 1998, Nature, {\bf 340}, 111.
\bibitem{}
Poulet F., and B. Sicardy, 1996, Bull. Astr. Am. Soc., {\bf 28}, 1124.
\bibitem{}
Pourbaix D., 2000, Astron. Astrophys. Suppl., {\bf 145}, 215.
\bibitem{}
Prieur J., E. Oblak, P. Lampens, M. Kurpinska-Winiarska, E. Aristidi, L. 
Koechlin, and G. Ruymaekers, 2001, Astron. Astrophys., {\bf 367}, 865. 
\bibitem{}
Puetter R., and A. Yahil, 1999, astro-ph/9901063.
\bibitem{}
Quirrenbach A., 2001, Annu. Rev. Astron. Astrophys., {\bf 39}, 353. 
\bibitem{}
Quirrenbach A., D. Mozurkewich, D. Buscher, C. Hummel, and J. Armstrong, 1996,
Astron. Astrophys., {\bf 312}, 160. 
\bibitem{}
Quirrenbach A., J. Roberts, K. Fidkowski, W. de Vries, and W. van Breugel,
2001, Astrophys. J. (to appear). 
\bibitem{}
Rabbia Y., D. Mekarnia, and J. Gay, 1990, SPIE, {\bf 1341}, 172.
\bibitem{}
Racine R., 1984, IAU Colloq. 79, eds., M. Ulrich \& Kj\"ar, 235.
\bibitem{}
Racine R., G. Herriot, and R. McClure, 1996, Proc. `Adaptive Optics' ed., M. 
Cullum, ESO, Germany, 35.
\bibitem{}
Ragazzoni R., and D. Bonaccini, 1996, Proc. `Adaptive Optics', ed., M. Cullum, 
17.
\bibitem{}
Ragazzoni R., E. Marchetti, and G. Valente, 2000, Nature, {\bf 403}, 54.
\bibitem{}
Reinheimer T., and G. Weigelt, 1987, Astron. Astrophys., {\bf 176}, L17.
\bibitem{}
Richardson W. H., 1972, J. Opt. Soc. Am., {\bf 62}, 55.
\bibitem{}
Ridgway S. T., and F. Roddier, 2000, SPIE, {\bf 4006}, 940.
\bibitem{}
Rimmele T. R., 2000, SPIE, {\bf 4007}, 218.
\bibitem{}
Robbe S., B. Sorrente, F. Cassaing, Y. Rabbia, and G. Rousset, 1997, Astron. 
Astrophys. Suppl. {\bf 125}, 367.
\bibitem{}
Robertson N. A., 2000, Class. Quantum. Grav., {\bf 17}, 19.
\bibitem{}
Roddier C., and F. Roddier, 1988, Proc. NATO-ASI workshop,  
eds., D. M. Alloin \& J. -M. Mariotti, 221.
\bibitem{}
Roddier C., F. Roddier, M. J. Northcott, J. E. Graves, and K. Jim, 1996, 
Astrophys. J, {\bf 463}, 326.
\bibitem{}
Roddier F., 1981, Progress in Optics, {\bf XIX}, 281.
\bibitem{}
Roddier F., 1988, Phys. Rep., {\bf 170}, 97.
\bibitem{}
Roddier F., 1999, `Adaptive Optics in Astronomy', ed., F. Roddier, Cambridge
Univ. Press.
\bibitem{}
Roddier F., C. Roddier, A. Brahic, C. Dumas, J. Graves, M. Northcott, 
and T. Owen, 1997, Planet Sp. Sci., {\bf 45}, 1031.
\bibitem{}
Roddier F., C. Roddier, J. E. Graves, and M. J. Northcott, 1995, Astrophys. J, 
{\bf 443}, 249.
\bibitem{}
Roggemann M. C., B. M. Welsh, and R. Q. Fugate, 1997, Rev. Modern Phys., 
{\bf 69}, 437.
\bibitem{}
Rouan D., D. Field, J. -L. Lemaire, O. Lai, G. P. For\"ets, E. Falgarone, 
and J. -M. Deltorn, 1997, Mon. Not. R. Astron. Soc., {\bf 284}, 395.
\bibitem{}
Rouan D., P. Riaud, A. Boccaletti, Y. Cl\'enet, and A. Labeyrie, 2000,
Pub. Astron. Soc. Pac., {\bf 112}, 1479.
\bibitem{}
Rouan D., F. Rigaut, D. Alloin, R. Doyon, O. Lai, D. Crampton, E. Gendron,
and R. Arsenault, 1998, Astron. Astrophys., {\bf 339}, 687.
\bibitem{}
Rousset G., 1999, `Adaptive Optics in Astronomy', ed., F. Roddier, Cambridge
Univ. Press, 91.
\bibitem{}
Rousset G., J. C. Fontanella, P. Kem, P. Gigan, F. Rigaut, P. L\'ena, P. Boyer, 
P. Jagourel, J. P. Gaffard, and F. Merkle, 1990, Astron. Astrophys., {\bf 230}, 
L29.
\bibitem{}
Rousselet-Perraut K., F. Vakili, and D. Mourard, 1996, Opt. Eng., {\bf 35}, 
2943.
\bibitem{}
Ryan S., and P. Wood, 1995, Pub. Astron. Soc. Austr., {\bf 12}, 89.
\bibitem{}
Saha S. K., 1999, Bull. Astron. Soc. Ind., {\bf 27}, 443.
\bibitem{}
Saha S. K., and V. Chinnappan, 1999, Bull. Astron. Soc. Ind., {\bf 27}, 327.
\bibitem{}
Saha S. K., and D. Maitra, 2001, Ind. J. Phys., {\bf 75B}, 391.
\bibitem{}
Saha S. K., and S. Morel, 2000, Bull. Astron. Soc. Ind., {\bf 28}, 175.
\bibitem{}
Saha S. K., R. Rajamohan, P. Vivekananda Rao, G. Som Sunder, R. Swaminathan,
and B. Lokanadham, 1997, Bull. Astron. Soc. Ind., {\bf 25}, 563.
\bibitem{}
Saha S. K., R. Sridharan, and K. Sankarasubramanian, 1999, `Speckle image 
reconstruction of binary stars', Presented at XIX ASI meeting, Bangalore.
\bibitem{}
Saha S. K., and P. Venkatakrishnan, 1997, Bull. Astron. Soc. Ind., {\bf 25}, 
329.
\bibitem{}
Sams B. J., K. Schuster, and B. Brandl, 1996, Astrophys. J., {\bf 459}, 491.
\bibitem{}
Sato K. et al., 1998, SPIE, {\bf 3350}, 212.
\bibitem{}
Schertl D., K. -H. Hofmann, W. Seggewiss, and G. Weigelt, 1996, Astron. 
Astrophys., {\bf 302}, 327.
\bibitem{}
Sch\"oller M., W. Brandner, T. Lehmann, G. Weigelt, and H. Zinnecker, 1996, 
Astron. Astrophys., {\bf 315}, 445.
\bibitem{}
Seldin J., R. Paxman, and C. Keller, 1996, SPIE., {\bf 2804}, 166.
\bibitem{}
Seldin J., and R. Paxman, 1994, SPIE., {\bf 2302}, 268.
\bibitem{}
Serabyn E., 2000, SPIE, {\bf 4006}, 328.
\bibitem{}
Shannon C. J., 1949, Proc. IRE, {\bf 37}, 10.
\bibitem{}
Shao M., and M. M. Colavita, 1992, Astron. Astrophys., {\bf 262}, 353.
\bibitem{}
Shao M., and M. M. Colavita, 1994, Proc. IAU Symp. 158,  
eds., J. G. Robertson and W. J. Tango, 413.
\bibitem{}
Shao M., and D. Staelin, 1977, J. Opt. Soc. Am., {\bf 67}, 81.
\bibitem{}
Shao M. et al., 1988, Astron. Astrophys., {\bf 193}, 357.
\bibitem{}
Sicardy B., F. Roddier, C. Roddier, E. Perozzi, J. E. Graves, O. Guyon,
and M. J. Northcott, 1999, Nature, {\bf 400}, 731.
\bibitem{}
Simon M., L. Close, and T. Beck, 1999, Astron. J., {\bf 117}, 1375.
\bibitem{}
Stee P., de Ara\'ujo, F. Vakili, D. Mourard, I. Arnold, D. Bonneau, F. Morand,
and I. Tallon-Bosc, 1995, Astron. Astrophys., {\bf 300}, 219.
\bibitem{}
Stee P., F. Vakili, D. Bonneau, and D. Mourard, 1998, Astron. Astrophys., 
{\bf 332}, 268.
\bibitem{}
Tallon M., R. Foy, and A. Blazit, 1988, Proc. ESO Conf. ed., M. -H. Ulrich,
ESO, FRG, 743.
\bibitem{}
Tango W. J., and R. Q. Twiss, 1980, Prog. Opt., {\bf 17}, 239.
\bibitem{}
Tatarski V. I., 1967, `Wave Propagation in a Turbulent Medium', Dover, New York.
\bibitem{}
Tatarski V. I., 1993, J. Opt. Soc. Am. A, {\bf 56}, 1380.
\bibitem{}
Taylor G. L., 1921, in `Turbulence', eds., S. K. Friedlander \&
L. Topper, 1961, Wiley-Interscience, New York, 1.
\bibitem{}
Thom C., P. Granes, and F. Vakili, 1986, Astron. Astrophys., {\bf 165}, L13.
\bibitem{}
Thompson L. A., and C. S. Gardner, 1988, Nature, {\bf 328}, 229.
\bibitem{}
Timothy J. G., 1993, SPIE., {\bf 1982}, 4.
\bibitem{}
Torres G., R. Stefanik, and D. Latham, 1997, Astrophys. J, {\bf 485}, 167.
\bibitem{}
Townes C. H., M. Bester, W. Danchi, D. Hale, J. Monnier, E. Lipman,
A. Everett, P. Tuthill, M. Johnson, and D. Walters, 1998, SPIE, 
{\bf 3350}, 908.
\bibitem{}
Traub W. A., 1986, Appl. Opt., {\bf 25}, 528. 
\bibitem{}
Traub W. A., 2000, Course notes on `Principles of long baseline 
interferometry', ed., P. R. Lawson, 31.
\bibitem{}
Traub W. A. et al., 2000, SPIE, {\bf 4006}, 715.
\bibitem{}
Traub W. A., R. Millam-Gabet, and M. Garcia, 1998, Bull. Astr. Am. Soc., 
{\bf 193}, 52.06.
\bibitem{}
Troxel S. E., B. M. Welsh, and M. C. Roggemann, 1994, J. Opt. Soc. Am. A, 
{\bf 11}, 2100.
\bibitem{}
Tuthill P., C. Haniff, and J. Baldwin, 1997, Mon. Not. R. Astron. Soc., 
{\bf 285}, 529.
\bibitem{}
Tuthill P., C. Haniff, and J. Baldwin, 1999, Mon. Not. R. Astron. Soc., 
{\bf 306}, 353.
\bibitem{}
Tuthill P. G., J. D. Monnier, and W. C. Danchi, 1999, Nature, {\bf 398}, 487.
\bibitem{}
Tuthill P. G., J. D. Monnier, and W. C. Danchi, 2001, Nature, {\bf 409}, 1012.
\bibitem{}
Tuthill P. G., J. D. Monnier, W. C. Danchi, and Wishnow, 2000, 
Pub. Astron. Soc. Pac., {\bf 116}, 2536.
\bibitem{}
Ulrich M., L. Maraschi, and C. Urry, 1997, Annu. Rev. Astron. Astrophys., 
{\bf 35}, 445.
\bibitem{}
Unwin S., S. Turyshev, and M. Shao, 1998, SPIE, {\bf 3350}, 551.
\bibitem{}
Vakili F., D. Mourard, D. Bonneau, F. Morand, and P. Stee, 1997, Astron. 
Astrophys., {\bf 323}, 183.
\bibitem{}
Vakili F., D. Mourard, P. Stee, D. Bonneau, P. Berio, O. Chesneau, N. Thureau,
F. Morand, A. Labeyrie, and I. Tallon-Bosc, 1998, Astron. Astrophys., {\bf 335}, 261.
\bibitem{}
van Belle G. et al., 1999, Astron. J., {\bf 117}, 521.
\bibitem{}
van Belle G., D. Ciardi, R. Thompson, R. Akeson, and E. A. Lada,
2001, astro-ph/0106184.
\bibitem{}
van Belle G., H. Dyck, J. Benson, and M. Lacasse, 1996, Astron. J., {\bf 112}, 
2147.
\bibitem{}
van Belle G., H. Dyck, R. Thomson, J. Benson, and S. Kannappan, 1997, 
Astron. J., {\bf 114}, 2150.
\bibitem{}
Von der L\"uhe O., 1984, J. Opt. Soc. Am. A, {\bf 1}, 510. 
\bibitem{}
Walkup J. F., and J. W. Goodman, 1973, J. Opt. Soc. Am, {\bf 63}, 399.
\bibitem{}
Wallace J. et al., 1998, SPIE, {\bf 3350}, 864.
\bibitem{}
Wehinger P. A., 2001, Private communication.
\bibitem{}
Weigelt G., 1977, Opt. Communication, {\bf 21}, 55.
\bibitem{}
Weigelt G., and G. Bair, 1985, Astron. Astrophys., {\bf 150}, L18.
\bibitem{}
Weigelt G., Y. Balega, T. Bl\"ocker, A. Fleischer, R. Osterbart, and J. 
Winters, 1998, Astron. Astrophys., {\bf 333}, L51.
\bibitem{}
Weigelt G., Y. Balega, K. -H. Hofmann, and M. Scholz, 1996, Astron. Astrophys.,
{\bf 316}, L21.
\bibitem{}
Weigelt G., and J. Ebersberger, 1986, Astron. Astrophys., {\bf 163}, L5.
\bibitem{}
Weinberger A., G. Neugebauer, and K. Matthews, 1999, Astron. J., {\bf 117}, 
2748.
\bibitem{}
Wilken V., C. R. de Boer, C. Denker, and F. Kneer, 1997, Astron. Astrophys.,
{\bf 325}, 819.
\bibitem{}
Wilson R., J. Baldwin, D. Busher, and P. Warner, 1992, Mon. Not. R. Astron. 
Soc., {\bf 257}, 369.
\bibitem{}
Wittkowski M., Y. Balega, T. Beckert, W. Duschi, K. Hofmann, and G. Weigelt,
1998, Astron. Astrophys., {\bf 329}, L45.
\bibitem{}
Wittkowski M., Y. Balega, K. Hofmann, and G. Weigelt, 1999, Mitteilungen der
Astronomischen Gesellschaft (AGM)., {\bf 15}, 83.
\bibitem{}
Wittkowski M., C. Hummel, K. Johnston, D. Mozurkewich, A. Hajian, and N. White,
2001, Astron. Astrophys., {\bf 377}, 981. 
\bibitem{}
Wittkowski M., N. Langer, and G. Weigelt, 1998, Astron. Astrophys., {\bf 340}, 
L39.
\bibitem{}
Wood P., M. Bessel, and M. Dopita, 1986, Astrophys. J, {\bf 311}, 632.
\bibitem{}
Wood P., P. Nulsen, P. Gillingham, M. Bessel, M. Dopita, 
and C. McCowage, 1989, Astrophys. J, {\bf 339}, 1073.
\bibitem{}
Worden S. P., C. R. Lynds, and J. W. Harvey, 1976, J. Opt. Soc. Am., {\bf 66}, 
1243.
\bibitem{}
Wyngaard J. C., Y. Izumi, and S. A. Collins, 1971, J. Opt. Soc. Am., {\bf 60}, 
1495.
\bibitem{}
Young A. T., 1974, Astrophys. J., {\bf 189}, 587.
\bibitem{}
Young J. et al., 2000, Mon. Not. R. Astron. Soc., {\bf 315}, 635.
\bibitem{}
Zago L., 1995, http://www.eso.org/gen-fac/pubs/astclim/lz-thesis/node4-html.
\end{references}
\end{document}